\documentstyle[aps,eqsecnum]{revtex}
\begin{document}
\draft
\title
{
Two-dimensional conformal field theory for disordered systems at criticality
}
\author{Christopher Mudry, Claudio Chamon, and Xiao-Gang Wen}
\address{Department of Physics, Massachusetts Institute of Technology,
77 Massachusetts Avenue, Cambridge, Massachusetts 02139}
\date{\today}
\maketitle

\begin{abstract}
Using a Kac-Moody current algebra with $U(1/1)\times U(1/1)$ graded
symmetry, we describe a class of (possibly disordered) critical points in
two spatial dimensions. The critical points are labelled by
the triplets $(l,m,k^{\ }_j)$, where $l$ is an odd integer,
$m$ is an integer, and $k^{\ }_j$ is real.
For most such critical points,
we show that there are infinite hierarchies of relevant operators
with negative scaling dimensions. To interpret this result, we show
that the line of critical points $(1,1,k^{\ }_j>0)$ is realized by
a field theory of massless Dirac fermions in the presence of $U(N)$
vector gauge-like static impurities.
Along the disordered critical line $(1,1,k^{\ }_j>0)$, we find
an infinite hierarchy of relevant operators with negative scaling dimensions
$\{\Delta^{\ }_q|q\in {\rm I}\hskip -0.08 true cm{\bf N}\}$,
which are related to the disorder average over
the $q$-th moment of the single-particle Green function.
Those relevant operators
can be induced by non-Gaussian moments of the probability
distribution of a mass-like static disorder.

\end{abstract}



\section{Introduction}
\label{sec:Introduction}

The plateau transition in the integer quantum Hall effect perhaps provides
the most compelling experimental evidence in support of the scaling
hypothesis in the theory of localization close to the mobility edge.
Scaling with temperature, sample size, as well as
dynamical scaling have been observed in two-dimensional
degenerate electronic systems subject to a strong magnetic field
\cite{Huckestein 1995}.
(How strong the magnetic field must be is controlled by the ratio
of the magnetic length to the mean free path of the system.)
Assuming that there exists one and only one relevant diverging length scale
$\xi$ at the transition, the experimentally observed scaling is consistent
with a power law divergence of $\xi$ on the energy separation {}from the
center of the Landau band (the mobility edge) with scaling exponent
$\nu^{\ }_{\rm exp}\approx{7\over3}$ \cite{Huckestein 1995}.
The same exponent has also
been measured in numerical simulations
of the plateau transition for two-dimensional {\it non-interacting} electronic
system in the presence of a static (quenched) weakly correlated disorder
and a strong magnetic field \cite{Huckestein 1995}.
It is tempting to interpret
these experimental and numerical observations as
evidences for the existence of a quantum critical point for a disordered
non-interacting two-dimensional system at which, as a
function of a strong magnetic field, a zero temperature continuous phase
transition between two insulating quantum Hall states takes place.
We take this point of view in this paper.

The present understanding of the plateau transition {}from a theoretical point
of view is quite unsatisfactory for two basic reasons.
First, it has not been possible to construct a field theory
describing a second order phase transition governed by only one diverging
length scale with scaling exponent $\nu^{\ }_{\rm ft}={7\over3}$.
On the one hand, a semi-classical
picture based on the continuum model of percolation yields, in the
infinite strong magnetic field limit, the critical exponent
$\nu^{\ }_{\rm per}={4\over3}$ \cite{Trugman 1983}.
Although the classical continuum model of
percolation does not reproduce the correct scaling exponent for the plateau
transition, it inspired Chalker and Coddington to propose a network model
incorporating in a simplified way quantum effects such as tunneling
and quantum interferences \cite{Chalker 1988}.
Numerical studies of the Chalker-Coddington (CK)
network model suggest that the physics essential to the understanding of
the plateau transition has been incorporated in it, since, among others, they
predict the scaling exponent $\nu^{\ }_{\rm CK}\approx{7\over3}$
for the localization length in agreement
with ``first-principle'' numerical simulations \cite{Chalker 1988}.
On the other hand, motivated by the mapping of the problem of the
Anderson metal-insulator transition onto an effective non-linear
$\sigma$ model,
Pruisken and collaborators proposed
a non-linear $\sigma$ model with the addition of a new topological term
to describe criticality at the plateau transition
\cite{Levine 1983}.
This approach has been partially successful in that it predicts
the existence of critical points describing transitions between
some insulating phases. However, it is not clear whether those
critical points correspond to the plateau transition. Moreover,
it has not been possible to calculate their critical properties.
Further attempts to identify
the effective field theory describing the critical properties of
the plateau transition have consisted in approximating the replicated
(supersymmetric) non-linear $\sigma$ model with topological term by a
generalized replicated (supersymmetric) spin chain
\cite{Affleck 1985}
($\!$\cite{Zirnbauer 1994}),
or in approximating the CK network model by a replicated Hubbard chain
\cite{dhLee 1995}.
Unfortunately, all these effective field theories have remained quite
intractable to analytical methods.
The intrinsic difficulty in calculating the critical indices of
the plateau transition is that this transition
is between two insulating phases. This should
be contrasted with the conventional Anderson metal-insulator transition
for which critical properties can be calculated perturbatively
{}from the metallic regime \cite{Wegner 1976,Abrahams 1979}.
All attempts to calculate
the critical properties of an effective field theory presumed to describe
the plateau transition are of a non-perturbative nature and have
thus far failed.

Second, a highly non-trivial property of the critical field theory
describing the plateau transition is that its operator content must be
consistent with the observed phenomenon of multifractality
\cite{Janssen 1994}.
Multifractality, in the context of the plateau transition, means that
infinitely many universal exponents, which are not linearly related,
are needed to characterize
the scaling of the generalized inverse participation ratios
at criticality.
\footnote
{
This is to be contrasted with gap scaling in usual critical phenomena.
An example of gap scaling occurs in the
two-dimensional Ising model. If the scaling of
the susceptibility for the two-dimensional Ising model
is controlled by the exponent $\eta$, then the scaling
of the $n$-th power of the susceptibility is controlled by the
exponent $n\eta$.
\label{footnoteongapscaling}
}
It reflects the fact that the inverse participation ratio
is not a selfaveraging observable, whereas its logarithm is.
Duplantier and Ludwig have shown how difficult it is to find
{}from a stable Lagrangian field theory
a critical point characterized by
an infinite hierarchy of operators related to the phenomenon of
multifractality \cite{Duplantier 1991}.
There are some perturbative calculations
in $2+\epsilon$ dimensions and asymptotically close to
the mobility edge of
{\it disorder averaged} generalized inverse participation ratios
\cite{Wegner 1980,Wegner 1985,Wegner 1987}.
They are interpreted as evidences for multifractality
\cite{Castellani 1986}.
It would be highly desirable to obtain {\it non-perturbatively} a hierarchy
of scaling exponents for disorder
averaged generalized inverse participation ratios,
and to show how they are related to the universal
scaling exponents observed numerically {}from the scaling with system size
of the generalized inverse participation ratios for {\it fixed}
realizations of the disorder.

It was recently shown that a periodic potential can
induce a continuous transition between two quantum Hall states
or {}from a quantum Hall state to a Mott insulator
in an anyonic system (at zero temperature) subject to a magnetic field
\cite{Wen 1993,Chen 1993}.
It is then natural to probe the fate of such a (pure) quantum critical
point upon perturbation by some weak disorder.
In fact, such a procedure was used by Ludwig in a
different context to show the existence of a new random fixed point
for the two-dimensional $q$-state Potts model ($q\geq2$) with weak quenched
bond randomness \cite{Ludwig 1987}.
Ludwig also showed that the random fixed point
for the two-dimensional $q$-state Potts model is characterized by an
infinite hierarchy of relevant operators implying a complex scaling
behavior related to multifractality \cite{Ludwig 1990}.
A convenient model undergoing an integer quantum Hall effect in the absence
of disorder was soon thereafter constructed by Ludwig et al.
\cite{Ludwig 1994}. The critical properties of the
pure system can be described by one species of massless Dirac fermions
in $2+1$ space-time.
In the absence of disorder, the mass $m$ of the Dirac fermions plays
the role of a relevant coupling inducing, as it changes sign,
a quantized jump in the transverse static conductivity.
Ludwig at al. then studied the effect of three
types of quenched (static) Gaussian randomness on the properties of the
pure critical point:
$(i)$ an Abelian vector gauge-like impurity potential
$A^{\ }_{\mu}$, $\mu=1,2$,
$(ii)$ an Abelian scalar-like impurity potential $V$,
$(iii)$ and a mass-like impurity potential $M$.
The independent strengths of the three impurity potentials are
measured by the variances $g^{\ }_{A,V,M}$ of the white noise  probability
distributions ${\cal P}[A]$, ${\cal P}[V]$, and ${\cal P}[M]$, respectively.
To leading order in the impurities strength,
the three randomness represent marginal
perturbations to the pure critical point.
In the presence of only one type of impurities,
it was shown in Ref. \cite{Ludwig 1994} that, up to second order,
$g^{\ }_{M(V)}$ is marginally irrelevant (relevant),
and that $g^{\ }_A$ is exactly marginal to all orders, thus
generating a line of critical points starting at the pure limit of
the Dirac fermion model.
The line $g^{\ }_A>0,g^{\ }_V=0,g^{\ }_M=0$
of critical points is in fact {\it multicritical} since
Gaussian randomness in $V$ and $M$ induce relevant perturbations
\cite{Ludwig 1994}. Another important property along the multicritical line
$g^{\ }_A>0,g^{\ }_V=0,g^{\ }_M=0$ is that, after proper infrared
and ultraviolet regularizations, the amplitude of the
typical (with respect to the randomness)
critical wave function in a large box
(eigenstate obeying periodic boundary conditions with vanishing energy)
is  expected to be
selfsimilar down to a microscopic length scale (the ultraviolet cutoff)
and to scale like a multifractal measure with the system size
(the infrared cutoff).
Because the CK model can be viewed as
a random tight-binding Hamiltonian, there must exist a lattice
regularization of a field theory of random massless Dirac fermions
describing the plateau transition.
Ludwig et al. then conjectured that there is,
in the three-dimensional coupling space $\{g^{\ }_{A,V,M}>0\}$,
a stable fixed point which describes the critical properties of the
plateau transition in the integer quantum Hall effect. However,
if this fixed point exists, it cannot be reached perturbatively {}from
any point of the multicritical line $g^{\ }_A>0,g^{\ }_V=0,g^{\ }_M=0$.

In this paper, we study models of random massless Dirac fermions
in two spatial dimensions.
These effective field theories describe,
in the presence of static disorder, the physics of
the low energy and long wave-length single-particle excitations of
quasi two-dimensional electronic systems which are
characterized by {\it isolated} Fermi points.
Examples for which models of random Dirac fermions are appropriate
include degenerate semi-conductors \cite{Fradkin 1986},
two-dimensional graphite sheets \cite{Semenoff 1984},
tight-binding Hamiltonians in the flux phase \cite{Fisher 1985},
and dirty $d$-wave superconductors in two dimensions
\cite{Lee 1993,Nersesyan 1994}.
Models of random massless Dirac fermions are closely related to models
on the effect of randomness on criticality in statistical mechanics
(e.g., two-dimensional Ising \cite{Dotsenko 1983,Shankar 1987}
and $q$-states random Potts model \cite{Ludwig 1987,Ludwig 1990}),
since, at the energy of interest, the pure Dirac system is {\it already}
critical. In the pure system and at the critical (vanishing) energy,
the spatial decay of Green functions is algebraic.
Our goal is threefold. First, the study of two-dimensional models of
random Dirac fermions might yield a systematic classification of
some new disordered critical points. Second, their study might shed light
on the field theoretical realization of complex scaling behavior
(such as multifractality) not present in the pure limit.
Third, it might be possible to infer some unconventional properties
of the field theory describing the plateau transition {}from the properties
of the field theory describing the multicritical points
$g^{\ }_A>0,g^{\ }_V=0,g^{\ }_M=0$, although, as we will see below, they
must differ in an important way.

\subsection{Summary of results}
\label{subsection:Summary of results}

We will assume only one type of randomness:
impurities interacting with $N$ independent species of massless
Dirac fermions like $U(N)=U(1)\times SU(N)$ (static) vector gauge fields
would do. The $U(1)$ case was studied by Ludwig et al. in relation to
the plateau transition \cite{Ludwig 1994}.
The $SU(N)$ case was studied by
Nersesyan et al. in relation to $d$-wave superconductivity
\cite{Nersesyan 1994}.
Both groups used the replica trick to treat the disorder.
We use instead the graded supersymmetric representation of single-particle
Green functions \cite{Efetov 1983}
to treat the $U(1)\times SU(N)$ vector gauge-like randomness
(see also Ref. \cite{Bernard 1995}).
The advantage of the graded supersymmetric representation
is that with it the operator content at criticality is much easier to obtain
than by using the replica trick.
The disadvantage is that it is limited to non-interacting systems.
In addition to the Abelian randomness
$A^{\ }_{\mu}\in U(1)$ with Gaussian probability
distribution of variance $g^{\ }_A$,
the non-Abelian randomness is represented by the
static vector gauge field $B^{\ }_{\mu}\in SU(N)$ with Gaussian probability
distribution of variance $g^{\ }_B$.
We make the important assumption that both $A^{\ }_{\mu}$ and
$B^{\ }_{\mu}$ are topologically trivial. For example, in the Abelian case,
this means that $A^{\ }_{\mu}$ is smooth everywhere and thus is not
associated with a configuration of magnetic monopoles.

We first prove that the generating function for the disorder average of any
given product of $m$ retarded and $n$ advanced
single-particle Green functions factorizes into $m+n$ pieces
which are each independent of disorder,
provided all energy scales in the single-particle Green functions vanish,
i.e., at the critical energy.
This factorization is unique to vector gauge-like
randomness. For example, it already fails for scalar-like $V$
or mass-like $M$ randomness.
Because of this factorization, we can relate the operator
content of the effective field theory
for the averaged single-particle Green function
at vanishing energy to the operator contents of averaged products of
retarded and advanced single-particle Green functions at vanishing energies.
Moreover, we show that,
when all energy scales vanish, averaged products
of single-particle Green functions are algebraic functions for arbitrary
strength $g^{\ }_A>0$ of the Abelian randomness, and in the limits
$g^{\ }_B=0,\infty$ for the strength of the non-Abelian randomness.
In other words, the generating functions for averaged products
of single-particle Green functions describe field theories at criticality
along the lines $g^{\ }_A\geq0,g^{\ }_B=0,\infty$,
whenever all energy scales vanish.
It is the unconventional properties of these critical
effective field theories that are at the heart of this paper.

For the disordered critical points $g^{\ }_A\geq0,g^{\ }_B=0,\infty$,
we calculate the averaged local density of states asymptotically close to
the critical energy.
We reproduce the results of Ludwig et al. \cite{Ludwig 1994}
for the Abelian case,
and Nersesyan et al. for the non-Abelian case \cite{Nersesyan 1994}.
It is noteworthy to notice that in the Abelian case,
the averaged local density of states can be constant
or even diverge for strong
enough randomness (compare with \cite{Gade 1991} and \cite{Hikami 1993}).
Since the order parameter for the averaged local density of states breaks
a continuous symmetry of the effective field theory (the chiral symmetry),
any finite or diverging averaged local density of states at criticality
(vanishing energy) signifies the breakdown of the Mermin-Wagner theorem.
The physical origins of the breakdown of the Mermin-Wagner theorem
are two-fold. On the one hand,
the pure system of massless Dirac fermions is critical.
As a result there exists a continuous chiral symmetry in the effective field
theory for averaged product of single-particle Green functions at
vanishing energies \cite{Fisher 1985,Fradkin 1986}.
On the other hand, the effective field theory for
the disordered system is non-unitary, whereas the Mermin-Wagner
theorem is proven for {\it unitary} field theories with
{\it compact} symmetries.
A power law dependency on energy for the averaged local density of states
of disordered Dirac fermions is unusual
compared to the generic case of a non-critical averaged single-particle
Green functions in the problem of localization at the mobility edge
such as, say, the plateau transition \cite{Wegner 1983,Brezin 1984}.
Note that there are other examples
of diverging averaged local density of states
in two-dimensional disordered systems
\cite{Gade 1991,Hikami 1993}.

The new result of this work is that we can calculate {\it exactly} the
dominant scaling dimensions $\Delta^{\ }_q$
of arbitrary integer powers $q$ of the local order parameter corresponding
to the averaged local single-particle Green function. They are given by
\begin{equation}
\Delta^{\ }_q\ =\
q\ -\ {g^{\ }_A\over\pi}q^2\ -\ {N-1\over N^2}\left(q^2\ +\ Nq\right).
\label{Delta_q}
\end{equation}
This result is non-perturbative. It yields a spectrum of exponents
which are not linearly related, form a {\it concave} set, are always relevant,
and can be negative: $\Delta^{\ }_q\approx -q^2$ for large $q$.
The physical meaning of Eq. (\ref{Delta_q})
is that the local single-particle
Green function, if considered as a random variable,
is not selfaveraging, rather it is log-normal distributed
at criticality. Eq. (\ref{Delta_q}) was already obtained
in Ref. \cite{Chamon 1995a} for the Abelian case.

The existence of an infinite hierarchy of operators which are relevant at
criticality calls into question the stability of the disordered critical
points
$g^{\ }_A\geq0,g^{\ }_B=0,\infty$. As was pointed out by Ludwig et al.
\cite{Ludwig 1994},
impurities inducing a scalar-like randomness $V$
or a mass-like randomness $M$ with Gaussian probability distributions
generate relevant perturbations with scaling exponents $\Delta^{\ }_2$.
They then conclude that the points $g^{\ }_A\geq0,g^{\ }_B=0,\infty$
are multicritical \cite{Ludwig 1994}.
Our result, however, shows that the points
$g^{\ }_A\geq0,g^{\ }_B=0,\infty$ (excluding the pure limit)
are unstable in a very special way, namely that there are infinitely many
relevant perturbations
$\{g^{(n)}_{V,M}|n\in {\rm I}\hskip -0.08 true cm{\bf N}\}$
which are generated by all non-Gaussian cumulants of the
scalar-like or mass-like probability distributions.
(For disorder with Gaussian probability distribution, only the
variance, say $g^{(2)}_{A,B,V,M}\equiv g^{\ }_{A,B,V,M}$,
is non-vanishing.)
Unless there exists a symmetry of the underlying lattice
regularization of the random Dirac fermions which forbids
randomness in the scalar-like or mass-like potentials, the search for
the disordered critical points of random Dirac fermions must
take place in an infinite dimensional disorder induced coupling space.

It is instructive to contrast our results on the operator content
for the disordered critical points $g^{\ }_A\geq0,g^{\ }_B=0,\infty$
to the $O(n)$ non-linear $\sigma$ model in $2+\epsilon$ dimensions
for which there exists a non-trivial critical point
$(t^*,h^*)=(t^{\ }_c(\epsilon)>0,0)$ in a two-dimensional coupling space.
Both the temperature $t$ and the magnetic field $h$ are relevant perturbations,
and both have to be tuned simultaneously to reach the critical point.
The fact that the critical point is unstable with
respect to $t$ does not mean that there is no physics associated with
the $h$ direction: correlation functions such as the magnetization
near the critical point will depend on both $t$ and $h$.
In our model of random Dirac fermions,
there are infinitely many relevant directions,
each additional direction corresponding to a new class of randomness
(characterized by the number of non-vanishing cumulants).
If one were to compute some correlation function (numerically, for example)
close enough to the, say, multicritical line $g^{\ }_A>0,g^{(n)}_{V,M}=0$,
the correlation function would depend on the {\it form} of the
probability distribution for the disorder (all $g^{(n)}_{V/M}$)
not just on its {\it strength} ($g^{(2)}_{V/M}$). The multicritical line
of Ludwig et al. \cite{Ludwig 1994}
is unstable in the sense that it is not possible to find
in its vicinity a simple scaling form
(i.e., depending on a finite number of universal exponents)
for correlation functions.
It is interesting to note that the non-trivial critical point $(t^*,h^*)$
of the $O(n)$ non-linear $\sigma$ model in $2+\epsilon$ dimensions
also has infinitely many relevant operators \cite{Brezin 1976}.
However, they break the $O(n)$ symmetry and cannot be realized
physically by switching on an external magnetic field. Hence, they
do not destabilize the critical point $(t^*,h^*)$.
Finally, the validity of the $2+\epsilon$ expansion itself has been questioned,
due to delicate issues of convergence \cite{Wegner 1990,Castilla 1993}.
In Eq. (\ref{Delta_q}), the scaling exponents
$\{\Delta^{\ }_q|q\in{\rm I}\hskip -0.08 true cm {\bf N}\}$
are obtained non-perturbatively,
a highly non-trivial result if one attempts to derive it using
the replica trick.

We recall that in the theory of localization, the multifractal
analysis consisting in first generating,
for a {\it given} realization of the disorder,
so-called box probabilities
{}from the squared amplitude of a normalized energy eigenstate,
and then performing a finite size scaling analysis on
their $q$-th moment, yields a hierarchy of scaling exponents
$\{\tau(q)=D(q)(q-1)|q\in{\rm I}\hskip -0.08 true cm {\bf N}\}$
\cite{Janssen 1994}.
When the $\tau(q)$'s are not linearly related, the wave function
is said to be multifractal. Multifractal wave functions are
observed for energies close to the mobility edge in two spatial
dimensions. More importantly, the $\tau(q)$'s are
observed to be independent of the realization of disorder \cite{Janssen 1994}.
It is then tempting to conjecture that $(i)$ the scaling
exponents
$\{\tau(q)=D(q)(q-1)|q\in{\rm I}\hskip -0.08 true cm {\bf N}\}$
are universal, and $(ii)$ can be calculated {}from some
critical effective field theory
describing the critical wave function in the thermodynamic limit.
It turns out that we can relate the spectrum of scaling exponents
$\{\Delta^{\ }_q|q\in{\rm I}\hskip -0.08 true cm {\bf N}\}$
for powers of the {\it local} order parameter corresponding to the
averaged local density of states
to scaling of the
{\it averaged} generalized inverse participation ratios
introduced by Wegner \cite{Wegner 1980}.
This is so because the critical behavior of averaged products of
advanced and retarded single-particle Green functions is controlled
by the averaged single-particle Green function.
Again random Dirac fermions differ in this respect {}from
the conventional critical behavior at the mobility edge
in the theory of localization, since in the latter it is the
averaged product of one retarded and one advanced single-particle
Green function, in other words the amplitude and not the phase of
the single-particle Green function, which controls criticality.

We find that, along the critical line $g^{\ }_A>0, g^{\ }_B=\infty$,
the scaling of the averaged generalized inverse participation ratios
with respect to the energy separation {}from the critical energy and
asymptotically close to it is controlled by the scaling exponents
\begin{equation}
\varpi(q)\ =\
{D^*(q)(q-1)\over z},\quad
D^*(q)\ =\ 2-\left({g^{\ }_A\over\pi}+{N-1\over N^2}\right)q,\quad
q\in{\rm I}\hskip -0.08 true cm {\bf N},
\label{varpi(q)}
\end{equation}
where $z$ is the dynamical exponent:
\begin{equation}
z\ =\ 1\ +\ {g^{\ }_A\over\pi}\ +\ {N^2-1\over N^2}.
\label{z}
\end{equation}
The non-linear contribution in $q$ to the right-hand-side of
Eq. (\ref{varpi(q)}) comes {}from $\Delta^{\ }_q$. Consequently,
the function $\tau^*(q)\equiv z\varpi(q)$, when continued to real $q$, is
non-linear, concave and unbounded {}from below for large $q$.
The strict concaveness of $\tau^*(q)$ reflects the non-selfaveraging nature of
the squared amplitude of wave functions with energy close to the
critical energy. The unboundness of $\tau^*(q)$ arises because wave
functions have not been normalized. Rare events, i.e.,
realizations of the disorder such that amplitudes of the wave functions
show ``giant'' spikes, can then dominate the averaging procedure
for large $q$.

We would like to stress that
the function $\tau^*(q)$ only equals
$\tau(q)$ for $q=0$ and $q=1$. It is thus too naive to use
the sole information contained in the hierarchy of local operators
with scaling dimensions
$\{\Delta^{\ }_q|q\in{\rm I}\hskip -0.08 true cm {\bf N}\}$
(see Eq. (\ref{Delta_q}))
to calculate $\tau(q)$ {}from the critical field theory.
Moreover, it is far {}from obvious that the
strict concavity of $\tau^*(q)$ implies the multifractal
scaling of the probability measure constructed {}from the
critical wave function for a given realization of the disorder,
i.e., the strict concavity of $\tau(q)$, as is implicitly assumed in
Refs.
\cite{Castellani 1986,Wegner 1987,Duplantier 1991,Ludwig 1994}.
The important issue of how to calculate the universal scaling exponents
$\{\tau(q)|q\in{\rm I}\hskip -0.08 true cm {\bf N}\}$
within our critical field theory remains open, since
the scaling exponents
$\{\Delta^{\ }_q|q\in{\rm I}\hskip -0.08 true cm {\bf N}\}$
are {\it not} simply related to the $\tau(q)$'s.
An attempt to relate the $\Delta^{\ }_q$'s
to the $\tau(q)$'s will be published elsewhere \cite{Chamon}.

On the other hand, scaling with respect to spatial separation
{\it and} effective system size of the
averaged two-point correlation between the
$q$-th generalized inverse participation ratio and the
$r$-th generalized inverse participation ratio agrees in form
with the one proposed for non-normalizable box-observables
obeying multifractal scaling \cite{Cates 1987,Duplantier 1991,Janssen 1994}.
Since the existence of negative dimensional operators is crucial
to the former scaling, it is tempting to conjecture that
multifractality of the critical wave function might be related to the
existence of negative dimensional operators in the effective
critical field theory.

To uncover the nature of the critical field theories with
$g^{\ }_A\geq0,g^{\ }_B=0,\infty$, we study their current algebras.
Without loss of generality, we restrict ourselves to Dirac fermions
with Abelian vector gauge-like disorder. We show that any disordered critical
point $g^{\ }_A>0$ realizes a Virasoro algebra with vanishing central
charge. Moreover, we show that this Virasoro algebra
is obtained {}from currents obeying a $U(1/1)\times U(1/1)$
graded Kac-Moody algebra. The vanishing of the Virasoro central
charge follows {}from using an effective partition function which is
chosen to be unity in order to perform averages over the disorder.
The Kac-Moody currents must obey a $U(1/1)\times U(1/1)$ graded
algebra, since, on the one hand, the Hamiltonian for our random Dirac fermions
can be interpreted as a Lagrangian in two-dimensional
Euclidean space with $U(1)\times U(1)$ chiral symmetry,
and, on the other hand,
bosonic copies of the Dirac fermions have been introduced
to perform averages over disorder. After having introduced
bosonic copies of Dirac fermions, the effective field theory cannot
be unitary anymore. All unusual properties of the disordered critical point
$g^{\ }_A>0$ can be traced to the bosonic sector of the
effective field theory.

Having identified the underlying current algebras for the
disordered critical points $g^{\ }_A>0$, we find new
critical points by considering more general current algebras.
We simply require that the critical points be
described by two-dimensional Conformal Field Theories (CFT's)
\cite{Belavin 1984}
with vanishing central charges and obeying
$U(1/1)\times U(1/1)$ Kac-Moody current algebras.
All such critical points are classified in
terms of three numbers $(l,m,k^{\ }_j)$.
For odd integer $l$, the spin of Dirac spinors,
which is ${1\over2}$ in the absence of disorder,
remains half-integer as the $U(1/1)\times U(1/1)$
symmetric interaction is switched on.
For any real $m$ and $k^{\ }_j$,
the Virasoro algebra constructed {}from the currents
generating the $U(1/1)\times U(1/1)$ Kac-Moody algebra
has vanishing central charge: a necessary condition to describe
disordered criticality.
Finally, the quantization of $m$
follows {}from the requirement that the CFT be local.
The Dirac fermions with Abelian vector gauge-like disorder
provide a Lagrangian realization of the CFT with
\begin{equation}
(l,m,k^{\ }_j)\ =\ (1,1,{4g^{\ }_A\over\pi}).
\label{114gpi}
\end{equation}
We also construct Lagrangian realizations for all other CFT's.
However, whether or not the remaining critical points
describe criticality in systems of disordered non-interacting electrons
is still an open question.

For each critical point $(l,m,k^{\ }_j)$,
we obtain the operator content {}from the current algebra.
We show, for example, that there exists an infinite hierarchy of local
operators transforming like $U(1/1)\times U(1/1)$ singlets
with the spectrum of scaling exponents
\begin{equation}
\Delta^{(l,m,k^{\ }_j)}_q\ =\
mq\ +\
q^2(l-m)\ -\
q^2\left[l-m^2\left(1-{k^{\ }_j\over4}\right)\right],\quad
q\in {\rm I}\hskip -0.08 true cm{\bf N}.
\label{xlmkj_q}
\end{equation}
Note that Eq. (\ref{Delta_q}) is recovered for $(l,m,k^{\ }_j)$ given by
Eq. (\ref{114gpi}).
Conditions on $l$, $m$, and $k^{\ }_j$ to insure that no
operator with negative scaling dimensions be present at criticality
are easy to obtain.
When operators with negative scaling
dimensions are present in the operator content
and are not forbidden by some additional symmetry requirement,
the critical point is
unstable in the special way described above.

\subsection{Outline}
\label{subsec:outline}

The paper is organized as follows.
We briefly review the supersymmetric
representation of single-particle Green functions in the theory of
localization in section \ref{sec:AveragedGreenfunctions}.
The operator content of  Dirac fermions in the presence
of Abelian and non-Abelian vector gauge-like randomness is derived in
sections \ref{sec:Abelianvectorgaugerandomness} and
\ref{sec:Non-Abelianvectorgaugerandomnes}, respectively.
The scaling with energy of the averaged local density of states,
averaged generalized inverse participation ratios,
and of the averaged spatial correlations of
generalized inverse participation ratios
is derived in section \ref{sec:Scalingoffcriticality}.
In section
\ref{sec:Current Algebra for Graded Symmetries: $U(1/1)$ current algebra},
we study the
algebraic structure of Conformal Field Theories related to
Dirac fermions in the presence of vector gauge-like disorder.
This section can be read independently {}from the previous ones.
We conclude with a discussion of implications {}from our work
for the plateau transition and the random $XY$ model.
In appendix \ref{sec:U(1/1) Suguwara construction}
we give a detailed proof of
the Sugawara construction of section
\ref{sec:Current Algebra for Graded Symmetries: $U(1/1)$ current algebra}.
Our last appendix
\ref{sec:Lagrangian realization of U(1/1) x U(1/1) current algebra}
is devoted to  Lagrangian realizations of the CFT's
constructed in section
\ref{sec:Current Algebra for Graded Symmetries: $U(1/1)$ current algebra}.


\section{Averaged product of single-particle Green functions}
\label{sec:AveragedGreenfunctions}

We are concerned with calculating the disorder average over arbitrary
products of matrix elements of the resolvent
\begin{equation}
G(x,y,\omega\pm{\rm i}\eta;V)\ \equiv\
\langle x|
{1\over \omega-H^{\ }_0-V\pm{\rm i}\eta}
|y \rangle,\quad \eta>0.
\end{equation}
Here, $H^{\ }_0$ is any single-particle Hamiltonian in the absence of disorder,
whereas $V$ is any static single-particle potential representing disorder.
Both are assumed to be local in space.
The limit $\eta\rightarrow0^+$ gives the retarded ($+$) and advanced ($-$)
single-particle Green functions in the energy representation for a given
realization $V$ of
the disorder in a statistical ensemble with probability distribution
${\cal P}[V]$ . The coordinates $x,y$ always denote spatial coordinates.
They also implicitly denote degenerate internal degrees of freedom,
unless explicitly stated.

Impurity average over products of single-particle Green functions is
defined by
\begin{eqnarray}
&&\Gamma^{(m,n)}(\{x^{\ }_k,y^{\ }_k,\omega^{\ }_k\pm{\rm i}\eta\})
\ \equiv\
\nonumber\\
&&\overline{
\left(
\prod_{i=1}^m G(x^{\ }_i,y^{\ }_i,\omega^{\ }_i+{\rm i}\eta;V)
\right)
\left(
\prod_{j=1}^n G(x^{\ }_{m+j},y^{\ }_{m+j},\omega^{\ }_{m+j}-{\rm i}\eta;V)
\right)
}
\ \equiv\
\nonumber\\
&&
\int{\cal D}[V]{\cal P}[V]\
\left(
\prod_{i=1}^m G(x^{\ }_i,y^{\ }_i,\omega^{\ }_i+{\rm i}\eta;V)
\right)
\left(
\prod_{j=1}^n G(x^{\ }_{m+j},y^{\ }_{m+j},\omega^{\ }_{m+j}-{\rm i}\eta;V)
\right).
\label{averagedproductgreenfunctions}
\end{eqnarray}
We will be mostly interested in the averaged density of states
per volume and per energy
\begin{equation}
\rho(\omega)\ \equiv\
\lim_{\eta\rightarrow0^+}
{1\over2\pi{\rm i}}\
\overline
{
{\rm tr}
\left[
G(x,x,\omega-{\rm i}\eta;V)-
G(x,x,\omega+{\rm i}\eta;V)
\right]
},
\end{equation}
the impurity average over the $q$-th power of the
{\it smeared} density of states per
volume and per energy
\begin{equation}
A^{(q)}(\omega,\eta)\ \equiv\
(2\pi{\rm i})^{-q}\
\overline
{
\left\{{\rm tr}
\left[
G(x,x,\omega-{\rm i}\eta;V)-
G(x,x,\omega+{\rm i}\eta;V)
\right]\right\}^q
},
\label{qteAomegaeta}
\end{equation}
and the impurity average over the product
\begin{equation}
\Gamma^{(1,1)}(x,y,\omega+{\rm i}\eta;y,x,\omega-{\rm i}\eta)\ \equiv\
\overline{
G(x,y,\omega+{\rm i}\eta;V)G(y,x,\omega-{\rm i}\eta;V)
}.
\end{equation}
The trace is performed over any internal degrees of freedom such as spin,
orbital degeneracy, etc. Averaging over randomness is
assumed to restore any spatial homogeneity which can be broken by a particular
member $V$ of the statistical ensemble. Hence, $\rho(\omega)$ and
$A^{(q)}(\omega,\eta)$ are assumed not to depend on $x$.
There exists an important Ward identity \cite{Kane 1981}
\begin{mathletters}
\label{ward}
\begin{eqnarray}
&&
\int d^dx
\lim_{\eta\rightarrow0^+}
\eta\
\overline{|{\rm tr}\ G(x,y,\omega+{\rm i}\eta;V)|^2}\ =\
\pi\rho^{\ }_{\rm loc}(\omega),
\label{ward1}
\\
&&
\lim_{\eta\rightarrow0^+}\eta
\int d^dx\
\overline{|{\rm tr}\ G(x,y,\omega+{\rm i}\eta;V)|^2}\ =\
\pi\rho(\omega).
\label{ward2}
\end{eqnarray}
In Eq. (\ref{ward1}),
the limit $\eta\rightarrow0^+$ is taken {\it before}
integrating over space. States which diffuse at infinity
do not contribute to the right-hand-side of Eq. (\ref{ward1})
for any finite separation $|x-y|$. These are the {\it extended} states.
The only contributions in Eq. (\ref{ward1}) come {}from {\it localized} states.
On the other hand,
both localized and extended states contribute in Eq. (\ref{ward2}).
We will always be working in the thermodynamic limit, i.e.,
we will be concerned with the order of limits of Eq. (\ref{ward2}).
\end{mathletters}

\subsection{Graded supersymmetric treatment of the disorder}
\label{subsec:Supersymmetrictreatmentofthedisorder}

In this subsection, we recall how the representation of the single-particle
Green functions in terms of a path integral over anticommuting (Grassmann)
and commuting variables allows to perform the impurity average over the
resolvent {\it before} taking its matrix elements \cite{Efetov 1983}.
To this end, we simply
use the Gaussian integrals
\begin{mathletters}
\begin{eqnarray}
&&\int d\psi^* d\psi\ e^{-\psi^* a \psi}\ =\ a,
\quad \forall a\ {\rm complex},
\\
&&\int d\psi^* d\psi\ \psi \psi^*\ e^{-\psi^* b \psi}\ =\ 1,
\quad \forall b\ {\rm complex},
\\
&&
{
\int d\psi^* d\psi\ \psi \psi^*\ e^{-\psi^* b \psi}
\over
\int d\psi^* d\psi\ e^{-\psi^* a \psi}
}
\ =\ {1\over a},
\quad \forall a,b\ {\rm complex},
\end{eqnarray}
for the pair $\psi^*,\psi$ of Grassmann variables, and
\end{mathletters}
\begin{mathletters}
\begin{eqnarray}
&&\int d\varphi^* d\varphi\ e^{-\varphi^* a\varphi}\ =\
{1\over a},
\quad {\rm Re}\ a>0,
\\
&&\int d\varphi^* d\varphi\ \varphi^*\varphi\ e^{-\varphi^* a\varphi}\ =\
{1\over a}\int d\varphi^* d\varphi\ e^{-\varphi^* a\varphi},
\quad {\rm Re}\ a>0,
\\
&&{
\int d\varphi^* d\varphi\  \varphi^*\varphi\ e^{-\varphi^* a \varphi}
\over
\int d\varphi^* d\varphi\ e^{-\varphi^* a \varphi}
}
\ =\ {1\over a},
\quad\ {\rm Re}\ a>0,
\end{eqnarray}
for the pair $\varphi^*,\varphi$ of complex variables
and generalize them to Gaussian integrals for fermionic coherent states and
bosonic coherent states, respectively.
\end{mathletters}

In this way, the denominator in the fermionic path integral representation
\begin{mathletters}
\label{grassmannpath}
\begin{equation}
G(x,y,\omega\pm{\rm i}\eta;V)\ =\
\mp{\rm i}
{
\int{\cal D}\psi^*{\cal D}\psi\
\psi(x)\psi^*(y)\
e^{+{\rm i}\int
\psi^*\left[(\pm)\left(\omega-H^{\ }_0-V\right)+{\rm i}\eta\right]
\psi}
\over
\int{\cal D}\psi^*{\cal D}\psi\
e^{+{\rm i}\int
\psi^*\left[(\pm)\left(\omega-H^{\ }_0-V\right)+{\rm i}\eta\right]
\psi}
}
\end{equation}
can be rewritten as a Gaussian integrals over bosonic coherent states:
\begin{eqnarray}
&&
G(x,y,\omega\pm{\rm i}\eta;V)\ =\
\nonumber\\
&&
\mp{\rm i}
\int{\cal D}\psi^*{\cal D}\psi\
\int{\cal D}\varphi^*{\cal D}\varphi\
\psi(x)\psi^*(y)\
e^{
+{\rm i}\int
\psi^*\left[(\pm)\left(\omega-H^{\ }_0-V\right)+{\rm i}\eta\right]\psi
+{\rm i}\int
\varphi^*\left[(\pm)\left(\omega-H^{\ }_0-V\right)+{\rm i}\eta\right]\varphi
}.
\label{super1}
\end{eqnarray}
We could have equally well chosen the bosonic path integral representation
\end{mathletters}
\begin{mathletters}
\label{bosonicpath}
\begin{equation}
G(x,y,\omega\pm{\rm i}\eta;V)\ =\
\mp{\rm i}
{
\int{\cal D}\varphi^*{\cal D}\varphi\
\varphi(x)\varphi^*(y)\
e^{+{\rm i}\int
\varphi^*\left[(\pm)\left(\omega-H^{\ }_0-V\right)+{\rm i}\eta\right]\varphi}
\over
\int{\cal D}\varphi^*{\cal D}\varphi\
e^{+{\rm i}\int
\varphi^*\left[(\pm)\left(\omega-H^{\ }_0-V\right)+{\rm i}\eta\right]\varphi}
},
\end{equation}
which can be rewritten as
\begin{eqnarray}
&&
G(x,y,\omega\pm{\rm i}\eta;V)\ =\
\nonumber\\
&&
\mp{\rm i}
\int{\cal D}\psi^*{\cal D}\psi\
\int{\cal D}\varphi^*{\cal D}\varphi\
\varphi(x)\varphi^*(y)\
e^{
+{\rm i}\int
\psi^*\left[(\pm)\left(\omega-H^{\ }_0-V\right)+{\rm i}\eta\right]\psi
+{\rm i}\int
\varphi^*\left[(\pm)\left(\omega-H^{\ }_0-V\right)+{\rm i}\eta\right]\varphi
}.
\label{super2}
\end{eqnarray}
Eqs. (\ref{super1}) and (\ref{super2}) can be combined into
\end{mathletters}
\begin{eqnarray}
G(x,y,\omega\pm{\rm i}\eta;V)\ &=&\
\mp{\rm i}
\int{\cal D}\psi^*{\cal D}\psi\
\int{\cal D}\varphi^*{\cal D}\varphi\
{\left[\psi(x)\psi^*(y)+\varphi(x)\varphi^*(y)\right]\over2}
\nonumber\\
&\times&\
e^{
+{\rm i}\int
\psi^*\left[(\pm)\left(\omega-H^{\ }_0-V\right)+{\rm i}\eta\right]\psi
+{\rm i}\int
\varphi^*\left[(\pm)\left(\omega-H^{\ }_0-V\right)+{\rm i}\eta\right]\varphi
}.
\label{super3}
\end{eqnarray}
Eqs. (\ref{super1},\ref{super2},\ref{super3}) are {\it graded supersymmetric}
representations of the matrix elements of the resolvent.
They require $H^{\ }_0+V$ to be quadratic in the fermionic
($\psi$) or bosonic ($\varphi$) coherent states and thus can only be used
when the pure system is non-interacting.

In the graded supersymmetric representation, average over
the statistical ensemble of impurity potential
can be performed before calculating the matrix elements
of the resolvent, and by so doing, averaged matrix elements of the
resolvent are obtained {}from Green functions of an effective
interacting theory for fermionic and bosonic coherent states.
The effective interaction resulting {}from performing first the
impurity average depends on two distinct properties of the
impurities. One property is the nature of the coupling between the
impurities and the single-particle eigenstates of $H^{\ }_0$.
It is the same for all impurity potentials in the statistical ensemble.
The second property is the probability distribution
${\cal P}[V]$ for the impurity potential.
The scaling hypothesis in the theory of
localization assumes that given one type of impurity potential, there can exist
a critical energy $\omega^{\ }_c$, the so-called mobility edge.
Upon approaching this critical energy,
the scaling hypothesis assumes the existence of one relevant diverging
length scale. Whereas the existence of the mobility edge is determined by
the nature of the coupling between the impurities and the single-particle
eigenstates of $H^{\ }_0$, the {\it microscopic} properties of the probability
distribution ${\cal P}[V]$ should be unobservable close to the mobility edge.
In other words, properties of ${\cal P}[V]$ on length scales much smaller
than the diverging relevant length scale should not be observable.
It is then natural to consider white noise probability distributions,
i.e., statistical ensembles such that
\begin{equation}
\overline{V^{\ }_a(x)V^{\ }_b(y)}\ =\ v^2\delta^{\ }_{ab}\delta(x-y),
\end{equation}
where $a,b$ collectively denote degrees of freedom such as spin, orbital
degeneracy, and so on, to describe a statistical ensemble with
short-range spatial correlations.
The energy scale $v$ then alone characterizes
the impurity strength. In the spirit of the scaling hypothesis,
it is usually assumed that close to the mobility edge the choice of a
Gaussian probability distribution,
\begin{equation}
{\cal P}[V]\ \propto\
\prod_{x} e^{- {1\over 2v^2} {\rm tr}\left[ V^2(x)\right]},
\end{equation}
can be made without loss of generality. Averaging over the impurity potentials
then amounts to simple Gaussian integrals inducing an effective interaction
which is quartic in the coherent fermionic and bosonic fields.
In the graded supersymmetric representation, the average defined in
Eq. (\ref{averagedproductgreenfunctions})
becomes the $2(m+n)$-point Green function
\begin{mathletters}
\label{2(m+n)-pointGreenfunction}
\begin{equation}
\Gamma^{(m,n)}(\{x^{\ }_k,y^{\ }_k,\omega^{\ }_k\pm{\rm i}\eta\})=
{\rm i}^{n-m}
\left\langle
\left(\prod_{i=1}^m
\psi^{\ }_i(x^{\ }_i)\psi^{* }_i(y^{\ }_i)
\right)
\left(\prod_{j=1}^n
\psi^{\ }_{m+j}(x^{\ }_{m+j})\psi^{* }_{m+j}(y^{\ }_{m+j})
\right)
\right\rangle.
\label{gammamnsupersymmetric}
\end{equation}
Here, the expectation value is taken with respect to the partition
function ${\cal Z}^{(m,n)}_{\rm eff}$ defined by the weight
\begin{eqnarray}
&&
e^{{\rm i}S^{(m,n)}_{\rm eff}(\{\omega^{\ }_k\pm{\rm i}\eta\})}\ \equiv\
\int{\cal D}[V]{\cal P}[V]\
e^{{\rm i}S^{(m)}_{\rm r}+{\rm i}S^{(n)}_{\rm a}-S_{\eta}^{(m+n)}},
\label{effweight}\\
&&
S^{(m)}_{\rm r}\ =\
+\sum_{i=1}^m
\int d^dx\
\left[
\psi^{* }_i\left(\omega^{\ }_i-H^{\ }_0-V\right)\psi^{\ }_i
+
\varphi^{* }_i\left(\omega^{\ }_i-H^{\ }_0-V\right)\varphi^{\ }_i
\right],
\label{+mweight}\\
&&
S^{(n)}_{\rm a}\ =\
-\sum_{j=1}^n
\int d^dx\
\left[
\psi^{* }_{m+j}\left(\omega^{\ }_{m+j}-H^{\ }_0-V\right)\psi^{\ }_{m+j}
+
(\psi^*,\psi)\rightarrow(\varphi^*,\varphi)
\right],
\label{-nweight}\\
&&
S_{\eta}^{(m+n)}\ =\
\eta
\sum_{k=1}^{m+n}
\int d^dx\
\left[
\psi^{* }_k\psi^{\ }_k+\varphi^{* }_k\varphi^{\ }_k
\right],
\label{etaweight}\\
\end{eqnarray}
and the measure
\begin{equation}
{\cal D}[\psi^*,\psi,\varphi^*,\varphi]
\ =\
\prod_x
\prod_{k=1}^{m+n}
d\psi   ^{* }_k(x)
d\psi   ^{\ }_k(x)
d\varphi^{* }_k(x)
d\varphi^{\ }_k(x).
\label{measure}
\end{equation}
Notice that by construction ${\cal Z}^{(m,n)}_{\rm eff}=1$.
\end{mathletters}

The action $S_{\eta}^{(m+n)}$ insures the convergence of the path integral.
It also has the largest symmetry, being diagonal in all degrees of freedom.
In particular,
it possesses a local $U(m+n)$ symmetry where $m$ and $n$ represent
the number of retarded and advanced single-particle Green functions over
which the impurity average is performed. The action for the retarded
sector $S^{(m)}_{\rm r}$ has a smaller global $U(m)$ symmetry
when all energies $\omega^{\ }_{i},\ i=1,\cdots,m$
are equal. The same holds in the advanced sector $S^{(n)}_{\rm a}$ with
$n$ replacing $m$. Hence, in the special case
\begin{equation}
\omega^{\ }_1=\cdots=\omega^{\ }_m\ =\ \omega^{\ }_{\rm r},
\quad
\omega^{\ }_{m+1}=\cdots=\omega^{\ }_{m+n}\ =\ \omega^{\ }_{\rm a},
\end{equation}
the effective action $S^{(m,n)}_{\rm eff}$ is symmetric under global
$U(m)\times U(n)$ transformations
acting on the index $k=1,\cdots,m+n$ of the coherent states.
Another important property of the effective action $S^{(m,n)}_{\rm eff}$
is that when all energies
are set to zero in the single-particle Green functions:
\begin{equation}
0\ =\ \omega^{\ }_1\ =\ \cdots\ =\ \omega^{\ }_{m+n},
\end{equation}
then $S^{(m)}_{\rm r}+S^{(n)}_{\rm a}$ is invariant
under the group $U(m,n)$ which leaves
\begin{equation}
{\rm diag}\
(\underbrace{+1,\cdots,+1}_{m},\underbrace{-1,\cdots,-1}_{n})
\end{equation}
unchanged.

\section{Abelian vector gauge randomness}
\label{sec:Abelianvectorgaugerandomness}

{}From now on,
we restrict the generality of our discussion to {\it two} spatial dimensions
and to the single-particle Hamiltonian describing one species
of massless Dirac fermions in the presence of static Abelian vector gauge-like
randomness \cite{Ludwig 1994}. In other words, the Hamiltonian
of the pure system is the two by two matrix
\begin{equation}
H^{\ }_0\ =\ -{\rm i}\gamma^{\ }_{\mu}\partial^{\ }_{\mu}.
\end{equation}
Here, $\mu=1,2$ denote the two spatial directions and
$\gamma^{\ }_1,\gamma^{\ }_2$
are any two of the three Pauli matrices,
$\gamma^{\ }_5=-{\rm i}\gamma^{\ }_1\gamma^{\ }_2$
being the third. The coupling of the impurities to the single-particle
states is through the dimensionfull vector gauge-like
potential $A^{\ }_{\mu}(x)$:
\begin{equation}
V(x)\ =\ -\sqrt{g^{\ }_A} A^{\ }_{\mu}(x)\gamma^{\ }_{\mu}.
\end{equation}
The probability distribution for the random vector gauge-like disorder
is Gaussian:
\begin{equation}
{\cal P}[A^{\ }_{\mu}]\ \propto\ e^{-{1\over 2}\int d^2 x\ A^{2 }_{\mu}(x)}.
\end{equation}
In the graded supersymmetric path integral representation of the Green
functions,
the fermionic coherent states are denoted
$(\bar\psi,\psi)$, while the bosonic coherent states are denoted
$(\bar\varphi,\varphi)$.
Both fermionic and bosonic coherent states transform like Dirac spinors in
two-dimensional Euclidean space, and thus implicitly carry spinor indices.
Eq. (\ref{2(m+n)-pointGreenfunction}) becomes
\begin{mathletters}
\begin{eqnarray}
&&
\Gamma^{(m,n)}(\{x^{\ }_k,y^{\ }_k,\omega^{\ }_k\pm{\rm i}\eta\})=
{\rm i}^{n-m}
\left\langle
\left(\prod_{i=1}^m
\psi^{\ }_i(x^{\ }_i)\bar\psi^{\ }_i(y^{\ }_i)
\right)\!\!
\left(\prod_{j=1}^n
\psi^{\ }_{m+j}(x^{\ }_{m+j})\bar\psi^{\ }_{m+j}(y^{\ }_{m+j})
\right)
\right\rangle,\!
\label{2(m+n)pointGrenfunctionabelianrandomness}\\
&&
e^{{\rm i}S^{(m,n)}_{\rm eff}(\{\omega^{\ }_k\pm{\rm i}\eta\})}\ \equiv\
\int{\cal D}[A^{\ }_{\mu}]\
e^{
-{1\over2}\int d^2x A^2_{\mu}(x)\ +\
{\rm i}S^{(m)}_{\rm r}\ +\ {\rm i}S^{(n)}_{\rm a}\ -\ S_{\eta}^{(m+n)}
},
\label{effweightabelianrandomness}\\
&&
S^{(m)}_{\rm r}\ =\
+\sum_{i=1}^m
\int d^2x
\left\{
\bar\psi^{\ }_i
\left[
{\rm i}\gamma^{\ }_{\mu}
(\partial^{\ }_{\mu}-{\rm i}\sqrt{g^{\ }_A}A^{\ }_{\mu})+
\omega^{\ }_i
\right]
\psi^{\ }_i
+
(\bar\psi,\psi)\rightarrow(\bar\varphi,\varphi)
\right\},
\label{+mweightabelianrandomness}\\
&&
S^{(n)}_{\rm a}\ =\
-\sum_{j=1}^n
\int d^2x
\left\{
\bar\psi^{\ }_{m+j}
\left[
{\rm i}\gamma^{\ }_{\mu}
(\partial^{\ }_{\mu}-{\rm i}\sqrt{g^{\ }_A}A^{\ }_{\mu})+
\omega^{\ }_{m+j}
\right]
\psi^{\ }_{m+j}
+
(\bar\psi,\psi)\rightarrow(\bar\varphi,\varphi)
\right\},
\label{-nweightabelianrandomness}\\
&&
S_{\eta}^{(m+n)}\ =\
\eta
\sum_{k=1}^{m+n}
\int d^2x
\left(
\bar\psi^{\ }_k\psi^{\ }_k+\bar\varphi^{\ }_k\varphi^{\ }_k
\right),
\label{etaweightabelianrandomness}\\
&&
{\cal D}[\bar\psi,\psi,\bar\varphi,\varphi]
\ =\
\prod_{k=1}^{m+n}
{\cal D}\bar\psi   ^{\ }_k
{\cal D}    \psi   ^{\ }_k
{\cal D}\bar\varphi^{\ }_k
{\cal D}    \varphi^{\ }_k.
\label{measureabelianrandomness}
\end{eqnarray}
Integrating over the random vector gauge-like
potential induces a quartic coupling
mixing all the indices labelling the $m+n$ single-particle Green functions.
It would therefore appear that the effective action $S^{(m+n)}_{\rm eff}$
is hopelessly complicated. This is not so however in a very special case,
namely when all energy scales are set to zero:
\end{mathletters}
\begin{equation}
\omega^{\ }_1\ =\ \cdots \ =\ \omega^{\ }_{m+n}\ =\ 0,
\quad
\eta\ =\ 0.
\label{conditionforcriticalm+ngreenfunction}
\end{equation}
Indeed it is then possible to
perform a very simple trick to decouple entirely the Dirac
spinors {}from the random vector gauge-like potential.
To see this, we first make use of the fact that,
since $x$ belongs to a two-dimensional manifold taken to be the
Euclidean plane and since the vector potential $
(A^{\ }_1(x),A^{\ }_2(x))$ has only two components,
$A^{\ }_\mu$ can always be decomposed into a transversal $\Phi^{\ }_1(x)$ and
a longitudinal $\Phi^{\ }_2(x)$ component,
provided $A^{\ }_\mu$ is free of singularities associated with configurations
of magnetic monopoles. Accordingly, we can use the Hodge decomposition
\begin{equation}
A^{\ }_{\mu}(x)\ =\
\tilde\partial^{\ }_{\mu}\Phi^{\ }_1(x)
\ +\
\partial^{\ }_{\mu}\Phi^{\ }_2(x),
\quad\tilde\partial^{\ }_{\mu}\ \equiv\
\epsilon^{\ }_{\mu\nu}\partial^{\ }_{\nu}.
\label{abeliangaugechangeofvariables1}
\end{equation}
We have used the completely antisymmetric second rank (Levi-Cevita) tensor
$\epsilon^{\ }_{\mu\nu}$ with $\epsilon^{\ }_{12}=1$.
Parametrizing the random vector gauge-like fields
in terms of their transversal and longitudinal components
induces two changes in their probability distribution.
The first one comes {}from the
Jacobian induced by the change of variables. The second one comes {}from the
change in their probability distribution. We thus have \cite{Furuya 1982}
\begin{eqnarray}
{\cal D}[A^{\ }_{\mu}]\ {\cal P}[A^{\ }_{\mu}]\ &=&\
{\cal D}[\Phi^{\ }_{\mu}]\ {\cal P}[\Phi^{\ }_{\mu}]\
{\rm Det}\ [\partial^2]
\nonumber\\
\ &=&\
{\cal D}[\Phi^{\ }_{\mu}]\
e^
{
-{1\over2}\int d^2x\
\left[
\left(
\partial^{\ }_{\mu}\Phi^{\ }_1
\right)^2
\ +\
\left(
\partial^{\ }_{\mu}\Phi^{\ }_2
\right)^2
\right]
}
\int{\cal D}[\bar\alpha,\alpha]\
e^
{
-\int d^2x\
\bar\alpha\ \partial^2\ \alpha
}.
\label{abeliangaugechangeofvariables2}
\end{eqnarray}
We are implicitly assuming that the vector gauge fields vanish at infinity.
On the last line, the Jacobian of the transformation is rewritten
in terms of a path integral over fermionic coherent states
$(\bar\alpha,\alpha)$. We will refer to these degrees of freedom as ghost
fields. They do not play an important role for an Abelian vector gauge
impurity potential, but they must be accounted for in the case
of non-Abelian vector gauge impurity potentials as we will see later on.
The reason for which Abelian ghosts are not important is that the Jacobian
in Eq. (\ref{abeliangaugechangeofvariables2}) is independent of the
Abelian vector gauge-like configuration $A^{\ }_{\mu}$.

After the change of variables Eq. (\ref{abeliangaugechangeofvariables1}),
the minimal coupling between the Dirac spinors and the impurity potential
takes the form
\begin{mathletters}
\begin{equation}
\bar\psi^{\ }_k
\gamma^{\ }_{\mu}
\left[
\partial^{\ }_{\mu}
-{\rm i}\sqrt{g^{\ }_A}
\left(
(\tilde\partial^{\ }_{\mu}\Phi^{\ }_1)+(\partial^{\ }_{\mu}\Phi^{\ }_2)
\right)
\right]
\psi^{\ }_k,
\quad k=1,\cdots,m+n,
\end{equation}
in the fermionic coherent state sector, and
\begin{equation}
\bar\varphi^{\ }_k
\gamma^{\ }_{\mu}
\left[
\partial^{\ }_{\mu}
-{\rm i}\sqrt{g^{\ }_A}
\left(
(\tilde\partial^{\ }_{\mu}\Phi^{\ }_1)+(\partial^{\ }_{\mu}\Phi^{\ }_2)
\right)
\right]
\varphi^{\ }_k,
\quad k=1,\cdots,m+n,
\end{equation}
in the bosonic coherent state sector.
Again we make use of the fact that we are effectively working with
covariant gauge fields defined on a two-dimensional manifold
so that the important identity
\end{mathletters}
\begin{equation}
\gamma^{\ }_{\mu}\ \gamma^{\ }_5\ =\
{\rm i}\epsilon^{\ }_{\mu\nu}\ \gamma^{\ }_{\nu},
\quad\mu=1,2,
\end{equation}
holds. It is then easy to verify that for the $k$-th Green function
in the product $\Gamma^{(m,n)}$, the change of variables
\begin{mathletters}
\label{abeliandiracdecouplingtrsf}
\begin{eqnarray}
\bar\psi ^{\   }_k\ &=&\
\bar\psi'^{\   }_k\
e^{
\gamma^{\ }_5\sqrt{g^{\ }_A}\Phi^{\ }_1- {\rm i}\sqrt{g^{\ }_A}\Phi^{\ }_2}
,\quad
\psi ^{\   }_k\ =\
e^{
\gamma^{\ }_5\sqrt{g^{\ }_A}\Phi^{\ }_1+ {\rm i}\sqrt{g^{\ }_A}\Phi^{\ }_2}\
\psi'_k,
\label{abeliandiracdecouplingtrsffermionic}\\
\bar\varphi ^{\   }_k\ &=&\
\bar\varphi'^{\   }_k\
e^{
\gamma^{\ }_5\sqrt{g^{\ }_A}\Phi^{\ }_1- {\rm i}\sqrt{g^{\ }_A}\Phi^{\ }_2}
,\quad
\varphi^{\   }_k\ =\
e^{
\gamma^{\ }_5\sqrt{g^{\ }_A}\Phi^{\ }_1+ {\rm i}\sqrt{g^{\ }_A}\Phi^{\ }_2}\
\varphi'^{\   }_k,
\label{abeliandiracdecouplingtrsfbosonic}
\end{eqnarray}
decouples the Abelian vector gauge-like fields {}from the Dirac spinors
\cite{Belvedere 1979}.
This {\it local} transformation can be repeated for all $m+n$
single-particle Green
functions. It is {\it not} a pure gauge transformation since it involves
the longitudinal component $\Phi^{\ }_1$.
It is the product of a pure gauge transformation (the exponent $\Phi^{\ }_2$)
with an axial transformation (the exponent $\Phi^{\ }_1$).
\end{mathletters}

The axial factor of the decoupling transformation
Eq. (\ref{abeliandiracdecouplingtrsf})
is non-trivial in many respects.
First, in contrast to the pure gauge factor,
it does not leave the terms proportional to the energy scales
$\omega^{\ }_k,\eta$ unchanged:
\begin{equation}
\bar\psi^{\ }_k\psi^{\ }_k\ =\
\bar\psi'_k\ e^{2\gamma^{\ }_5\sqrt{g^{\ }_A}\Phi^{\ }_1}\ \psi'_k,
\quad
\bar\varphi^{\ }_k\varphi^{\ }_k\ =\
\bar\varphi'_k\ e^{2\gamma^{\ }_5\sqrt{g^{\ }_A}\Phi^{\ }_1}\ \varphi'_k,
\quad k=1,\cdots,m+n.
\label{axialtrsfmassterm}
\end{equation}
Second, it does induce a non-trivial Jacobian for both the
fermionic and bosonic coherent state measures,
as long as a fully gauge invariant regularization of these measures is used
\cite{Fujikawa 1979}.
However, the Jacobian induced {}from the fermionic measure \cite{Roskies 1981}
\begin{equation}
{\cal D}[\bar\psi,\psi]\ =\
{\cal D}[\bar\psi',\psi']\
e^{-{1\over2\pi}\int d^2x\ \left(\partial^{\ }_{\mu}\Phi^{\ }_1\right)^2 }
\label{abelianpolwieg}
\end{equation}
cancels that {}from the bosonic measure (this cancellation does not hold
for non-Abelian gauge fields due to the ghost sector).
Finally, the manifold defined by all axial transformations is not compact.

The impurity average of any product of single-particle Green functions
factorizes when all their energies are set to zero:
\begin{mathletters}
\label{abeliangaugefactorization}
\begin{eqnarray}
&&
\lim_{\eta\rightarrow0^+}
\Gamma^{(m,n)}(\{x^{\ }_k,y^{\ }_k,\pm{\rm i}\eta\})=
\nonumber\\
&&
{\rm i}^{n-m}
\left\langle
\left(\prod_{i=1}^m
\psi'_i(x^{\ }_i)\bar\psi'_i(y^{\ }_i)
\right)
\left(\prod_{j=1}^n
\psi'_{m+j}(x^{\ }_{m+j})\bar\psi'_{m+j}(y^{\ }_{m+j})
\right)
\right\rangle^{\ }_{\bar\psi',\psi',\bar\varphi',\varphi'}
\nonumber\\
&&\times
\left\langle
\left(\prod_{i=1}^m
e^{\gamma^{\ }_5\Phi^{\ }_1(x^{\ }_i)}
e^{\gamma^{\ }_5\Phi^{\ }_1(y^{\ }_i)}
\right)
\left(\prod_{j=1}^n
e^{\gamma^{\ }_5\Phi^{\ }_1(x^{\ }_{m+j})}
e^{\gamma^{\ }_5\Phi^{\ }_1(y^{\ }_{m+j})}
\right)
\right\rangle^{\ }_{\Phi^{\ }_1,\bar\alpha,\alpha}
\nonumber\\
&&\times
\left\langle
\left(\prod_{i=1}^m
e^{     +{\rm i}\Phi^{\ }_2(x^{\ }_i)}
e^{     -{\rm i}\Phi^{\ }_2(y^{\ }_i)}
\right)
\left(\prod_{j=1}^n
e^{     +{\rm i}\Phi^{\ }_2(x^{\ }_{m+j})}
e^{     -{\rm i}\Phi^{\ }_2(y^{\ }_{m+j})}
\right)
\right\rangle^{\ }_{\Phi^{\ }_2}.
\label{abeliangaugefactorization(a)}
\end{eqnarray}
The expectation value with respect to the spinor coherent states is
\footnote
{
Of course, integration over the bosonic
coherent states is not guaranteed to converge anymore having set $\eta$
to zero. Hence, Eq. (\ref{spinoreffmnactions})
is only a formal expression unless it is itself regularized.
}
\begin{eqnarray}
&&
\langle f\rangle^{\ }_{\bar\psi',\psi',\bar\varphi',\varphi'}\ =\
\int{\cal D}[\bar\psi',\psi',\bar\varphi',\varphi']\
e^
{
{\rm i} S'^{(m)}_{\rm r}
+
{\rm i} S'^{(n)}_{\rm a}
}
\ f,
\label{spinoreffmnactions}
\\
&&
S'^{(m)}_{\rm r} =
+\sum_{i=1}^m
\int d^2x
\left(
\bar\psi'_i
{\rm i}\gamma^{\ }_{\mu}\partial^{\ }_{\mu}
\psi'_i+
\bar\varphi'_i
{\rm i}\gamma^{\ }_{\mu}\partial^{\ }_{\mu}
\varphi'_i
\right),
\label{spinoreffmnactions1}
\\
&&
S'^{(n)}_{\rm a} =
-\sum_{j=1}^n
\int d^2x
\left(
\bar\psi'_{m+j}
{\rm i}\gamma^{\ }_{\mu}\partial^{\ }_{\mu}
\psi'_{m+j}
+
\bar\varphi'_{m+j}
{\rm i}\gamma^{\ }_{\mu}\partial^{\ }_{\mu}
\varphi'_{m+j}
\right).
\label{spinoreffmnactions2}
\end{eqnarray}
The expectation value with respect to the
longitudinal component of the Abelian gauge potential is
\begin{equation}
\langle f\rangle^{\ }_{\Phi^{\ }_1}\ =\
\int{\cal D}[\Phi^{\ }_1]\
e^{-S^{\ }_{\rm imp}[\Phi^{\ }_1]}\ f,
\quad
S^{\ }_{\rm imp}[\Phi^{\ }_1]\ =\
{1\over2}\int d^2x\ \left(\partial^{\ }_{\mu}\Phi^{\ }_1\right)^2.
\end{equation}
The expectation value with respect to the
the transversal component of the Abelian gauge potential is
\begin{equation}
\langle f\rangle^{\ }_{\Phi^{\ }_2}\ =\
\int{\cal D}[\Phi^{\ }_2]\
e^{-S^{\ }_{\rm imp}[\Phi^{\ }_2]}\ f,
\quad
S^{\ }_{\rm imp}[\Phi^{\ }_2]\ =\
{1\over2}\int d^2x\ \left(\partial^{\ }_{\mu}\Phi^{\ }_2\right)^2.
\end{equation}
Finally, there is an innocuous average over the ghost coherent states
\begin{equation}
\langle f\rangle^{\ }_{\bar\alpha,\alpha}\ =\
\int{\cal D}[\bar\alpha,\alpha]\
e^{{\rm i}S^{\ }_{\rm gh}}\ f,
\quad
S^{\ }_{\rm gh}\ =\ \int d^2x\ \bar\alpha{\rm i}\partial^2\alpha.
\end{equation}
Thus, the impurity average of any product of single-particle
Green functions can be obtained {}from the correlation functions
generated by the corresponding {\it critical} Euclidean effective action
\end{mathletters}
\begin{equation}
S^{(m,n)}_{\rm cr}\ \equiv\
S^{\ }_{\rm imp}[\Phi^{\ }_1]\ +\
S^{\ }_{\rm imp}[\Phi^{\ }_2]\ -\
{\rm i}S^{\ }_{\rm gh}\ -\
{\rm i}S'^{(m)}_{\rm r}\ -\
{\rm i}S'^{(n)}_{\rm a}.
\label{abeliancriticalaction}
\end{equation}
The impurity strength does not appear explicitly, being hidden
in the rotation {}from the original fields to the new ones.
The effective action is critical in the sense that all correlation
functions are algebraic as we will see below.

{}From the factorization in Eq. (\ref{abeliangaugefactorization})
follows immediately that given
$m+n$ pairs of points $x^{\ }_k,y^{\ }_k$ on the Euclidean plane,
the $2(m+n)$-point correlation function
$\Gamma^{(m,n)}(\{x^{\ }_k,y^{\ }_k,0\})$
can be calculated by using any effective action
$S^{(p,q)}_{\rm cr}$
as long as $p\geq m$ and $q\geq n$.
For example, the average local density of states
$\lim_{\omega\rightarrow0}\rho(\omega)$
can be calculated {}from all effective actions
$S^{(p,q)}_{\rm cr}$
with either $p\geq 1$ or $q\geq 1$.
We now turn to the operator content of the effective actions
$S^{(m,n)}_{\rm cr}$.

\subsection{Operator content at criticality for Abelian vector gauge
randomness}
\label{subsec:Operatorcontent1}

To study the operator content of the critical action in
Eq. (\ref{abeliancriticalaction}), it is useful to introduce the so-called
chiral basis for Dirac spinors $\bar\psi,\psi,\bar\varphi,\varphi$.
The chiral basis is defined by
\begin{mathletters}
\label{chiralbasis}
\begin{eqnarray}
\psi^{\dag}\ \equiv\ \bar\psi\gamma^{\ }_1,\quad
\psi^{\dag}_{\pm}\ \equiv\
\psi^{\dag}{1\over2}\left(1\pm\gamma^{\ }_5\right),\quad
\psi^{\   }_{\pm}\ \equiv\
{1\over2}\left(1\pm\gamma^{\ }_5\right)\ \psi,
\\
\varphi^{\dag}\ \equiv\ \bar\varphi\gamma^{\ }_1,\quad
\varphi^{\dag}_{\pm}\ \equiv\
\varphi^{\dag}{1\over2}\left(1\pm\gamma^{\ }_5\right),\quad
\varphi^{\ }_{\pm}\ \equiv\
{1\over2}\left(1\pm\gamma^{\ }_5\right)\ \varphi.
\end{eqnarray}
In the chiral basis, the decoupling transformation,
Eq. (\ref{abeliandiracdecouplingtrsf}),
becomes
\end{mathletters}
\begin{mathletters}
\begin{eqnarray}
\psi ^{\dag}_{k\pm}\ &=&\
\psi'^{\dag}_{k\pm}\
e^{\mp\sqrt{g^{\ }_A}\Phi^{\ }_1- {\rm i}\sqrt{g^{\ }_A}\Phi^{\ }_2},
\quad
\psi ^{\   }_{k\pm}\ =\
e^{\pm\sqrt{g^{\ }_A}\Phi^{\ }_1+ {\rm i}\sqrt{g^{\ }_A}\Phi^{\ }_2}\
\psi'_{k\pm},\quad k=1,\cdots,m+n,
\label{abeliandiracdecouplingtrsffermionicchiral}
\\
\varphi ^{\dag}_{k\pm}\ &=&\
\varphi'^{\dag}_{k\pm}\
e^{\mp\sqrt{g^{\ }_A}\Phi^{\ }_1- {\rm i}\sqrt{g^{\ }_A}\Phi^{\ }_2},
\quad
\varphi^{\   }_{k\pm}\ =\
e^{\pm\sqrt{g^{\ }_A}\Phi^{\ }_1+ {\rm i}\sqrt{g^{\ }_A}\Phi^{\ }_2}\
\varphi'^{\   }_{k\pm},\quad k=1,\cdots,m+n.
\label{abeliandiracdecouplingtrsfbosonicchiral}
\end{eqnarray}
The terms in Eq. (\ref{axialtrsfmassterm}) which couple to the energy
in denominators of single-particle Green functions are represented by
\end{mathletters}
\begin{mathletters}
\begin{eqnarray}
&&
\bar\psi^{\ }_k\psi^{\ }_k\ =\
\psi'^{\dag}_{k+}e^{-2\sqrt{g^{\ }_A}\Phi^{\ }_1}\psi'^{\ }_{k-}\ +\
\psi'^{\dag}_{k-}e^{+2\sqrt{g^{\ }_A}\Phi^{\ }_1}\psi'^{\ }_{k+},
\quad k=1,\cdots,m+n,
\\
&&
\bar\varphi^{\ }_k\varphi^{\ }_k\ =\
\varphi'^{\dag}_{k+}e^{-2\sqrt{g^{\ }_A}\Phi^{\ }_1}\varphi'^{\ }_{k-}\ +\
\varphi'^{\dag}_{k-}e^{+2\sqrt{g^{\ }_A}\Phi^{\ }_1}\varphi'^{\ }_{k+},
\quad k=1,\cdots,m+n,
\end{eqnarray}
provided  $(\gamma^{\ }_1,\gamma^{\ }_2,\gamma^{\ }_5)$
are chosen to be the usual three Pauli matrices.
The effective actions in the retarded and advanced Dirac spinor sectors take
a simple form in the chiral basis.
To see this, we first think of the Euclidean plane
$\{x=(x^{\ }_1,x^{\ }_2)|x^{\ }_1,x^{\ }_2\in
{\rm I}\hskip -0.08 true cm {\bf R}\}$
as being a real section
of the two-dimensional complex manifold with local coordinates
\end{mathletters}
\begin{equation}
     z\ =\ x^{\ }_1\ +\ {\rm i}x^{\ }_2
,\quad
\bar z\ =\ x^{\ }_1\ -\ {\rm i}x^{\ }_2,
\end{equation}
where $x^{\ }_1$ and $x^{\ }_2$ are now arbitrary complex numbers.
The effective actions in
Eqs. (\ref{spinoreffmnactions1},\ref{spinoreffmnactions2})
now become
\begin{mathletters}
\begin{eqnarray}
S'^{(m)}_{\rm r}&=&
+2{\rm i}\sum_{i=1}^m
\left(
\psi'^{\dag}_{i+}
\ \partial^{\ }_{\bar z}\
\psi'_{i+}
\ +\
\psi'^{\dag}_{i-}
\ \partial^{\ }_z\
\psi'_{i-}
\ +\
\varphi'^{\dag}_{i+}
\ \partial^{\ }_{\bar z}\
\varphi'_{i+}
\ +\
\varphi'^{\dag}_{i-}
\ \partial^{\ }_z\
\varphi'_{i-}
\right),
\\
S'^{(n)}_{\rm a}&=&
-2{\rm i}\sum_{j=1}^n
\left(
\psi'^{\dag}_{(m+j)+}
\partial^{\ }_{\bar z}
\psi'_{(m+j)+}
+
\psi'^{\dag}_{(m+j)-}
\partial^{\ }_z
\psi'_{(m+j)-}
+
(\psi'^{\dag}_{\pm},\psi'_{\pm})
\rightarrow
(\varphi'^{\dag}_{\pm},\varphi'_{\pm})
\right),
\end{eqnarray}
respectively. The advantage of the chiral basis is that the two-point
correlation functions in the decoupled spinor sectors are very simple.
The only non-vanishing ones in the retarded sector are
$(k,l=1,\cdots,m)$:
\end{mathletters}
\begin{mathletters}
\begin{eqnarray}
\left\langle
\psi'^{\ }_{k+}(z,\bar z)\ \psi'^{\dag}_{l+}(0,0)
\right\rangle^{\ }_
{\psi'^{\dag}_{\pm},\psi'_{\pm},\varphi'^{\dag}_{\pm},\varphi'_{\pm}}\ &=&\
\left\langle
\varphi'^{\ }_{k+}(z,\bar z)\ \varphi'^{\dag}_{l+}(0,0)
\right\rangle^{\ }_
{\psi'^{\dag}_{\pm},\psi'_{\pm},\varphi'^{\dag}_{\pm},\varphi'_{\pm}}\ =\
{\delta^{\ }_{kl}\over 2\pi z},
\label{holo2pointretarded}\\
\left\langle
\psi'^{\ }_{k-}(z,\bar z)\ \psi'^{\dag}_{l-}(0,0)
\right\rangle^{\ }_
{\psi'^{\dag}_{\pm},\psi'_{\pm},\varphi'^{\dag}_{\pm},\varphi'_{\pm}}\ &=&\
\left\langle
\varphi'^{\ }_{k-}(z,\bar z)\ \varphi'^{\dag}_{l-}(0,0)
\right\rangle^{\ }_
{\psi'^{\dag}_{\pm},\psi'_{\pm},\varphi'^{\dag}_{\pm},\varphi'_{\pm}}\ =\
{\delta^{\ }_{kl}\over2\pi{\bar z}},
\label{antiholo2pointretarded}
\end{eqnarray}
In the advanced sector, they are the same except for a
relative sign $(k,l=m+1,\cdots,m+n)$:
\end{mathletters}
\begin{mathletters}
\begin{eqnarray}
\left\langle
\psi'^{\ }_{k+}(z,\bar z)\ \psi'^{\dag}_{l+}(0,0)
\right\rangle^{\ }_
{\psi'^{\dag}_{\pm},\psi'_{\pm},\varphi'^{\dag}_{\pm},\varphi'_{\pm}} &=&
\left\langle
\varphi'^{\ }_{k+}(z,\bar z)\ \varphi'^{\dag}_{l+}(0,0)
\right\rangle^{\ }_
{\psi'^{\dag}_{\pm},\psi'_{\pm},\varphi'^{\dag}_{\pm},\varphi'_{\pm}} =
-{\delta^{\ }_{kl}\over 2\pi z},
\label{holo2pointadvanced}\\
\left\langle
\psi'^{\ }_{k-}(z,\bar z)\ \psi'^{\dag}_{l-}(0,0)
\right\rangle^{\ }_
{\psi'^{\dag}_{\pm},\psi'_{\pm},\varphi'^{\dag}_{\pm},\varphi'_{\pm}} &=&
\left\langle
\varphi'^{\ }_{k-}(z,\bar z)\ \varphi'^{\dag}_{l-}(0,0)
\right\rangle^{\ }_
{\psi'^{\dag}_{\pm},\psi'_{\pm},\varphi'^{\dag}_{\pm},\varphi'_{\pm}} =
-{\delta^{\ }_{kl}\over2\pi{\bar z}},
\label{antiholo2pointadvanced}
\end{eqnarray}
Two-point functions between retarded and advanced spinor fields vanish.
In the chiral basis,
the two-point correlation functions are thus either holomorphic as in
Eqs. (\ref{holo2pointretarded},\ref{holo2pointadvanced})
or antiholomorphic as in
Eqs. (\ref{antiholo2pointretarded},\ref{antiholo2pointadvanced}).
\end{mathletters}

Another important set of two-point correlation functions that we will need is
\begin{mathletters}
\label{scalingdimprimaryfieldsphi1}
\begin{equation}
\left\langle
e^{+{\rm i}f\Phi^{\ }_{\mu}(z,\bar z)}\
e^{-{\rm i}f\Phi^{\ }_{\nu}(0,0)}
\right\rangle^{\ }_{\Phi^{\ }_1,\Phi^{\ }_2}\ \propto
\delta^{\ }_{\mu\nu}\
z^{-2h}\bar z^{-2\bar h},
\quad\mu,\nu=1,2,
\label{freescalarcorrelator}
\end{equation}
where the conformal weights $h$ and $\bar h$ are given by
\begin{equation}
h\ =\ \bar h\ =\ {f^2\over8\pi}.
\label{conformalweightsfreescalarfield}
\end{equation}
The proportionality constant on the right-hand-side of
Eq. (\ref{freescalarcorrelator})
depends on the short distance cutoff
used to regulate the free scalar field theory. The parameter $f$
is analytically continued {}from the real numbers to arbitrary complex numbers.
In particular, we will be interested in purely imaginary values of $f$
owing to the non-compact nature of the decoupling transformation in the
longitudianal ($\Phi^{\ }_1$) vector gauge  sector.
\end{mathletters}

For any given Euclidean effective action
$S^{(m,n)}_{\rm cr}$ in Eq. (\ref{abeliancriticalaction}),
we can calculate the scaling dimensions of all possible local operators,
thus explicitly showing that all correlation functions are algebraic.
To illustrate how this is done,
we begin with the effective action $S^{(1,0)}_{\rm cr}$
for the averaged retarded single-particle Green function.
One verifies that
\begin{mathletters}
\begin{eqnarray}
\left\langle    \psi_+(z,\bar z)    \psi^{\dag}_+(0,0) \right\rangle\ &=&\
\left\langle \varphi_+(z,\bar z) \varphi^{\dag}_+(0,0) \right\rangle\ =\
{1\over2\pi z},
\label{propagator+}
\\
\left\langle \psi_-(z,\bar z) \psi^{\dag}_-(0,0) \right\rangle\ &=&\
\left\langle \varphi_-(z,\bar z) \varphi^{\dag}_-(0,0) \right\rangle\ =\
{1\over2\pi\bar z},
\label{propagator-}
\end{eqnarray}
as follows immediately {}from, say,
\end{mathletters}
\begin{eqnarray}
\left\langle
\psi^{\dag}_{\pm}(z,\bar z)
\psi^{\   }_{\pm}(0,0)
\right\rangle
\ &=&\
\left\langle
\psi'^{\dag}_{\pm}(z,\bar z)
\psi'^{\   }_{\pm}(0,0)
\right\rangle
\nonumber\\
&\times&\
\left\langle
e^{ {\rm i} (\pm {\rm i} \sqrt{g^{\ }_A})\Phi^{\ }_1(z,\bar z)}
e^{ {\rm i} (\mp {\rm i} \sqrt{g^{\ }_A})\Phi^{\ }_1(0,0     )}
\right\rangle
\nonumber\\
&\times&\
\left\langle
e^{ {\rm i} (-       \sqrt{g^{\ }_A})\Phi^{\ }_2(z,\bar z)}
e^{ {\rm i} (+       \sqrt{g^{\ }_A})\Phi^{\ }_2(0,0     )}
\right\rangle.
\label{producttwopoint}
\end{eqnarray}
The bracket denotes an average performed with the effective action
\begin{equation}
S^{(1,0)}_{\rm cr}\ =\
S^{\ }_{\rm imp}[\Phi^{\ }_1]\ +\
S^{\ }_{\rm imp}[\Phi^{\ }_2]\ -\
{\rm i}S^{\ }_{\rm gh}\ -\
{\rm i}S'^{(1)}_{\rm r}.
\label{s(10)cr}
\end{equation}
We see that any impurity ($g^{\ }_A$)
dependency coming {}from averaging over $\Phi^{\ }_1$ precisely
cancels that coming {}from averaging over $\Phi^{\ }_2$.
Hence the scaling properties of the fields
$\psi^{\ }_{\pm}$ and $\varphi^{\ }_{\pm}$
are not affected by impurity averaging.
This is not so, however, when calculating the two-point
function for the composite operator
\begin{equation}
{\cal O}(z,\bar z)\ =\
\left(
\psi   ^{\dag m^{\ }_1}_+
\psi   ^{\dag m^{\ }_2}_-
\psi   ^{     m^{\ }_3}_+
\psi   ^{     m^{\ }_4}_-
\varphi^{\dag m^{\ }_5}_+
\varphi^{\dag m^{\ }_6}_-
\varphi^{     m^{\ }_7}_+
\varphi^{     m^{\ }_8}_-
\right)(z,\bar z),
\label{compositeoperator10}
\end{equation}
which is defined through point-splitting ($m^{\ }_1,\cdots,m^{\ }_8$
are arbitrary positive integers). Indeed,
\begin{eqnarray}
\left\langle
{\cal O}^{\dag}(z,\bar z) {\cal O}(0,0)
\right\rangle
&\propto&
\left\langle
\psi'^{\dag m^{\ }_4}_-(z,\bar z)
\psi'^{     m^{\ }_4}_-(0,     0)
\right\rangle
\left\langle
\psi'^{\dag m^{\ }_3}_+(z,\bar z)
\psi'^{     m^{\ }_3}_+(0,     0)
\right\rangle
\nonumber\\
&\times&
\left\langle
\psi'^{     m^{\ }_2}_-(z,\bar z)
\psi'^{\dag m^{\ }_2}_-(0,     0)
\right\rangle
\left\langle
\psi'^{     m^{\ }_1}_+(z,\bar z)
\psi'^{\dag m^{\ }_1}_+(0,     0)
\right\rangle
\nonumber\\
&\times&
\left\langle
\varphi'^{\dag m^{\ }_8}_-(z,\bar z)
\varphi'^{     m^{\ }_8}_-(0,     0)
\right\rangle
\left\langle
\varphi'^{\dag m^{\ }_7}_+(z,\bar z)
\varphi'^{     m^{\ }_7}_+(0,     0)
\right\rangle
\nonumber\\
&\times&
\left\langle
\varphi'^{     m^{\ }_6}_-(z,\bar z)
\varphi'^{\dag m^{\ }_6}_-(0,     0)
\right\rangle
\left\langle
\varphi'^{     m^{\ }_5}_+(z,\bar z)
\varphi'^{\dag m^{\ }_5}_+(0,     0)
\right\rangle
\nonumber\\
&\times&\
\left\langle
e^{ +{\rm i} f^{\ }_1\sqrt{g^{\ }_A} \Phi^{\ }_1(z,\bar z)}
e^{ -{\rm i} f^{\ }_1\sqrt{g^{\ }_A} \Phi^{\ }_1(0,0     )}
\right\rangle
\nonumber\\
&\times&\
\left\langle
e^{ +{\rm i} f^{\ }_2\sqrt{g^{\ }_A} \Phi^{\ }_2(z,\bar z)}
e^{ -{\rm i} f^{\ }_2\sqrt{g^{\ }_A} \Phi^{\ }_2(0,0     )}
\right\rangle.
\end{eqnarray}

The conformal weight $h$ ($\bar h$) of
$\psi'^{m}_{+(-)}$ scales quadratically with $m$
in contrast to that of $\varphi'^{m}_{+(-)}$ which scales linearly with $m$.
This is perhaps most easily seen with the help of Abelian bosonization
when calculating correlation functions of $\psi'_{\pm}$'s. Exponentials in the
longitudinal and transversal components of the random gauge fields scale as in
Eq. (\ref{scalingdimprimaryfieldsphi1})
with
\begin{mathletters}
\label{fin10}
\begin{eqnarray}
&&
f_1\ =\
-\ {\rm i}\
\left(
m^{\ }_1-m^{\ }_2-m^{\ }_3+m^{\ }_4
+
m^{\ }_5-m^{\ }_6-m^{\ }_7+m^{\ }_8
\right),
\label{f110}
\\
&&
f_2\ =\
+\
\left(
m^{\ }_1+m^{\ }_2-m^{\ }_3-m^{\ }_4
+
m^{\ }_5+m^{\ }_6-m^{\ }_7-m^{\ }_8
\right),
\label{f210}
\end{eqnarray}
respectively. Notice the difference in the relative sign of the $m$'s in
$f^{\ }_1$ and $f^{\ }_2$.
We thus conclude \cite{Chamon 1995a} that ${\cal O}$ scales like
\end{mathletters}
\begin{mathletters}
\label{scalingofmostgeneral10}
\begin{equation}
\left\langle
{\cal O}^{\dag}(z,{\bar z})\
{\cal O}       (0,0)
\right\rangle^{\ }_{S^{(1,0)}_{\rm cr}}
\ \propto\
       z^{-2h}
{\bar z}^{-2{\bar h}}
\label{scalingofvertexoperator10},
\end{equation}
where the  conformal weights are given by
\begin{eqnarray}
&&
h\ =\
{1\over2}
\left[
m^2_1+m^2_3+m^{\ }_5+m^{\ }_7
+
\left(
f^2_2
-
|f^{\ }_1|^2
\right)
{g^{\ }_A\over4\pi}
\right],
\label{holomoweight10}
\\
&&
{\bar h}\ =\
{1\over2}
\left[
m^2_2+m^2_4+m^{\ }_6+m^{\ }_8
+
\left(
f^2_2
-
|f^{\ }_1|^2
\right)
{g^{\ }_A\over4\pi}
\right].
\label{antiholomoweight10}
\end{eqnarray}
Impurity averaging does not change the scaling dimensions of local operators
when $f^2_2=|f^{\ }_1|^2$. On the other hand,
the non-compactness of the decoupling transformation allows for
the possibility that local operators have negative scaling dimensions
when $|f^{\ }_1|^2\gg f^2_2$.
\end{mathletters}

Negative dimensional operators are now easily obtained
in the theory with effective action $S^{(1,0)}_{\rm cr}$.
For example, consider the composite operator
$\Psi^{\ }_{n^{\ }_1n^{\ }_2}$
defined by
\begin{mathletters}
\begin{equation}
\Psi^{\ }_{n^{\ }_1n^{\ }_2}=
\cases{
\varphi^{n^{\ }_1}_+ \varphi^{n^{\ }_2}_-,&$n^{\ }_1>0,n^{\ }_2>0$,\cr
\varphi_+^{\dag-n_1} \varphi^{n^{\ }_2}_-,&$n^{\ }_1<0,n^{\ }_2>0$,\cr
\varphi^{n_1}_+      \varphi_-^{\dag-n_2},&$n^{\ }_1>0,n^{\ }_2<0$,\cr
\varphi_+^{\dag-n_1} \varphi_-^{\dag-n_2},&$n^{\ }_1<0,n^{\ }_2<0$,\cr}
\label{Psi}
\end{equation}
and use Eqs. (\ref{fin10}-\ref{scalingofmostgeneral10}).
The conformal weights of these
operators are
\begin{equation}
      h =\frac{1}{2}|n^{\ }_1|+\frac{g^{\ }_A}{2\pi}n^{\ }_1n^{\ }_2,
\quad
{\bar h}=\frac{1}{2}|n^{\ }_2|+\frac{g^{\ }_A}{2\pi}n^{\ }_1n^{\ }_2.
\label{weightsofPsi}
\end{equation}
This demonstrates that for {\it any given} value of $g^{\ }_A$,
there are infinitely many local composite operators with
negative conformal weights. This is very different {}from
unitary two-dimensional conformal field theory, say
the $U(1)$ Thirring model \cite{Thirring}
(the action for the Euclidean $U(1)$ Thirring model differs {}from
$S^{(1,0)}_{\rm cr}$ only in that the $\varphi$ sector is absent),
which cannot support operators with negative dimensions.
Operators like  $\Psi^{\ }_{-|n||n|}$ are not mere curiosities.
For one thing, they are generated in
the effective graded supersymmetric model if non-Gaussian moments in the
probability distribution of a mass-like random perturbation of the
Dirac Hamiltonian are present. Furthermore, they yield direct information
on the impurity average of products of single-particle Green functions
as we now show.
\end{mathletters}

The generalization to the calculation of the scaling dimension of any
local operator in the theory with effective action $S^{(m,n)}_{\rm cr}$
is now obvious. The most general local composite operator
\begin{mathletters}
\label{mostgeneralcompositeoperator}
\begin{equation}
{\cal O}(z,\bar z)\ =\
\prod_{k=1}^{m+n}
\left(
\psi   ^{\dag m^{\ }_{1k}}_{k+}
\psi   ^{\dag m^{\ }_{2k}}_{k-}
\psi   ^{     m^{\ }_{3k}}_{k+}
\psi   ^{     m^{\ }_{4k}}_{k-}
\varphi^{\dag m^{\ }_{5k}}_{k+}
\varphi^{\dag m^{\ }_{6k}}_{k-}
\varphi^{     m^{\ }_{7k}}_{k+}
\varphi^{     m^{\ }_{8k}}_{k-}
\right)(z,\bar z),
\label{compositeoperatormn}
\end{equation}
where point-splitting is implied and all $m$'s are positive integers,
scales like
\begin{eqnarray}
&&
\left\langle
{\cal O}^{\dag}(z,\bar z)\
{\cal O}       (0,     0)
\right\rangle^{\ }_{S^{(m,n)}_{\rm cr}}
\ \propto\
     z^{-2     h}
\bar z^{-2\bar h},
\label{scalingofvertexoperatormn}
\\
&&
h\ =\
{1\over2}\sum_{k=1}^{m+n}
\left[
m^2_{1k}+m^2_{3k}+m^{\ }_{5k}+m^{\ }_{7k}
\right]+
\left(
f^2_2
-
|f^{\ }_1|^2
\right)
{g^{\ }_A\over8\pi},
\label{holomoweightmn}
\\
&&
{\bar h}\ =\
{1\over2}\sum_{k=1}^{m+n}
\left[
m^2_{2k}+m^2_{4k}+m^{\ }_{6k}+m^{\ }_{8k}
\right]+
\left(
f^2_2
-
|f^{\ }_1|^2
\right)
{g^{\ }_A\over8\pi},
\label{antiholomoweightmn}
\\
&&
f_1\ =\
-\ {\rm i}\sum_{k=1}^{m+n}
\left(
m^{\ }_{1k}-m^{\ }_{2k}-m^{\ }_{3k}+m^{\ }_{4k}
+
m^{\ }_{5k}-m^{\ }_{6k}-m^{\ }_{7k}+m^{\ }_{8k}
\right),
\label{f1mn}
\\
&&
f_2\ =\
+\sum_{k=1}^{m+n}
\left(
m^{\ }_{1k}+m^{\ }_{2k}-m^{\ }_{3k}-m^{\ }_{4k}
+
m^{\ }_{5k}+m^{\ }_{6k}-m^{\ }_{7k}-m^{\ }_{8k}
\right).
\label{f2mn}
\end{eqnarray}
An important consequence which follows {}from
Eqs. (\ref{scalingofmostgeneral10}-\ref{fin10})
and Eq. (\ref{mostgeneralcompositeoperator})
is that
\end{mathletters}
\begin{mathletters}
\label{m=m-j}
\begin{equation}
\left\langle
{\cal O}^{\dag}(z,\bar z)\
{\cal O}       (0,     0)
\right\rangle^{\ }_{S^{(1,0)}_{\rm cr}}
\ \sim\
\left\langle
{\cal P}^{\dag}(z,\bar z)\
{\cal P}       (0,     0)
\right\rangle^{\ }_{S^{(m-j,j)}_{\rm cr}},
\quad j=0,\cdots,m,
\label{10tom0}
\end{equation}
where
\begin{eqnarray}
&&
{\cal O}(z,\bar z)\ =\
\left(
\varphi^{\dag m^{\ }_5}_+
\varphi^{\dag m^{\ }_6}_-
\varphi^{     m^{\ }_7}_+
\varphi^{     m^{\ }_8}_-
\right)(z,\bar z),
\\
&&
{\cal P}(z,\bar z)\ =\
\left(
\prod_{i=1}^{m^{\ }_5}
\prod_{j=1}^{m^{\ }_6}
\prod_{k=1}^{m^{\ }_7}
\prod_{l=1}^{m^{\ }_8}
\varphi^{\dag}_{i+}
\varphi^{\dag}_{j-}
\varphi^{\   }_{k+}
\varphi^{\   }_{l-}
\right)(z,\bar z),
\\
&&
m\ =\ m^{\ }_5\ +\ m^{\ }_6\ +\ m^{\ }_7\ +\ m^{\ }_8.
\end{eqnarray}
Eq. (\ref{10tom0}) relates correlations in  a model with several copies
of the pair of spinors $(\psi,\varphi)$ to a model with a single pair.
Eq. (\ref{10tom0}) is only strictly true in the limit
$z,\bar z\rightarrow0$. There are subleading corrections
to this algebraic decay, in particular when $j\neq 0,m$.
It is an example of an operator product expansion.
Eq. (\ref{m=m-j}) allows to
relate the operator content in the theory with effective action
$S^{(1,0)}_{\rm cr}$ to impurity averaging of higher powers of
the single-particle Green functions. For example,
it tells us that the scaling dimension
of $\Psi^{\ }_{-mm}$, $m$ a positive integer, in the theory with
effective action $S^{(1,0)}_{\rm cr}$ is the same as the scaling dimension
of the operator whose expectation value in the theory with effective action
$S^{(m,0)}_{\rm cr}$ yields the impurity average
$
\overline{[{\rm tr}\ G(x,x,0;V)]^m}.
$
Indeed, we recall that
\end{mathletters}
\begin{equation}
\overline{\left[{\rm tr}\ G(x,x,0;V)\right]^m}
\ \propto\
\left\langle
\prod_{k=1}^m
\left[
\psi^{\   }_{k+}
\psi^{\dag}_{k-}
+
\psi^{\   }_{k-}
\psi^{\dag}_{k+}
\ +\
\varphi^{\   }_{k+}
\varphi^{\dag}_{k-}
+
\varphi^{\   }_{k-}
\varphi^{\dag}_{k+}
\right](x)
\right\rangle^{\ }_{S^{(m,0)}_{\rm cr}}.
\label{DOSpowerm}
\end{equation}
After expanding the product on the right-hand-side of Eq. (\ref{DOSpowerm}),
we can calculate the scaling dimensions of any of the resulting
$4^m$ terms using Eq. (\ref{mostgeneralcompositeoperator}).
They all share the scaling dimension of
\begin{mathletters}
\label{sharingofscalingdim}
\begin{equation}
{\cal P}(x)\ =\
\prod_{k=1}^m\left(\varphi^{\ }_{k+}\varphi^{\dag}_{k-}\right)(x),
\end{equation}
which in turn can be extracted {}from the algebraic decay of
\begin{equation}
\left\langle
{\cal P}^{\dag}(x){\cal P}^{\ }(y)
\right\rangle^{\ }_{S^{(m,0)}_{\rm cr}}
\ \propto\
\left\langle
\left(
\varphi^{\ }_-\varphi^{\dag}_+
\right)^m(x)
\left(
\varphi^{\ }_+\varphi^{\dag}_-
\right)^m(y)
\right\rangle^{\ }_{S^{(1,0)}_{\rm cr}}.
\end{equation}
\end{mathletters}

\section{Non-Abelian vector gauge randomness}
\label{sec:Non-Abelianvectorgaugerandomnes}

Tight-binding one-body Hamiltonians, for which
the Fermi surface is made of a collection of isolated
points in the Brillouin zone, are well described by,
say, $N$ species of Dirac fermions,
if one is only interested in the long wavelength and low energy
properties of the system. In general, impurities induce
one-body interactions among the $N$ different species. In this section, we
consider such a two-dimensional system in which
impurities can be represented  by a static Abelian vector gauge-like
interaction on the one hand, and by a static non-Abelian vector gauge-like
interaction on the other hand. The Abelian vector gauge potential
$A^{\ }_{\mu}$ does not induce scattering events between different species.
The non-Abelian vector gauge potential $B^a_{\mu}$ does.
The index $\mu=1,2$ denotes the two spatial components of the gauge fields,
and $a=1,\cdots,N^2-1$ labels any basis $\{T^a\}$
of traceless Hermitean $N\times N$
matrices.
The Hamitonian of the pure system is the $2N\times2N$ matrix
\begin{equation}
H^{\ }_0\ =\ -{\rm i}\ I\gamma^{\ }_{\mu}\partial^{\ }_{\mu},
\end{equation}
where $I$ is the $N\times N$ unit matrix.
The impurity potential is given by the $2N\times2N$ matrix
\begin{equation}
V(x)\ =\
\gamma^{\ }_{\mu}
\left[
A^{\ }_{\mu}(x)I
\ +\
B^{\ }_{\mu}(x)
\right],
\label{nonabelianpot}
\end{equation}
where
\begin{equation}
B^{\ }_{\mu}\ =\ B^a_{\mu}\ T^a,
\end{equation}
transforms like the adjoint of the Lie algebra $su(N)$.
Our conventions for the generators of $su(N)$ are
\begin{equation}
{\rm tr} \left(T^aT^b\right)\ =\ {\delta^{ab}\over2},\quad
[T^a,T^b]\ =\ {\rm i}f^{abc}T^c,\quad
f^{acd}f^{bcd}\ =\ N\delta^{ab},\quad
a,b,c\ =\ 1,\cdots,N^2-1.
\label{su(n)conv}
\end{equation}

Any impurity average over a product of single-particle Green
functions can be obtained {}from an effective partition function.
We are only interested in the
single-particle Green functions in which all energies have been
set to zero. Without loss of generality, we will concentrate on the
averaged single-particle Green function.
It can be obtained {}from the expectation value of
$\psi(x)\bar\psi(y)+\varphi(x)\bar\varphi(y)$
with respect to the partition function
\begin{mathletters}
\begin{eqnarray}
{\cal Z}\ &=&\
\int{\cal D}[\bar\psi,\psi,\bar\varphi,\varphi]
\int{\cal D}[A^{\ }_{\mu}]\int{\cal D}[B^{\ }_{\mu}]\
e^{-\int d^2x {\cal L}},
\label{partitionabelianonabelian}
\\
{\cal L}\ &=&\
\bar\psi
\gamma^{\ }_{\mu}
\left(
\partial^{\ }_{\mu}
+{\rm i}A^{\ }_{\mu}I + {\rm i}B^{\ }_{\mu}
\right)
\psi
+
\bar\varphi
\gamma^{\ }_{\mu}
\left(
\partial^{\ }_{\mu}
+{\rm i}A^{\ }_{\mu}I + {\rm i}B^{\ }_{\mu}
\right)
\varphi
\nonumber\\
&+&
{1\over2g^{\ }_A}A^{\ }_{\mu}A^{\ }_{\mu}
+
{1\over g^{\ }_B}{\rm tr}\left(B^{\ }_{\mu}B^{\ }_{\mu}\right).
\label{lagranabelianonabelian}
\end{eqnarray}
The Dirac spinor $\psi$ denotes a Grassmann coherent state, $\varphi$ a
bosonic one.
Both transform like the fundamental representation of $su(N)$.
The vector gauge fields are still dimensionfull,
although they have been rescaled compared to section
\ref{sec:Abelianvectorgaugerandomness}.
The positive impurity strengths $g^{\ }_A$
and  $g^{\ }_B$ act like inverse masses for the Abelian and
non-Abelian vector gauge fields, respectively.
We show in this section that the partition function
in Eq. (\ref{partitionabelianonabelian}) factorizes into four
independent sectors. However,
in contrast to the case of Abelian vector gauge impurity only,
we show that ${\cal Z}$ does not describe a theory at criticality
for all but one value of $g^{\ }_B>0$. Criticality is proved
for the special case $g^{\ }_B=\infty$. In this limit,
${\cal L}$ in Eq. (\ref{lagranabelianonabelian})
is shown to be conformally invariant, and to yield an energy-momentum
tensor with a vanishing central charge in its operator product
expansion in agreement with Ref. \cite{Tsvelik 1995}.
The vanishing of the central charge follows {}from the fact
that by construction ${\cal Z}=1$,
and thus serves as a consistency check.
The operator content of ${\cal Z}$ is constructed for
$g^{\ }_A=0$, $g^{\ }_B=\infty$ and we obtain the new result
that an infinite hierarchy of negative dimensional operators
is present at the (non-Abelian) disordered critical point.
\end{mathletters}

\subsection{Operator content at criticality for non-Abelian vector gauge
randomness}
\label{subsec:Operatorcontent2}

We begin by showing that it is possible to decouple the impurities
{}from the spinor coherent states. To this end, let
\begin{mathletters}
\begin{eqnarray}
&&
A^{\ }_{\bar z}\ \equiv\
A^{\ }_1-{\rm i}A^{\ }_2,\quad
A^{\ }_{     z}\ \equiv\
A^{\ }_1+{\rm i}A^{\ }_2,
\\
&&
B^{\ }_{\bar z}\ \equiv\
B^{\ }_1-{\rm i}B^{\ }_2,\quad
B^{\ }_{     z}\ \equiv\
B^{\ }_1+{\rm i}B^{\ }_2.
\end{eqnarray}
The antiholomorphic components $A^{\ }_{\bar z}$ and $B^{\ }_{\bar z}$
of the gauge fields are {\it independent} {}from the holomorphic components
$A^{\ }_{     z}$ and $B^{\ }_{     z}$. Both antiholomorphic and holomorphic
components of the gauge fields belong to the complex extensions
$u^{c*}(N)$ and $u^{c}(N)$, respectively, of the real Lie Alebra $u(N)$.
The antiholomorphic and holomorphic components are a convenient choice
when working with the chiral basis for the spinors since
\end{mathletters}
\begin{eqnarray}
{\cal L}\ &=&\
\psi^{\dag}_+
\left(
2\partial^{\ }_{\bar z}
+{\rm i}A^{\ }_{\bar z}I + {\rm i}B^{\ }_{\bar z}
\right)
\psi^{\ }_+
+
\psi^{\dag}_-
\left(
2\partial^{\ }_{     z}
+{\rm i}A^{\ }_{     z}I + {\rm i}B^{\ }_{     z}
\right)
\psi^{\ }_-
\nonumber\\
&+&\
\varphi^{\dag}_+
\left(
2\partial^{\ }_{\bar z}
+{\rm i}A^{\ }_{\bar z}I + {\rm i}B^{\ }_{\bar z}
\right)
\varphi^{\ }_+
+
\varphi^{\dag}_-
\left(
2\partial^{\ }_{     z}
+{\rm i}A^{\ }_{     z}I + {\rm i}B^{\ }_{     z}
\right)
\varphi^{\ }_-
\nonumber\\
&+&\
{1\over2g^{\ }_A}A^{\ }_+A^{\ }_-
\ +\
{1\over g^{\ }_B}{\rm tr} \left(B^{\ }_+B^{\ }_-\right).
\end{eqnarray}
We parametrize the components of the gauge fields in the Lie algebras
$u^{c*}(N)$ and $u^{c}(N)$ by fields in the complex extensions
$U^{c*}(N)$ and $U^{c}(N)$ of the Lie group $U(N)$:
\begin{mathletters}
\begin{eqnarray}
&&
A^{\ }_{\bar z}I\ =\
+{\rm i}
\left(2\partial^{\ }_{\bar z} H^{\ }_{\bar z}\right)H^{-1}_{\bar z},
\quad H^{\ }_{\bar z}\ =\ e^{-{\rm i}I\phi^{\ }_{\bar z}},
\quad \phi^{\ }_{\bar z}\ =\ \Phi^{\ }_2 -{\rm i}\Phi^{\ }_1,
\\
&&
A^{\ }_{     z}I\ \equiv\
+{\rm i}
\left(2\partial^{\ }_{     z} H^{\ }_{     z}\right)H^{-1}_{     z},
\quad H^{\ }_{     z}\ =\ e^{-{\rm i}I\phi^{\ }_{     z}},
\quad \phi^{\ }_{     z}\ =\ \Phi^{\ }_2 +{\rm i}\Phi^{\ }_1,
\\
&&
B^{\ }_{\bar z}\ \; \equiv\
+{\rm i}
\left(2\partial^{\ }_{\bar z} G^{\ }_{\bar z}\right)G^{-1}_{\bar z},
\quad G^{\ }_{\bar z}\in\ SU^{c*}(N),
\\
&&
B^{\ }_{     z}\ \; \equiv\
+{\rm i}
\left(2\partial^{\ }_{     z} G^{\ }_{     z}\right)G^{-1}_{     z},
\quad G^{\ }_{     z}\in\  SU^{c}(N).
\end{eqnarray}
This change of variables costs the Jacobian
\end{mathletters}
\begin{equation}
{\cal J}^{\ }_{\rm gh}\ =\
{\rm Det}\ (2\partial^{\ }_{\bar z})
{\rm Det}\ (2\partial^{\ }_{     z})
{\rm Det}\ (\nabla^{\ }_{\bar z})
{\rm Det}\ (\nabla^{\ }_{     z}),
\end{equation}
where
$\nabla^{\ }_{\bar z}$
and
$\nabla^{\ }_{     z}$
are the covariant derivatives in the adjoint
representations of $SU^{c*}(N)$ and $SU^{c}(N)$, respectively. For example
\begin{eqnarray}
\delta A^{\ }_{\bar z}\ &=&\
{\rm i}\ \delta
\left(
(2\partial^{\ }_{\bar z}\ G^{\ }_{\bar z})\
G^{-1}_{\bar z}
\right)
\nonumber\\
&=&\
\Big\{
2\partial^{\ }_{\bar z}\ \cdot\
+\ {\rm i}\
\left[
{\rm i}\
(2\partial^{\ }_{\bar z}\ G^{\ }_{\bar z})\
G^{-1}_{\bar z}\ ,\
\cdot\
\right]
\Big\}
\left(
{\rm i}\
(\delta\ G^{\ }_{\bar z})\
G^{-1}_{\bar z}
\right).
\end{eqnarray}
We represent this Jacobian as a path integral over auxiliary Grassmann
fields (ghosts):
\begin{mathletters}
\begin{eqnarray}
&&
{\rm Det}(2\partial^{\ }_{\bar z})\ =\
\int{\cal D}\ [\beta^{0 }_+,\alpha^{0 }_+]\
e^{
+{\rm i}\int\ d^2x\
2
{\rm tr}(\beta^0_+T^0\ {\rm i}2\partial_{\bar z} T^0\alpha^0_+)
},
\\
&&
{\rm Det}(2\partial^{\ }_{     z})\ =\
\int{\cal D}\ [\beta^{0 }_-,\alpha^{0 }_-]\
e^{
+{\rm i}\int\ d^2x\
2
{\rm tr}(\beta^0_-T^0\ {\rm i}2\partial_{     z} T^0\alpha^0_-)
},
\\
&&
{\rm Det}(\nabla_{\bar z})\ =\
\int{\cal D}\ [\beta^{a }_+,\alpha^{a }_+]\
e^{
+{\rm i}\int\ d^2x\
2
{\rm tr}(\beta^a_+T^a\ {\rm i}\nabla_ {\bar z} T^b\alpha^b_+)
},
\\
&&
{\rm Det}(\nabla_{     z})\ =\
\int{\cal D}\ [\beta^{a }_-,\alpha^{a }_-]\
e^{
+{\rm i}\int\ d^2x\
2
{\rm tr}(\beta_-  ^aT^a\ {\rm i}\nabla_{    z} T^b\alpha^b_-)
}.
\end{eqnarray}
The unit matrix $I$ has been rescaled to
$T^0\equiv{1\over\sqrt{2N}}I$ for notational convenience.
\end{mathletters}

We are now ready to decouple the vector gauge fields
{}from all other fields. For example, the chiral transformation
\begin{mathletters}
\begin{eqnarray}
&&
\psi^{\dag}_+\ =\
\psi'^{\dag}_+\ G^{-1}_{\bar z}H^{-1}_{\bar z},
\qquad
\psi^{\ }_+\ =\
H^{\ }_{\bar z}G^{\ }_{\bar z}\ \psi'_+,
\\
&&
\psi^{\dag}_-\ =\
\psi'^{\dag}_-\ G^{-1}_{     z}H^{-1}_{     z},
\qquad
\psi^{\ }_-\ =\
H^{\ }_{     z}G^{\ }_{     z}\ \psi'_-,
\end{eqnarray}
decouples the $\psi$ sector {}from the vector gauge fields at the price of
the Jacobian \cite{Fujikawa 1979}
\end{mathletters}
\begin{equation}
{\cal J}^{\ }_{\psi}\ =\
{
{\rm Det}
\left[
{\rm i}\gamma^{\ }_{\mu}\left(\partial^{\ }_{\mu}+{\rm i}A^{\ }_{\mu}I\right)
\right]
\over
{\rm Det}
\left(
{\rm i}\gamma^{\ }_{\mu}\partial^{\ }_{\mu}
\right)
}
\times
{
{\rm Det}
\left[
{\rm i}\gamma^{\ }_{\mu}\left(\partial^{\ }_{\mu}+{\rm i}B^{\ }_{\mu}\right)
\right]
\over
{\rm Det}
\left(
{\rm i}\gamma^{\ }_{\mu}\partial^{\ }_{\mu}
\right)
}.
\end{equation}
Similarly, the chiral transformation
\begin{mathletters}
\begin{eqnarray}
&&
\beta^a_+T^a\ =\
G^{\ }_{\bar z}
\ \beta'^a_+T^a\
G^{-1}_{\bar z},
\qquad
\alpha^a_+T^a\ =\
G^{\ }_{\bar z}
\ \alpha'^a_+T^a\
G^{-1}_{\bar z},
\\
&&
\beta^a_-T^a\ =\
G^{\ }_{     z}
\ \beta'^a_-T^a\
G^{-1}_{     z},
\qquad
\alpha^a_-T^a\ =\
G^{\ }_{     z}
\ \alpha'^a_-T^a\
G^{-1}_{     z},
\end{eqnarray}
decouples the ghosts {}from the vector gauge fields at the price
of the Jacobian \cite{Polyakovhouches}
\end{mathletters}
\begin{equation}
{\cal J}^{\ }_{\alpha^1,\cdots,\alpha^N}\ =\
\left\{
{
{\rm Det}
\left[
{\rm i}\gamma^{\ }_{\mu}\left(\partial^{\ }_{\mu}+{\rm i}B^{\ }_{\mu}\right)
\right]
\over
{\rm Det}
\left(
{\rm i}\gamma^{\ }_{\mu}\partial^{\ }_{\mu}
\right)
}
\right\}^{2N}.
\end{equation}
The exponent $2N$ is twice the quadratic Casimir eigenvalue $c^{\ }_v=N$
in the adjoint representation as is apparent {}from Eq. (\ref{su(n)conv}).
Finally,
by performing exactly the same chiral transformation on the $\varphi$'s
as we did on the $\psi$'s, we can entirely decouple the spinor sector {}from
the
vector gauge sector. Decoupling the $\varphi$'s {}from the gauge fields costs
a Jacobian
\begin{equation}
{\cal J}^{\ }_{\varphi}\ =\
{\cal J}^{-1}_{\psi},
\end{equation}
which cancels the one coming {}from the $\psi$ sector.
We are then left with the Jacobian
${\cal J}^{\ }_{\alpha^1,\cdots,\alpha^{N}}$
{}from the ghost sector.

Polyakov and Wiegmann have calculated
${\cal J}^{\ }_{\alpha^1,\cdots,\alpha^{N}}$.
It is given by the exponential of the Euclidean Wess-Zumino-Witten action
\cite{Polyakov 1983}:
\begin{equation}
{\cal J}^{\ }_{\alpha^1,\cdots,\alpha^{N}}\ =\
e^{-(-2N)\Gamma[G^{-1}_{     z}G^{\ }_{\bar z}]},
\end{equation}
where the Euclidean Wess-Zumino-Witten action is given by
\begin{eqnarray}
\Gamma[G]\ &=&\
{1\over8\pi}\
\int\ dx^{\ }_1dx^{\ }_2\
{\rm tr}
\left[
(\partial^{\ }_{\mu}\ G^{\ })\
(\partial^{\ }_{\mu}     \ G^{-1})
\right]
\nonumber\\
\ &-&\
{{\rm i}\over12\pi}\
\int_{ {\rm B},\ \partial{\rm B}={\rm S}^2 }\
dx^{\ }_1dx^{\ }_2dx^{\ }_3\
\epsilon^{\ }_{\mu\nu\lambda}\
{\rm tr}
\left[
(\partial^{\ }_{\mu}\ G)G^{-1}\
(\partial^{\ }_{\nu}\ G)G^{-1}\
(\partial^{\ }_{\lambda}\ G)G^{-1}\
\right].
\end{eqnarray}

Before collecting all contributions to the effective Lagrangian density,
we rewrite the probability distributions for the vector gauge fields in
a more useful form. In the Abelian sector, the Gaussian
weight is
\begin{equation}
+{1\over2g^{\ }_A}A^{\ }_{\bar z}A^{\ }_{     z}\ =\
-{1\over2g^{\ }_AN}{\rm tr}\
\left[
\left(2\partial^{\ }_{\bar z}H^{\ }_{\bar z}\right)H^{-1}_{\bar z}
\left(2\partial^{\ }_{     z}H^{\ }_{     z}\right)H^{-1}_{     z}
\right].
\end{equation}
In the non-Abelian sector, the Gaussian weight is
\begin{equation}
+{1\over g^{\ }_B}{\rm tr}\ (B^{\ }_{\bar z}B^{\ }_{     z})\ =\
-{1\over g^{\ }_B}{\rm tr}\
\left[
\left(2\partial^{\ }_{\bar z}G^{\ }_{\bar z}\right)G^{-1}_{\bar z}
\left(2\partial^{\ }_{     z}G^{\ }_{     z}\right)G^{-1}_{     z}
\right].
\end{equation}
With the help of the Polyakov-Wiegmann identity \cite{Polyakov 1984}
\begin{eqnarray}
&&
\Gamma[G^{-1}H]\ =\
\Gamma[G^{-1}]+\Gamma[H]+
{1\over4\pi}
\int dx^{\ }_1dx^{\ }_2\
{\rm tr}\
\left[
(2\partial^{\ }_{     z}G^{\ })G^{-1}\
(2\partial^{\ }_{\bar z}H^{\ })H^{-1}
\right],
\end{eqnarray}
the Gaussian weights for the Abelian and non-Abelian vector gauge fields
are rewritten
\begin{equation}
+{1\over2g^{\ }_A}A^{\ }_{\bar z}A^{\ }_{     z}
\ =\
-{\pi\over g^{\ }_AN}
\left(
\Gamma[H^{-1}_{     z}H^{\ }_{\bar z}]
-
\Gamma[H^{\ }_{     z}H^{\ }_{\bar z}]
\right),
\end{equation}
and
\begin{equation}
+{1\over g^{\ }_B}{\rm tr}\ (B^{\ }_{\bar z}B^{\ }_{     z})
\ =\
-{4\pi\over g^{\ }_B}
\left(
\Gamma[G^{-1}_{     z}G^{\ }_{\bar z}]
-
\Gamma[G^{-1}_{     z}]
-
\Gamma[G^{\ }_{\bar z}]
\right),
\end{equation}
respectively.
In the Abelian case, the mass term separates into two independent sectors.
In the non-Abelian case, due to the topological term in the
Wess-Zumino-Witten action, the mass term does not separate
into two independent sectors.

We conclude that the trace of the
averaged single-particle Green function in which
all energies have been set to zero can be obtained {}from the
expectation value of
\begin{eqnarray}
&&
{\rm tr}\left[\psi(x)\bar\psi(y)+\varphi(x)\bar\varphi(y)\right]=
\nonumber\\
&&
{\rm tr}\left[
\left(H^{\ }_{\bar z}G^{\ }_{\bar z}\psi'_+\right)(x)
\left(\psi'^{\dag}_-G^{-1}_{     z}H^{-1}_{     z}\right)(y)
+
\left(H^{\ }_{     z}G^{\ }_{     z}\psi'_-\right)(x)
\left(\psi'^{\dag}_+G^{-1}_{\bar z}H^{-1}_{\bar z}\right)(y)
\right]+
\nonumber\\
&&
{\rm tr}\left[
\left(H^{\ }_{\bar z}G^{\ }_{\bar z}\varphi'_+\right)(x)
\left(\varphi'^{\dag}_-G^{-1}_{     z}H^{-1}_{     z}\right)(y)
\ +\
\left(H^{\ }_{     z}G^{\ }_{     z}\varphi'_-\right)(x)
\left(\varphi'^{\dag}_+G^{-1}_{\bar z}H^{-1}_{\bar z}\right)(y)
\right],
\end{eqnarray}
with respect to the partition function
\begin{mathletters}
\begin{eqnarray}
&&
{\cal Z} =
\int{\cal D}
[\psi'^{\dag}_{\pm},\psi'_{\pm},\varphi'^{\dag}_{\pm},\varphi'_{\pm}]
\int{\cal D}
[\beta'^{a }_{\pm},\alpha'^{a }_{\pm}]
\int{\cal D}
[H^{\ }_{\bar z},H^{\ }_{     z}]
\int{\cal D}
[G^{\ }_{\bar z},G^{\ }_{     z}]\
e^
{
-
\left(
S^{\ }_1+S^{\ }_2+S^{\ }_3+S^{\ }_4
\right)
},
\\
&&
S^{\ }_1\ =\
\int d^2 x
\left(
\psi   '^{\dag}_+2\partial^{\ }_{\bar z}\psi   '^{\ }_+
\ +\
\psi   '^{\dag}_-2\partial^{\ }_{     z}\psi   '^{\ }_-
\ +\
\varphi'^{\dag}_+2\partial^{\ }_{\bar z}\varphi'^{\ }_+
\ +\
\varphi'^{\dag}_-2\partial^{\ }_{     z}\varphi'^{\ }_-
\right),
\\
&&
S^{\ }_2\ =\
2
\sum_{a,b=0}^{N}
\int d^2 x\
{\rm tr}
\left(
\beta'^a_+T^a2\partial^{\ }_{\bar z}\alpha'^b_+T^b
+
\beta'^a_-T^a2\partial^{\ }_{     z}\alpha'^b_-T^b
\right),
\\
&&
S^{\ }_3\ =\
-{\pi\over g^{\ }_AN}\Gamma[H^{-1}_{     z}H^{\ }_{\bar z}]
+{\pi\over g^{\ }_AN}\Gamma[H^{\ }_{     z}H^{\ }_{\bar z}],
\\
&&
S^{\ }_4\ =\
-\left[{4\pi\over g^{\ }_B}+2N\right]
\Gamma[G^{-1}_{     z}G^{\ }_{\bar z}]
+
{4\pi\over g^{\ }_B}\Gamma[G^{-1}_{     z}]
+
{4\pi\over g^{\ }_B}\Gamma[G^{\ }_{\bar z}].
\end{eqnarray}
We can use the invariance of the Haar measure of $U^{c*}(N)$ ($U^{c}(N)$)
under left and right multiplication to reparametrize the Abelian
vector gauge components $H^{\ }_{\bar z}$ and $H^{\ }_z$ as
$H^{-1}_zH^{\ }_{\bar z}$ and $H^{\ }_zH^{\ }_{\bar z}$.
The partition function has factorized into {\it four independent} sectors.
The same factorization takes place for partition functions representing the
impurity average over an arbitrary product of single-particle Green functions,
as long as all energies in the Green functions are set to zero.
However, the factorization property does not guaranty criticality, i.e.,
algebraic decay of all correlation functions.
\end{mathletters}

The first sector describes a free
massless relativistic theory with $U(N/N)$ graded symmetry. All Dirac
spinors transform like the fundamental representation of $U(N)$.
This theory is conformally invariant with
primary fields $\psi^{\ }_{\pm}$ and $\varphi^{\ }_{\pm}$
carrying the conformal weights
\begin{equation}
(h,\bar h)\ =\ ({1\over2},0)\quad {\rm or}\quad (0,{1\over2}),
\end{equation}
depending on their chirality.
The first sector has vanishing central charge
\begin{equation}
c^{\ }_1=0,
\end{equation}
since the partition function restricted to  this sector is unity.

The second sector describes free massless relativistic ghosts
transforming like the adjoint of $U(N)$. This theory has a conformally
invariant energy-momentum tensor. The operator product expansion of the
energy-momentum components yields the {\it negative central charge}
\cite{Polyakovhouches}
\begin{equation}
c^{\ }_2\ =\ -2{\rm dim}\ U(N)\ =\ -2N^2.
\end{equation}

The third sector describes two independent and free scalar field theories
and is thus conformally invariant.
One has negative definite kinetic energy. Consequently,
its primary field
$H^{-1}_{     z}H^{\ }_{\bar z}\propto
e^{{\rm i}(2{\rm i}\Phi^{\ }_1)I}$
carries negative conformal weights
\begin{equation}
h\ =\ \bar h\ =\ -{g^{\ }_A\over8\pi}.
\end{equation}
The scalar field theory with negative definite kinetic energy
corresponds to the longitudinal component of the Abelian vector gauge field.
The second scalar field theory has the primary field
$H^{  }_{     z}H^{\ }_{\bar z}\propto e^{2{\rm i}\Phi^{\ }_2I}$
with positive conformal weight
\begin{equation}
h\ =\ \bar h\ =\ +{g^{\ }_A\over8\pi}.
\end{equation}
It corresponds to
the transversal component of the Abelian vector gauge field.
The central charge in the third sector is
\begin{equation}
c^{\ }_3\ =\ 2.
\end{equation}

The last sector describes the non-Abelian vector gauge fields.
In contrast to the Abelian vector gauge sector, it is not conformally
invariant for all values of the non-Abelian impurity strangth $g^{\ }_B$.
Only the limit $g^{\ }_B\rightarrow\infty$
\footnote
{
It is amusing to note that for negative values (and thus unphysical)
of the non-Abelian impurity strength $g^{\ }_B=-{2\pi\over N}$,
the non-abelian sector is also ``critical''.
}
yields a conformally invariant
theory, namely the $SU(N)^{\ }_k$ Wess-Zumino-Witten theory where
$k=-2N$ is the level of the underlying Kac-Moody algebra
\cite{Knizhnik 1984}.
The primary field $G^{-1}_{    z}G^{\ }_{\bar z}$
carries the conformal weight \cite{Knizhnik 1984}
\begin{equation}
h\ =\ \bar h\ =\
{1\over N+k} {N^2-1\over2N},\qquad k\ =\ -2N,
\end{equation}
which is thus negative. The associated central charge \cite{Knizhnik 1984}
\begin{equation}
c^{\ }_{SU(N)^{\ }_k }\ =\ {k\over k+N}(N^2-1),
\qquad k\ =\ -2N,
\end{equation}
is positive. In the limit $g^{\ }_B\rightarrow\infty$ ,
there exists a local non-Abelian gauge symmetry with
$G^{\ }_{    z}G^{\ }_{\bar z}$ a pure gauge
\cite{Karabali 1990}.
Integration over the pure gauge enforces the local constraint
\begin{equation}
J^a_{\mu}\ =\ 0,\qquad\mu=1,2,\quad a=1,\cdots,N^2-1,
\end{equation}
for the non-Abelian currents
\begin{equation}
J^a_{\mu}\ =\
\bar\psi T^a\gamma^{\ }_{\mu} \psi
\ +\
\bar\varphi T^a\gamma^{\ }_{\mu} \varphi,
\qquad\mu=1,2,\quad a=1,\cdots,N^2-1.
\end{equation}
Thus, only $SU(N)$ gauge singlets can acquire an expectation value
in this limit.

For all correlation functions calculated with the partition function in
Eq. (\ref{partitionabelianonabelian})
to be algebraic, one must take the limit
$g^{\ }_B\rightarrow\infty$. In this limit, the existence of a local gauge
symmetry requires a careful definition of the measure of the vector gauge
fields. Gauge fixing is in fact implicit in the chiral transformation which
decouples the non-Abelian vector gauge fields {}from the spinors and ghosts.
Indeed, one should think of the chiral transformation as the product of
a pure gauge transformation, which amounts to gauge fixing, and of an axial
transformation to achieve complete decoupling \cite{Karabali 1990}.
After gauge fixing, only
gauge inequivalent configurations must be summed over
in the partition function. Hence, the partition function restricted to
the non-Abelian vector gauge sector is given by
\begin{equation}
{\cal Z}^{\infty}_4\ =\
\int{\cal D}[G']\
e^
{
-
(-2N)\Gamma[G']
},
\label{purenonabelianpartition}
\end{equation}
in the limit $g^{\ }_B\rightarrow\infty$.
To verify the correctness of Eq. (\ref{purenonabelianpartition}),
we calculate the total
central charge:
\begin{equation}
c\ =\ c^{\ }_1\ +\ c^{\ }_2\ +\ c^{\ }_3\ +\ c^{\ }_{SU(N)^{\ }_{-2N}}\ =0.
\end{equation}
It vanishes as it should be.

The conformal weights of all primary field
$\{\phi^{\ }_l\}$ of the $SU(N)^{ }_k$
Wess-Zumino Witten theory have been calculated by
Knizhnik and Zamoklodchikov \cite{Knizhnik 1984}.
They are given by
\begin{equation}
     h\ =\ {c^{\ }_l       \over c^{\ }_v+k},\qquad
\bar h\ =\ {c^{\ }_{\bar l}\over c^{\ }_v+k}.
\label{allprimaryfieldsWZW}
\end{equation}
In the numerator,
$c^{\ }_l$ ($c^{\ }_{\bar l}$) stands for the quadratic Casimir eigenvalue of
the irreducible representation $l$ ($\bar l$).
Here, the representation $l$ ($\bar l$) is defined by the transformation law
obeyed by $\phi^{\ }_l$
under $SU(N)$ rotations generated by the holomorphic
(antiholomorphic) component of the level $k$ Kac-Moody currents.
Thus, by analytical continuation of $k$ to the value $k=-2N$,
we see that all primary fields originating {}from the non-Abelian vector gauge
impurity in the limit $g^{\ }_B=\infty$ have negative conformal weights
\begin{equation}
     h\ =\ -{c^{\ }_l       \over N},\qquad
\bar h\ =\ -{c^{\ }_{\bar l}\over N}.
\end{equation}
The eigenvalues $\{c^{\ }_l\}$ are known.
Examples are
\begin{eqnarray}
&&
c^{\ }_{\rm A}\ =\ {(N-2)(N+1)\over N},\quad
c^{\ }_{\rm fun}\ =\  {N^2-1\over2N},  \quad
c^{\ }_{\rm ad}\ =\ N,                 \quad
c^{\ }_{\rm S}\ =\ {(N+2)(N-1)\over N},
\end{eqnarray}
in the antisymmetric $\Box\atop\Box$, fundamental $\Box$, adjoint,
and symmetric $\Box\Box$ representations, respectively.
Here, $\Box$ represents a box in a Young tableau. We recall that all
irreducible representations of $U(N)$ and $SU(N)$ can be labelled by
Young tableaus made of $N$ rows with $f^{\ }_i$ boxes in the $i$-th
row, provided $f^{\ }_1>f^{\ }_2>\cdots>f^{\ }_{N-1}>f^{\ }_N\geq0$.
For $SU(N)$, $f^{\ }_N$ always vanishes. It is known
\cite{perelomov 1966}
that the $SU(N)$ irreducible representation with Young tableau
$\{ f^{\ }_i\}$ made of $f=\sum_{i=1}^N f^{\ }_i$ boxes, has the quadratic
Casimir eigenvalue
\begin{equation}
c^{\ }_{\{ f^{\ }_i\}}\ =\
{1\over2}\sum_{1=1}^N\left[f^2_i+(N+1-2i)f^{\ }_i\right]\ -\
{f^2\over2N}.
\label{casimiranyyoungtableau}
\end{equation}

We are now ready to compute the {\it most relevant scaling dimensions}
controlling the expectation value of the local density operator
$
\bar\psi\psi+\bar\varphi\varphi
$
and of higher powers thereof in the limit
$g^{\ }_A=0,\ g^{\ }_B\rightarrow\infty$.
Without loss of generality, we only need to consider powers
of $\varphi^{\dag}_-\varphi^{\ }_+$.
First and in agreement with Refs. \cite{Nersesyan 1994,Tsvelik 1995},
the conformal weights of
\begin{equation}
\left(\varphi^{\dag}_-\varphi^{\ }_+\right)(z,\bar z)\ =\
\left(\varphi'^{\dag}_-\ G'\ \varphi'_+\right)(z,\bar z),
\end{equation}
are given by the quadratic Casimir operator in the fundamental
representation $\Box$ of $SU(N)$:
\begin{equation}
h\ =\ \bar h\ =
\left(
{1\over2}+{1\over N+k}{N^2-1\over2N}
\right)^{\ }_{k=-2N}\ =\
{1\over2N^2}.
\label{hofbox}
\end{equation}
Next, higher powers of $\varphi^{\dag}_-\varphi^{\ }_+$ are defined through
point-splitting:
\begin{equation}
\left(\varphi^{\dag}_-\varphi^{\ }_+\right)^m(z,\bar z)\ =\
\lim_{{z^{\ }_i\rightarrow z\atop\bar z^{\ }_i\rightarrow \bar z}}
\prod_{i=1}^m \left(\varphi^{\dag}_-\varphi^{\ }_+\right)
(z^{\ }_i,\bar z^{\ }_i)\ =\
\lim_{{z^{\ }_i\rightarrow z\atop\bar z^{\ }_i\rightarrow \bar z}}
\prod_{i=1}^m \left(\varphi'^{\dag}_-G'\varphi'_+\right)
(z^{\ }_i,\bar z^{\ }_i).
\end{equation}
We have already dealt with the product of $\varphi'$'s in section
\ref{subsec:Operatorcontent1}.
The operator product expansion in the Wess-Zumino-Witten sector can be used
to replace the point-splitted product of $G'$'s by a sum over
primary fields $\phi^{\ }_l(z,\bar z)$. Each  $\phi^{\ }_l(z,\bar z)$
in the sum corresponds to an irreducible representation present in the
irreducible decomposition of the product representation
$\Box\otimes\cdots\otimes\Box= \Box\cdots\Box\oplus\cdots$.
It is thus the primary field $\phi^{\ }_m$ associated to the
irreducible representation with the largest quadratic Casimir eigenvalues
$c^{\ }_m$ and $c^{\ }_{\bar m}$ in the irreducible decomposition of
$\Box\otimes\cdots\otimes\Box$
which determines the conformal weights up to
subleading corrections:
\begin{equation}
     h\ =\ {m\over2}\ -\ {c^{\ }_m       \over N},\quad
\bar h\ =\ {m\over2}\ -\ {c^{\ }_{\bar m}\over N}.
\label{largecasimir}
\end{equation}
A first example is \cite{Tsvelik 1995}
\begin{mathletters}
\begin{eqnarray}
&&
\left\langle
\left(\varphi^{\dag}_-\varphi^{\ }_+\right)^2(z,\bar z)
\left(\varphi^{\dag}_+\varphi^{\ }_-\right)^2(0,0)
\right\rangle
\ \propto\
z^{-2h}\bar z^{-2\bar h},
\\
&&
h\ =\ \bar h\ =
\left(
1+{(N+2)(N-1)\over N(N+k)}
\right)^{\ }_{k=-2N}\ =\
-{N-2\over N^2}.
\label{symmetriconehalftimesonehalf}
\end{eqnarray}
The conformal weights are those of the primary field
transforming like the symmetric representation $\Box\Box$.
They vanish for $SU(2)$ and are negative when $N>2$.
Another example of a composite local operator with
vanishing conformal weights is \cite{Tsvelik 1995}
\end{mathletters}
\begin{mathletters}
\begin{eqnarray}
&&
\left\langle
\left(\varphi^{\dag}_-\varphi^{\ }_+
      \varphi^{\dag}_+\varphi^{\ }_-\right)(z,\bar z)
\left(\varphi^{\dag}_-\varphi^{\ }_+
      \varphi^{\dag}_+\varphi^{\ }_-\right)(0,     0)
\right\rangle
\ \propto\
z^{-2h}\bar z^{-2\bar h},
\\
&&
h\ =\ \bar h\ =
\left(
1+{N\over N+k}
\right)^{\ }_{k=-2N}\ =\
0.
\label{adjoint}
\end{eqnarray}
The conformal weights are those of the primary field transforming like the
adjoint representation.
One readily sees {}from Eq. (\ref{casimiranyyoungtableau}) that the
completely symmetric representation $\Box\cdots\Box=\{m,0\cdots,0\}$,
which corresponds to one row of $m$ boxes, has the largest
quadratic Casimir eigenvalue among all irreducible representations made
of $m$ boxes (for example, for $SU(2)$ it would correspond
to the representation with largest possible ``total spin'' ${m\over2}$).
Hence, we find the important result
\end{mathletters}
\begin{mathletters}
\begin{eqnarray}
&&
\left\langle
\left(\varphi^{\dag}_-\varphi^{\ }_+\right)^m(z,\bar z)
\left(\varphi^{\dag}_+\varphi^{\ }_-\right)^m(0,0)
\right\rangle
\ \propto\
z^{-2h}\bar z^{-2\bar h},
\\
&&
h\ =\ \bar h\ =\
{m\over2}\ -\
{N-1\over2N^2}\left(m^2+Nm\right).
\label{onerowmboxes}
\end{eqnarray}
As was the case along the critical line $g^{\ }_A>0,\ g^{\ }_B=0$,
there are infinitely many local composite operators with negative scaling
dimensions along the critical line $g^{\ }_A=0,\ g^{\ }_B=\infty$.
\end{mathletters}

\section{Scaling away {}from criticality}
\label{sec:Scalingoffcriticality}

In this section, we are going to calculate how some averaged
quantities such as the local density of states,
generalized inverse participation ratios, and their spatial correlations
scale with respect to energy and length.
We begin with the dynamical scaling exponent and with the local
density of states.

\subsection{Dynamical scaling exponent $z$ and scaling of $\rho(\omega)$
away {}from criticality}
\label{sub:Thedynamicalscalingexponentz}

The dynamical scaling exponent $z$ relates scaling
with energy to scaling with respect to length. We consider the local order
parameter ${\cal M}(x)$ defined by
\begin{equation}
{\cal M}(x)\ \equiv\ \left(\bar\psi\psi+\bar\varphi\varphi\right)(x),
\end{equation}
since it is coupled to the complex energy $E^{\ }_{\pm}=\omega\pm{\rm i}\eta$
in the retarded and advanced single-particle Green functions, respectively.
Notice that ${\cal M}(x)$ breaks the global $U(1/1)\times U(1/1)$ invariance.
We continue working in real space and choose a
renormalization group scheme defined by the scale transformation
\begin{equation}
\tilde x\ =\ a\ x, \quad a>0.
\label{tildex}
\end{equation}
The scaling dimension of $E$ or, for that matter, those of
$\psi$, $\varphi$, $A^{\ }_{\mu}$, $B^{\ }_{\mu}$,
are defined by the requirement that the effective
action in terms of the rescaled coordinates, fields, and coupling constants
has the same form as that in terms of the original coordinates, fields,
and coupling constants. The scaling dimensions of all local fields have been
calculated in sections \ref{sec:Abelianvectorgaugerandomness}
and \ref{sec:Non-Abelianvectorgaugerandomnes} along the disordered
critical line $g^{\ }_A\geq0$, $g^{\ }_B=\infty$.
The impurity strengths $g^{\ }_A$
and $g^{\ }_B$ are left unchanged by Eq. (\ref{tildex}),
since they couple conserved currents to vector gauge potentials.
The energy scale $E$ transforms like
\begin{equation}
\tilde E\ \equiv\ a^{-z}\ E,\quad
z\ =\ 1+{g^{\ }_A\over\pi}+{N^2-1\over N^2},
\label{scalingE}
\end{equation}
along the critical line $g^{\ }_A\geq0,\ g^{\ }_B=\infty$,
as follows {}from Eqs.  (\ref{hofbox}) and (\ref{weightsofPsi}) and
\begin{mathletters}
\begin{eqnarray}
&&
\tilde E\int d^2 \tilde x\
\left(\tilde\varphi^{\dag}_+\tilde\varphi^{\ }_-\right)(\tilde x)\ =\
\tilde E\ a^{2-(h+\bar h)}
\int d^2x\ \left(\varphi^{\dag}_+\varphi^{\ }_-\right)(x),
\\
&&
h+\bar h\ =\
1-{g^{\ }_A\over\pi}-
{N^2-1\over N^2}\ \equiv\
2-z.
\label{2-z}
\end{eqnarray}

The dynamical exponent $z$ allows
to turn scaling with respect to length into scaling with respect to energy.
When only Abelian vector gauge disorder
is present, $z=1+{g^{\ }_A\over\pi}$
is always larger than one and increases linearly with
the impurity strength. It reaches the value $z=2$
when $g^{\ }_A=\pi$. At this value, the scaling
dimension of the local symmetry breaking field ${\cal M}(x)$ vanishes,
a signal of spontaneous symmetry breaking.
The possibility of spontaneous symmetry breaking cannot be ruled out
by the Mermin-Wagner theorem as would be the case in two-dimensional
unitary conformal field theory,
since the disordered critical theory is not unitary.
On the other hand,
in the presence of non-Abelian vector gauge disorder only, the dynamical
exponent $z=2-{1\over N^2}$
is always smaller than two. Hence, the large impurity
strength limit appears to be very different depending on whether
the vector gauge impurity is Abelian or non-Abelian.
\end{mathletters}

We now perturb the effective action $S^{(1,0)}_{\rm cr}$ at the critical point
$g^{\ }_A\geq0$, $g^{\ }_B=\infty$ with $\omega\int d^2x\ {\cal M}(x)$.
To lowest order in $\omega$,
the averaged local density of states $\rho(\omega)$ scales like
\begin{equation}
\rho(\omega)\ \propto\
|\omega|^{\beta},\quad
\beta\ =\ {2-z\over z}\ +\ {\cal O}(\omega),
\label{scalingwithenergydos}
\end{equation}
since, according to  Eq. (\ref{2-z}),
the scaling dimension of $\rho(\omega)$ is $2-z$.
We see that in the presence of Abelian vector gauge-like disorder only,
the averaged local density of states is
finite at criticality for the impurity strength $g^{\ }_A=\pi$.
For larger values of $g^{\ }_A$, $\rho(\omega)$ diverges
in the limit $\omega\rightarrow0$.
In the presence of non-Abelian vector gauge disorder only, the
scaling exponent $\beta=(2N^2-1)^{-1}$
of the averaged local density of states decreases monotonically
as a function of
$N$ with the values $\beta={1\over7}$ when $N=2$ and
$\beta=0$ when $N\rightarrow\infty$.

If the system is not in the thermodynamic limit but in a finite volume $L^2$,
we recover Eq. (\ref{scalingwithenergydos}) with the Ansatz
\begin{equation}
\rho(\omega,L)\ =\
F^{\ }_{\beta}(\omega L^z)|\omega|^{\beta},
\quad \omega\ll L^{-z},
\label{scalingwithenergydosL}
\end{equation}
provided
\begin{equation}
\lim_{x\rightarrow\infty} F^{\ }_{\beta}(x)\ =\ K^{\ }_{\beta},
\quad
0\leq K^{\ }_{\beta}<\infty,
\end{equation}
holds. In a finite volume, the total number of energy eigenstates must
be finite. This is always so (except in the limit $g^{\ }_A\rightarrow\infty$),
since $\beta>-1$.

\subsection{Scaling of averaged generalized participation ratios}
\label{Scaling of averaged generalized participation ratios}

In this subsection, we turn our attention to the scaling
of averaged generalized inverse participation ratios for three reasons.
First, the scaling of averaged generalized inverse participation ratios
was calculated perturbatively in powers of $\epsilon$, where space is
$2+\epsilon$-dimensional, by Wegner for the non-linear $\sigma$ model
of the metal-insulator transition at the mobility edge
\cite{Wegner 1980,Wegner 1987}.
We will obtain non-perturbatively the dominant scaling exponents
of averaged generalized inverse participation ratios for Dirac fermions
with vector potential disorder.
Second, we will see that the negative dimensional operators found
in sections
\ref{sec:Abelianvectorgaugerandomness} and
\ref{sec:Non-Abelianvectorgaugerandomnes}
control the scaling of averaged generalized inverse participation ratios.
Thus, negative dimensional operators are not mere curiosities of the
disordered critical line $g^{\ }_A>0,g^{\ }_B=\infty$.
Finally, we reexamine arguments suggesting a relationship between
the scaling exponents of averaged generalized inverse participation ratios
and the existence of a typical multifractal wave function with vanishing
eigenenergy (critical wave function) \cite{Castellani 1986}.

The concept of the inverse participation ratio for the states of a
disordered system was introduced to distinguish localized {}from
extended states \cite{Thouless 1974}.
Let $\left|\Psi\right\rangle$ be any state defined in the Hilbert space
representing a single-particle tight-binding Hamiltonian. The Hamiltonian
describes, say, hopping on a finite $d$-dimensional lattice $\Lambda$
in the presence of
a random site potential (Anderson model). There are
$|\Lambda|$ sites labelled by $l=1,\cdots,|\Lambda|<\infty$.
All states are normalized to one:
\begin{equation}
\sum_{l\in\Lambda} |\langle l|\Psi\rangle|^2\ =\
\sum_{l\in\ {\rm supp}\Psi} |\langle l|\Psi\rangle|^2\ =\
1.
\end{equation}
Here, ${\rm supp}\Psi$ denotes all sites having a non-vanishing overlap
with $|\Psi\rangle$.
The inverse participation ratio is the functional
\begin{equation}
{\rm P}^{(2)}_{\Lambda}[\Psi]\ \equiv\
\sum_{l\in\ {\rm supp}\Psi}
\left|\langle l|\Psi\rangle\right|^4,
\end{equation}
taking values in $]0,1]$.
Let now $\Omega$ be any subset of $\Lambda$ and define the
normalized state $\left|\Psi^{\ }_{\Omega}\right\rangle$
by
\begin{equation}
\left\langle l|\Psi^{\ }_{\Omega}\right\rangle\ =\
\cases
{
|\Omega|^{-{1\over 2}},&$\forall l\in\Omega$,\cr
&\cr
0,&otherwise.\cr
}
\end{equation}
On the one hand, the inverse participation ratio for the state
$|\Psi^{\ }_{\Lambda}\rangle$
scales with the linear size $L=|\Lambda|^{1\over d}$ as
\begin{equation}
{\rm P}^{(2)}_{\Lambda}[\Psi^{\ }_{\Lambda}]\ =\ L^{-\tau(2)},
\quad\tau(2)\ =\ d.
\end{equation}
On the other hand, the inverse participation ratio for the state
$|\Psi^{\ }_{\Omega}\rangle$, where $\Omega$ has a finite number of sites
in the thermodynamic limit $L\rightarrow\infty$,
does not scale with the size of the system since
\begin{equation}
{\rm P}^{(2)}_{\Lambda}[\Psi^{\ }_{\Omega}]\ =\ |\Omega|^{-1}>0,
\label{psiomegaloc}
\end{equation}
is independent of $L$.

In practice, finite size scaling analysis of
${\rm P}^{(2)}_{\Lambda}$
does not distinguish extended {}from localized states close to the mobility
edge.
Indeed, in numerical simulations, one calculates
energy eigenstates whose wave functions have at most isolated zeroes
in contrast to, say,  $|\Psi^{\ }_{\Omega}\rangle$.
Moreover, for energies close to
the mobility edge, the localization length is so large that
the eigenstates, for all intent and purposes,
extend throughout the systems accessible to numerical simulations.
Close to the mobility edge, it is more appropriate to
ask what is the nature of the extended state in the finite size system:
is it a state with strongly fluctuating or uniform amplitude?
Generalized inverse participation ratios can address this issue.
Generalized inverse participation ratios are defined to be functionals
on the space of normalized wave functions $\{\Psi\}$ in a
volume of linear size $L$
\begin{equation}
{\rm P}^{(q)}_{L}[\Psi]\ \equiv\
\int_{{\rm supp}\Psi} {d^d x\over L^d}
\left|\Psi(x)\right|^{2q},
\quad
{\rm P}^{(1)}_{L}[\Psi]\ \equiv\
\int_{{\rm supp}\Psi}{d^d x\over L^d}\left|\Psi(x)\right|^{2}=1,
\label{generalizedparticipation}
\end{equation}
taking values in $]0,1]$ for positive real values of $q$,
and in $[1,\infty[$ for negative real values of $q$.
The usefulness of generalized inverse participation ratios
comes {}from the property that ${\rm P}^{(q)}_L[\Psi]$
is dominated by the largest
probability density $\sup_x\{|\Psi(x)|^{2}\}$ when $q$ is very
positive.
Conversely, when $q$ is very negative, the smallest probability density
$\inf_x\{|\Psi(x)|^{2}\}$ dominates in
${\rm P}^{(-|q|)}_L[\Psi]$.
Thus, the $q$-th generalized
inverse participation ratio for any extended state with {\it uniform}
amplitude scales with the linear size of the system as
\begin{equation}
{\rm P}^{(q)}_{\Lambda}
[\Psi^{\ }_{\Lambda}]\ =\ L^{-\tau(q)},
\quad \tau(q)\ =\ d(q-1).
\label{uniformlyextendedratios}
\end{equation}
This is an example of gap scaling.
\footnote
{
See footnote $^{\ref{footnoteongapscaling}}$ in the introduction.
}
Any scaling departing {}from Eq. (\ref{uniformlyextendedratios})
signals that the state is a simple fractal or multifractal.
This classification is based on writing
\begin{equation}
\tau(q)\ =\ D(q)(q-1).
\label{tau(q)}
\end{equation}
When $D(q)$ is independent of $q$, the state is said to be simple fractal,
in which case gap scaling still holds.
Otherwise, it is said to be multifractal and gap scaling breaks down.
For a multifractal wave function, the supports of
its maxima and minima scale with the system size with two different exponents
$D(-\infty)>D(+\infty)$.

The functional form of $\tau(q)$
is severely constrained by the fact that the wave function $\Psi(x)$
induces a normalized probability measure $|\Psi(x)|^2$.
By construction
$\tau(q)$ is strictly increasing and vanishes when $q=1$ due to the
normalization condition. For our purposes,
the most important property of the function $\tau(q)$ is that it is concave.
Multifractality of the wave function is equivalent to the
strict concavity of the function $\tau(q)$.
Other analytical properties of $\tau(q)$ can be found in Ref.
\cite{Janssen 1994}.

In numerical experiments on some two-dimensional
non-interacting tight-binding Hamiltonians
with static disorder \cite{Janssen 1994}, the collection of $q$-th
generalized inverse participation ratios
$
\{
P^{(q)}_{L}[\Psi^{\ }_{E,V}]
\}^{\ }_V
$
is calculated for some collection of normalized energy eigenfunctions
labelled by the impurity potential $V$, i.e.,
$
(H^{\ }_0+V) |\Psi^{\ }_{E,V}\rangle
= E |\Psi^{\ }_{E,V}\rangle
$.
It is observed that,
\begin{equation}
P^{(q)}_{L}[\Psi^{\ }_{E,V}]\approx L^{-\tau(q)},\quad \forall V,
\end{equation}
for energies close to the critical energy $E^{\ }_{\rm cr}$,
i.e., when $1\ll L\ll L^{\ }_E$ where $L^{\ }_E$
is the localization length.
We therefore make the hypothesis that at criticality and in the infinite
volume limit
$\ln P^{(q)}[\Psi^{\ }_{E^{\ }_{\rm cr},V}]$
is a selfaveraging random variable:
\begin{equation}
\lim_{L\rightarrow\infty}\
{
\overline{\ln P^{(q)}_{L}[\Psi^{\ }_{E^{\ }_{\rm cr},V}]}
\over -\ln L
}
\ =\
\lim_{L\rightarrow\infty}\
{
\ln P^{(q)}_{L}[\Psi^{\ }_{E^{\ }_{\rm cr},V}]
\over -\ln L
}
\ = \tau(q).
\label{truetau(q)}
\end{equation}
(The overbar denotes averaging over the random potential $V$.)
If Eq. (\ref{truetau(q)}) holds, it is not necessary to perform an averaging
over the disorder to calculate $\tau(q)$, as long as the system size $L$
is chosen
large enough. The challenge for a field theory describing the universal
properties of the multifractal critical wave function
$\Psi^{\ }_{E^{\ }_{\rm cr},V}(x)$ is to obtain the scaling
exponents $\{\tau(q)|q\in{\rm I}\hskip -0.08 true cm {\bf R}\}$.

To illustrate the difficulties involved in calculating {}from a critical field
theory the function $\tau(q)$ defined by Eq. (\ref{truetau(q)}),
we use the definition of Wegner \cite{Wegner 1980}
for disorder averaged generalized inverse participation ratios
\begin{equation}
{\cal P}^{(q)}(\omega)\ \equiv\
{
\overline
{
{\sum}^{\ }_{x,i}
|\left\langle x|\Psi^{\ }_{i,V}\right\rangle|^{2q}\
\delta(\omega-\omega^{\ }_i)
}
\over
\overline
{
{\sum}^{\ }_{x,i}
|\left\langle x|\Psi^{\ }_{i,V}\right\rangle|^{2 }\
\delta(\omega-\omega^{\ }_i)
}
},\quad q\geq0,
\label{inversepartratio}
\end{equation}
and apply it to Dirac fermions with random vector potentials.
Here, $i$ labels a complete basis of energy eigenstates
$\{|\Psi^{\ }_{i,V}\rangle\}^{\ }_i$
for a given realization of the static random vector potential $V$
defined in Eq (\ref{nonabelianpot}).
The denominator is the density of states per energy.
It insures that ${\cal P}^{(2)}(\omega)$ is normalized to one.
The right-hand-side of Eq. (\ref{inversepartratio}) equals
$\overline{P^{(q)}_{L^{\ }_{\omega^{\ }_i}}[\Psi^{\ }_{\omega^{\ }_i,V}]}$
if and only if the wave function $\Psi^{\ }_{\omega^{\ }_i,V}(x)$
has been normalized to one. Although for $q>0$
$\overline{P^{(q)}_{L^{\ }_{\omega^{\ }_i}}[\Psi^{\ }_{\omega^{\ }_i,V}]}$
takes values in $]0,1[$,
${\cal P}^{(q)}(\omega)$ takes values in $ ]0,\infty[$,
since the condition that the wave functions entering
Eq. (\ref{inversepartratio}) are normalized has been relaxed.
We now show that for energies asymptotically close to the
critical energy for Dirac fermions,
namely $\omega=0$, each ${\cal P}^{(q)}(\omega)$
scales like $|\omega|^{\varpi(q)}$. We will then discuss
the relationship between the set of scaling exponents
$\{\tau^*(q)\equiv z\varpi(q)|q\in {\rm I}\hskip -0.08 true cm{\bf N}\}$
calculated with the critical field theory of the previous sections and
the scaling exponents
$\{\tau(q)|q\in {\rm I}\hskip -0.08 true cm{\bf N}\}$
defined by Eq. (\ref{truetau(q)})
(i.e.,
derived {}from the normalized wave functions with vanishing energy
of Dirac fermions with random vector potentials).

We first assume that,
after averaging over disorder, each term in the numerator is
independent of the spatial coordinate $x$ in Eq. (\ref{inversepartratio}).
If so, the summation over the spatial coordinates can be dropped in
the numerator, if the denominator is replaced by the
averaged local density of states $\rho(\omega)$.
The usefulness of Eq. (\ref{inversepartratio}) then
comes {}from the identity \cite{Wegner 1980}
\begin{equation}\
{\cal P}^{(q)}(\omega) \rho(\omega)\ =\
{1\over C^{\ }_q}
\lim_{\eta\rightarrow0^+}
\left[
\eta^{q-1}
A^{(q)}(\omega,\eta)
\right],
\quad
q\in {\rm I}\hskip -0.08 true cm{\bf N},
\label{identityaveragedparticipationratio}
\end{equation}
since we can relate the right-hand-side to
a local operator in the critical field theory for Dirac fermions.
To see this, recall that according to Eq. (\ref{qteAomegaeta}),
$A^{(q)}(\omega,\eta)$ is the disorder average over the
$q$-th power of the smeared density of states per volume and per energy
for the complex energy $E=\omega+{\rm i}\eta$.
\footnote
{
The derivation of the combinatoric factor $C^{\ }_q$ can be found in
Ref. \cite{Wegner 1980}.
Multiplication by $\eta^{q-1}$ and taking the limit
$\eta\rightarrow0^+$ projects out of $A^{(q)}(\omega,\eta)$
contributions not present in ${\cal P}^{(q)}(\omega)$.
}
We now calculate the scaling exponent of ${\cal P}^{(q)}(\omega)$
with respect to $\omega$ as we did for the averaged local density of states.
We infer {}from Eqs.
(\ref{qteAomegaeta}),
(\ref{m=m-j}),
(\ref{sharingofscalingdim}),
(\ref{onerowmboxes}),
and (\ref{scalingE})
the transformation law obeyed by $\eta^{q-1}A^{(q)}(0,0)$
under real space rescaling on the critical line
$g^{\ }_A\geq0$, $g^{\ }_B=\infty$, namely
\begin{equation}
\tilde\eta^{q-1}\ \left(\tilde\varphi^{\dag}_+\tilde\varphi^{\ }_-\right)^q
(\tilde x)\ =\
a^
{
-z(q-1)-
q+
{g^{\ }_A\over\pi}q^2+
{N-1\over N^2}(q^2+Nq)
}
\eta^{q-1}
\left(\varphi^{\dag}_+\varphi^{\ }_-\right)^q(x), \quad a>0.
\label{localoperatorforP^q}
\end{equation}
With the help of
Eqs. (\ref{2-z}), and (\ref{scalingwithenergydos}),
we arrive to the scaling obeyed by ${\cal P}^{(q)}(\omega)$
to lowest order in $\omega$
\begin{mathletters}
\label{Pq(omega)}
\begin{eqnarray}
&&
{\cal P}^{(q)}(\omega)\ \propto\
|\omega|^{\varpi(q)},
\\
&&
\varpi(q)\ =\
{\tau^*(q)\over z},\quad
\tau^*(q)\ =\ D^*(q)(q-1),\quad
D^*(q)\ =\ 2-\left({g^{\ }_A\over\pi}+{N-1\over N^2}\right)q.
\label{NAtau*(q)}
\end{eqnarray}

It is remarkable that the same continuous spectrum of exponents
\end{mathletters}
\begin{equation}
\varpi(q)\ =\ {\tau^*(q)\over z},\quad
\tau^*(q)\ =\ D^*(q)(q-1),\quad
D^*(q)\ =\ 2-{g^{\ }_A\over\pi}q,
\label{Atau*(q)}
\end{equation}
along the critical line $g^{\ }_A\geq0$, $g^{\ }_B=0$, has been obtained
by Ludwig et al. in a very different manner \cite{Ludwig 1994}.
They first found
an eigenstate with vanishing energy for any realization of
the Abelian vector gauge-like impurity potential. They then normalized the
wave function in a square box of linear dimension $L$, thus obtaining the
real and strictly positive wave function $\Psi^{\ }_{0,V}(x)$.
Finally, they calculated how the {\it disorder average}
$\overline{|\Psi^{\ }_{0,V}(x)|^{2q}}$ scales with the
box size $L$ using the replica trick to treat the normalization constant,
and {\it assuming that this constant is selfaveraging}.
However, as we will discuss below, the normalization constant is not
selfaveraging. Thus, the calculation in Ref. \cite{Ludwig 1994} cannot
be viewed as a calculation of $\tau(q)$ defined by
Eq. (\ref{truetau(q)}).

Using the definition for averaged generalized inverse participation ratios
Eq. (\ref{inversepartratio}), we showed in Eq. (\ref{NAtau*(q)})
that they obey a scaling law controlled by
local operators in the critical field theory for Dirac fermions with
random vector potentials. The scaling exponents
$\{\tau^*(q)|q\in {\rm I}\hskip -0.08 true cm{\bf N}\}$
have been obtained non-perturbatively. By that we mean that
$\tau^*(q)$ is obtained {}from the exact
scaling dimensions $\Delta^{\ }_q$ of local operators
$(\varphi^{\dag}_+\varphi^{\ }_-)^q$
in the critical field theory (see Eq. (\ref{deltaq})).
Both the scaling exponents
$\{\tau^*(q)|q\in {\rm I}\hskip -0.08 true cm{\bf N}\}$
defined by Eq. (\ref{NAtau*(q)})
and the scaling exponents
$\{\tau  (q)|q\in {\rm I}\hskip -0.08 true cm{\bf N}\}$
defined by Eq. (\ref{truetau(q)})
characterize the statistical properties of the collection
of critical wave functions
$\{\Psi^{\ }_{0,V}(x)\}^{\ }_V$.
Hence, they should be closely related although they need not be equal.
In fact, they cannot be equal since Eq. (\ref{NAtau*(q)}) implies that
the scaling exponent $\tau^*(q)$ becomes negative for large enough $q$.
In contrast, $\tau(q)$ must be positive.
There are two reasons for this obvious difference. First,
the critical wave functions entering Eq. (\ref{truetau(q)})
are normalized whereas they are not in Eq. (\ref{NAtau*(q)}).
Second, the normalization factor for
unnormalized critical wave function is not a selfaveraging quantity
for Dirac fermions with random vector potentials:
\begin{equation}
\lim_{L\rightarrow\infty}
\overline{\int d^2 x |\Psi^{\ }_{0,V}(x)|^2}\ \neq\
\lim_{L\rightarrow\infty}\int d^2 x |\Psi^{\ }_{0,V}(x)|^2,
\quad\forall V.
\end{equation}
For large $q$, rare events, which are defined by $|\Psi^{\ }_{0,V}(x)|^2$
being much larger than $\overline{|\Psi^{\ }_{0,V}(x)|^2}$, dominate
the average in ${\cal P}^{(q)}(\omega)$.
Consequently, ${\cal P}^{(q)}(\omega)$ can diverge in the limit
$\omega\rightarrow0$.

It turns out that the issue of selfaveraging is far more important
than the issue of normalization. Indeed, even if we had required in
Eq. (\ref{inversepartratio})
that the wave functions be normalized to one, the corresponding
scaling exponents
$\{\tau^{**}(q)|q\in {\rm I}\hskip -0.08 true cm{\bf N}\}$
could only equal the scaling exponents
$\{\tau(q)|q\in {\rm I}\hskip -0.08 true cm{\bf N}\}$
if generalized inverse participation ratios were selfaveraging.
However, the parabolic shape of $\tau^*(q)$ strongly
suggests that this is not the case, since the parabolic shape
is characteristic of a log-normal distributed
random variable
\cite{Ludwig 1994,Janssen 1994}.
Conversely, if the critical wave function is multifractal
(i.e., $\tau(q)$ is strictly concave),
then all averaged generalized inverse participation ratios are necessarily
non-selfaveraging.
By enforcing the constraint of normalization of the wave functions
in Eq. (\ref{inversepartratio}),
the new set of scaling exponents $\tau^{**}(q)$ thus obtained
must be positive, but they need not equal $\tau(q)$.
\footnote
{
The Legendre transform $f^{**}(\alpha)\equiv\alpha q- \tau^{**}(q)$, where
$\alpha={d\tau^{**}(q)\over q}$, has been shown to take negative values
in a model of random resistors \cite{Fourcade 1987}. On the other hand,
$f(\alpha)\equiv\alpha q- \tau(q)$, where $\alpha={d\tau(q)\over q}$,
is always positive \cite{Janssen 1994}.
}
We must then conclude that for Dirac fermions
with random vector potentials
\begin{equation}
\{\tau^*   (q)|q\in {\rm I}\hskip -0.08 true cm{\bf N}\}\neq
\{\tau^{**}(q)|q\in {\rm I}\hskip -0.08 true cm{\bf N}\}\neq
\{\tau     (q)|q\in {\rm I}\hskip -0.08 true cm{\bf N}\},
\end{equation}
although we believe that the strict concavity of any one of
$\tau^*(q),\tau^{**}(q),\tau(q)$ imply that of the two others.

We close this subsection by reexamining an attempt made
in Ref. \cite{Duplantier 1991} to relate multifractal scaling exponents
to the scaling dimensions of powers of local operator
in critical field theories.
Let $\Psi_{0,V}(x)$ be a critical eigenfunction of the Dirac
Hamiltonian with random vector potential $V$ in a square box of linear
dimension $L$.
Define the so-called local random event
\begin{equation}
O^{\ }_{0,V}(x)\ =\
|\Psi^{\ }_{0,V}(x)|^2.
\end{equation}
We want to construct a normalized
probability measure out of local random events.
We therefore define the so-called local probability \cite{Duplantier 1991}
\begin{equation}
\mu^{\ }_{0,V}(x)\ =\
{O^{\ }_{0,V}(x)\over\int_{L^2}{d^2 x\over L^2}\ O^{\ }_{0,V}(x)}.
\end{equation}
We can always write
\begin{equation}
\int_{L^2} {d^2 x\over L^2}\ \mu^{q }_{0,V}(x)\ =\
\left({\varepsilon\over L}\right)^{\tau(q)}
F^{\ }_{0,V}(q,{\varepsilon\over L}),
\quad q\in{\rm I}\hskip -0.08 true cm {\bf R},
\end{equation}
where $\varepsilon$ is some microscopic length scale.
We made in Eq. (\ref{truetau(q)}) the assumption that
$\ln F^{\ }_{0,V}(q,{\varepsilon\over L})$
depends only weakly on $L$, i.e.,
\begin{eqnarray}
\lim_{L\rightarrow\infty}
{\ln F^{\ }_{0,V}(q,{\varepsilon\over L})\over\ln L}=
\lim_{L\rightarrow\infty}
{\overline{\ln F^{\ }_{0,V}(q,{\varepsilon\over L})}\over\ln L}=0,
\quad q\in{\rm I}\hskip -0.08 true cm {\bf R}.
\label{assumptionforlnselfa}
\end{eqnarray}
Note that Eq. (\ref{assumptionforlnselfa})
does not rule out the possibility that for any
$q\in{\rm I}\hskip -0.08 true cm {\bf R}$
averaging over randomness (denoted by an overbar) yields
\begin{equation}
\lim_{L\rightarrow\infty}
\overline{F^{\ }_{0,V}(q,{\varepsilon\over L})}
=
F^{\ }_{0}(q)
\lim_{L\rightarrow\infty}
\left({\varepsilon\over L}\right)^{\delta\tau(q)},
\quad
\delta\tau(q)\in{\rm I}\hskip -0.08 true cm {\bf R}.
\label{nonsaofF0V}
\end{equation}
Let us make instead of Eq. (\ref{assumptionforlnselfa})
the much stronger assumption that the amplitude
$F^{\ }_{0,V}(q,{\varepsilon\over L})$ does not depend on
the randomness $V$ and on the system size in the thermodynamic limit:
\begin{equation}
\lim_{L\rightarrow\infty}
F^{\ }_{0,V}(q,{\varepsilon\over L})\ =\
F^{\ }_{0  }(q),
\quad q\in{\rm I}\hskip -0.08 true cm {\bf R},
\label{selfaveragingpart1}
\end{equation}
i.e., $\mu^q_{0,V}(x)$ is selfaveraging.
This is the assumption made in Eqs. (1-3) of
Ref. \cite{Duplantier 1991}. Unfortunately, this assumption does not apply
to our model of Dirac fermions with random vector potentials
for which Eq. (\ref{nonsaofF0V})
with $\delta\tau(q)=\tau^*(q)-\tau(q)$ holds.
It is thus incorrect to use Eq. (3) of Ref. \cite{Duplantier 1991},
namely to equate (see Eq. (\ref{deltaq}))
$\Delta^{\ }_q-2-q(\Delta^{\ }_1-2)$ with
$\tau(q)$, as was done in Ref. \cite{Ludwig 1994}.
On the other hand, Eq. (\ref{selfaveragingpart1}) should yield a good
approximation to $\tau(q)$ in the vicinity of $q=0$, since
$\mu^q_{0,V}(x)=\exp[q\ln\mu^{\ }_{0,V}(x)]\approx 1+ q\ln\mu^{\ }_{0,V}(x)$
is very close to being selfaveraging for small $q$.

\subsection{Scaling of averaged spatial correlations of
generalized inverse participation ratios}
\label{Scaling of averaged spatial correlations of
generalized inverse participation ratios}

In the previous subsection, we have constructed a spectrum of exponents
$\{\tau^*(q)|q\in {\rm I}\hskip -0.08 true cm{\bf N}\}$ which characterizes
the scaling of averaged generalized inverse participation ratios.
There exists a linear relation between the $\tau^*(q)$'s
and the scaling dimensions
$\{\Delta^{\ }_q|q\in {\rm I}\hskip -0.08 true cm{\bf N}\}$,
Eq. (\ref{deltaq}),
of some local operators in the critical field theory.
In this subsection,
we study the interplay between the operator product expansion
and the strict concavity of $\Delta^{\ }_q$ by constructing
averaged spatial correlations of generalized inverse participation ratios.

Following Ref. \cite{Wegner 1985}, averaged spatial correlations of
generalized inverse participation ratios are defined by
\begin{equation}
{\cal Q}^{(q^{\ }_1,q^{\ }_2)}(x,y,\omega)\ \equiv\
{
\overline
{
{\sum}^{\ }_{i}
|\left\langle x|\Psi^{\ }_i\right\rangle|^{2q^{\ }_1}
|\left\langle y|\Psi^{\ }_i\right\rangle|^{2q^{\ }_2}
\
\delta(\omega-\omega^{\ }_i)
}
\over
\overline
{
{\sum}^{\ }_{i}
|\left\langle x|\Psi^{\ }_i\right\rangle|^{2 }\
\delta(\omega-\omega^{\ }_i)
}
},\quad q^{\ }_1,q^{\ }_2>0.
\end{equation}
The $q$-th generalized inverse participation ratio is recovered by choosing
$q^{\ }_1=q$, $q^{\ }_2=0$. Similarly to
Eq. (\ref{identityaveragedparticipationratioaq1q2}),
we use the identity \cite{Wegner 1985}
\begin{equation}
{\cal Q}^{(q^{\ }_1,q^{\ }_2)}(x,y,\omega)\ \rho(\omega)\ =\
{1\over C^{\ }_{q^{\ }_1+q^{\ }_2}}
\lim_{\eta\rightarrow0^+}
\left[
\eta^{q^{\ }_1+q^{\ }_2-1}
A^{(q^{\ }_1,q^{\ }_2)}(x,y,\omega,\eta)
\right].
\label{identityaveragedparticipationratioaq1q2}
\end{equation}
Here, $A^{(q^{\ }_1,q^{\ }_2)}(x,y,\omega,\eta)$, which is defined by
\begin{equation}
A^{(q^{\ }_1,q^{\ }_2)}(x,y,\omega,\eta)\ \equiv\
\overline
{
\left\{{{\rm Im}\over\pi}
{\rm tr}\left[ G(x,x,\omega-{\rm i}\eta;V)\right]\right\}
^{q^{\ }_1}
\left\{{{\rm Im}\over\pi}
{\rm tr}\left[ G(y,y,\omega-{\rm i}\eta;V)\right]\right\}
^{q^{\ }_2}
},
\end{equation}
should be compared to Eq. (\ref{qteAomegaeta}).
We calculate the scaling of
$A^{(q^{\ }_1,q^{\ }_2)}(x,y,\omega,\eta)$
with respect to {\it large} separation $|x-y|$ and with respect to
{\it small} $\omega\neq0$. In other words, we work in the regime
\begin{equation}
\varepsilon\ll |x-y| \ll L^{\ }_{\omega},
\quad L^{\ }_{\omega}\ \propto\ |\omega|^{-z}.
\end{equation}
Here, $\varepsilon$ is the ultraviolet cutoff, say, the lattice spacing.
Of course,  $\eta>0$ can be taken as small as needed.
Scaling is calculated as follows \cite{Wegner 1985}.
The first step is to perform a rescaling
which brings the separation down to the lattice spacing $\varepsilon$.
The second step makes use of the operator product expansion.
The final step is to perform a rescaling to energy
scales $\bar\omega$ and $\bar\eta$ which are arbitrarily chosen.
Under each rescaling, we use the scaling dimensions of
the appropriate operators. In particular, to lowest order in $\omega$, we
use the scaling dimensions at criticality
(see Eqs. (\ref{weightsofPsi},\ref{onerowmboxes}))
\begin{equation}
\Delta^{\ }_{q}\ =\
q- {g^{\ }_A\over\pi}q^2-
{N-1\over N^2}\left(q^2+Nq\right),
\label{deltaq}
\end{equation}
for the operators
$\left(\varphi^{\dag}_+\varphi^{\ }_-\right)^{q^{\ }_1}(x)$,
$\left(\varphi^{\dag}_+\varphi^{\ }_-\right)^{q^{\ }_2}(y)$,
and
$\left(\varphi^{\dag}_+\varphi^{\ }_-\right)^{q^{\ }_1+q^{\ }_2}
(\varepsilon)$,
respectively.

Under rescaling, $A^{(q^{\ }_1,q^{\ }_2)}(x,y,\omega,\eta)$ transforms like
\begin{equation}
\left\langle
\left(\varphi^{\dag}_+\varphi^{\ }_-\right)^{q^{\ }_1}(x)
\left(\varphi^{\dag}_+\varphi^{\ }_-\right)^{q^{\ }_2}(y)
\right\rangle^{\ }_{S^{(1,0)}_{\rm eff}(\omega-{\rm i}\eta)},
\end{equation}
according to section \ref{sec:Abelianvectorgaugerandomness}.
Hence, under the rescaling
\begin{equation}
\tilde x\ =\ ax,\quad
\tilde y\ =\ ay,\quad
\tilde\omega\ =\ a^{-z}\omega,\quad
\tilde\eta\ =\ a^{-z}\eta,\quad
a\ \equiv\ {\varepsilon\over|x-y|}\ll1,
\end{equation}
we find that
\begin{eqnarray}
A^{(q^{\ }_1,q^{\ }_2)}(x,y,\omega,\eta)
\ &\sim&\
\left({\varepsilon\over|x-y|}\right)^
{\Delta^{\ }_{q^{\ }_1}+\Delta^{\ }_{q^{\ }_2}}
\left\langle
\left(\tilde\varphi^{\dag}_+\tilde\varphi^{\ }_-\right)^{q^{\ }_1}(\tilde x)
\left(\tilde\varphi^{\dag}_+\tilde\varphi^{\ }_-\right)^{q^{\ }_2}(\tilde y)
\right\rangle^{\ }_{S^{(1,0)}_{\rm eff}(\tilde\omega-{\rm i}\tilde\eta)}
\nonumber\\
&\sim&\
\left({\varepsilon\over|x-y|}\right)^
{\Delta^{\ }_{q^{\ }_1}+\Delta^{\ }_{q^{\ }_2}}
\left\langle
\left(\tilde\varphi^{\dag}_+\tilde\varphi^{\ }_-\right)
^{q^{\ }_1+q^{\ }_2}(\tilde y)
\right\rangle^{\ }_{S^{(1,0)}_{\rm eff}(\tilde\omega-{\rm i}\tilde\eta)}.
\label{step12}
\end{eqnarray}
On the second line of Eq. (\ref{step12}), we have performed
an operator product expansion {}from which only the {\it most relevant}
operator of the expansion is kept.
Notice that the identity operator is not the most relevant operator in
the operator product expansion. It would have been, had we considered, say,
the pure Gaussian field theory with
the hierarchy of operators $\exp(q{\rm i}\phi)$
and the positive and convex spectrum of exponents
$\{q^2|q\in {\rm I}\hskip -0.08 true cm{\bf N}\}$.
The two factors on the right-hand-side of  Eq. (\ref{step12}) both
depends on the separation $|x-y|$. However, the rescaling to the
arbitrarily chosen energy $\bar\omega<\omega$
\begin{equation}
\bar y\ =\     b\tilde y,\quad
\bar\omega\ =\ b^{-z}\tilde\omega,\quad
\bar\eta\ =\   b^{-z}\tilde\eta,\quad
b\ =\ {1\over a}\left|{\bar\omega\over\omega}\right|^{-{1\over z}},
\end{equation}
yields
\begin{eqnarray}
A^{(q^{\ }_1,q^{\ }_2)}(x,y,\omega,\eta)
\ &\sim &\
\left({\varepsilon\over|x-y|}\right)^
{\Delta^{\ }_{q^{\ }_1}+\Delta^{\ }_{q^{\ }_2}-
\Delta^{\ }_{q^{\ }_1\!+q^{\ }_2}}
\left|{\omega\over\bar\omega}\right|^
{{1\over z}\Delta^{\ }_{q^{\ }_1\!+q^{\ }_2}}
\left\langle
\left(\bar\varphi^{\dag}_+\bar\varphi^{\ }_-\right)
^{q^{\ }_1+q^{\ }_2}(\bar y)
\right\rangle^{\ }_{S^{(1,0)}_{\rm eff}(\bar\omega-{\rm i}\bar\eta)}
\nonumber\\
&\equiv&\
\left({\varepsilon\over|x-y|}\right)^
{\Delta^{\ }_{q^{\ }_1}+\Delta^{\ }_{q^{\ }_2}-
\Delta^{\ }_{q^{\ }_1\!+q^{\ }_2}}
\left|{\omega\over\bar\omega}\right|^
{{1\over z}\Delta^{\ }_{q^{\ }_1\!+q^{\ }_2}}
F^{(q^{\ }_1,q^{\ }_2)}(\bar\omega,\bar\eta).
\label{step3}
\end{eqnarray}
The scaling function $F^{(q^{\ }_1,q^{\ }_2)}(\bar\omega,\bar\eta)$
is defined by Eq. (\ref{step3}). It is given by the expectation value
at energy $\bar\omega-{\rm i}\bar\eta$
of the most relevant operator in the operator product expansion in Eq.
(\ref{step12}).
By assumption,
it does not depend on $\bar y$. For given $q^{\ }_{1,2}$, it
therefore only depends on $\bar\omega$ and $\bar\eta$. Calculating
$F^{(q^{\ }_1,q^{\ }_2)}(\bar\omega,\bar\eta)$ is beyond the scope of
our approach. However, some of its limiting behaviours can be inferred {}from
the fact that $F^{(q^{\ }_1,q^{\ }_2)}(\bar\omega,\bar\eta)$
enters an averaged generalized inverse participation ratio.

We consider the limit $\omega>0$, $\eta\rightarrow0$.
Because we are working in the
thermodynamic limit at finite real energy $\omega$, the energy eigenstates
must be localized. Hence, ${\cal Q}^{(q^{ }_1,q^{ }_2)}(x,y,\omega)$
must be finite so that {}from Eq.
(\ref{identityaveragedparticipationratioaq1q2})
\begin{equation}
\lim_{\eta\rightarrow0}\
\eta^{q^{ }_1+q^{ }_2-1} F^{(q^{\ }_1,q^{\ }_2)}(\bar\omega,\bar\eta)
\ \sim\
\left({\eta\over\bar\eta}\right)^{q^{ }_1+q^{ }_2-1}
G^{(q^{\ }_1,q^{\ }_2)}(\bar\omega).
\end{equation}
This gives us the scaling
\begin{equation}
{\cal Q}^{(q^{ }_1,q^{ }_2)}(x,y,\omega)\ \rho(\omega)
\ \sim\
\left({\varepsilon\over|x-y|}\right)^
{\Delta^{\ }_{q^{\ }_1}+\Delta^{\ }_{q^{\ }_2}-
\Delta^{\ }_{q^{\ }_1\!+q^{\ }_2}}
\left|{\omega\over\bar\omega}\right|^
{
{1\over z}
\left[
\Delta^{\ }_{q^{\ }_1\!+q^{\ }_2}
+z\left(q^{\ }_1+q^{\ }_2-1\right)
\right]
}
G^{(q^{\ }_1,q^{\ }_2)}(\bar\omega).
\label{multifractalscalingofcorrelations}
\end{equation}
This scaling has two important features. First, the exponent
characterizing the scaling with separation is always positive
since
\begin{equation}
\Delta^{\ }_{q^{\ }_1}+\Delta^{\ }_{q^{\ }_2}-
\Delta^{\ }_{q^{\ }_1\!+q^{\ }_2}\ =\
2\left({g^{\ }_A\over\pi}+{N-1\over N^2}\right)q^{\ }_1q^{\ }_2,
\quad q^{\ }_1,q^{\ }_2>0.
\end{equation}
Secondly, the exponent characterizing the scaling with respect to $\omega$
is that of the $(q^{\ }_1+q^{\ }_2)$-th
generalized inverse participation ratio:
\begin{equation}
{1\over z}
\left[
\Delta^{\ }_{q^{\ }_1\!+q^{\ }_2}
+z\left(q^{\ }_1+q^{\ }_2-1\right)
\right]\ =\
\varpi(q^{\ }_1+q^{\ }_2)\ +\ {2-z\over z}.
\end{equation}
Consequently, we recover the scaling for
generalized inverse participation ratios
when either $q^{\ }_1$ or $q^{\ }_2$ vanishes as it should be.

Eq. (\ref{multifractalscalingofcorrelations}) tells us that the
property of strict concavity
$
\Delta^{\ }_{q^{\ }_1}+\Delta^{\ }_{q^{\ }_2}-\Delta^{\ }_{q^{\ }_1+q^{\ }_2}
>0,
$
$q^{\ }_1,q^{\ }_2>0$,
insures that spatial correlations {\it decrease} with separation.
Negative dimensional scaling exponents also control the scaling with
respect to energy, or, equivalently, with respect to the effective system size.
This latter scaling implies that the normalization of the local observable
can blow up with the system size in the thermodynamic limit.
In fact, Eq. (\ref{multifractalscalingofcorrelations})
reproduces the scaling for spatial correlations of so-called
non-normalizable box-observables
\cite{Cates 1987,Duplantier 1991,Janssen 1994}.
Also note that the identity operator never enters as
the most relevant operator in the operator product expansion between pairs
of local operators with scaling dimensions
$\Delta^{\ }_{q^{\ }_1}$ and $\Delta^{\ }_{q^{\ }_2}$,
and this, in turn, is guaranteed by $\Delta^{\ }_q$ becoming negative
for sufficiently large $q$. In contrast, it is much more difficult to
find a critical point characterized by an infinite
set of positive scaling exponents
$\{x^{\ }_q>0|q\in{\rm I}\hskip -0.08 true cm{\bf N}\}$,
which are not linearly related, and such that
$x^{\ }_{q^{\ }_1+q^{\ }_2}>0$ is generated {}from $x^{\ }_{q^{\ }_1}$
and $x^{\ }_{q^{\ }_2}$ as the most relevant operator in the operator
product expansion of their associated operators \cite{Duplantier 1991}.

In summary, several scaling properties of the multifractal measure induced
by the critical wave function of Dirac fermions with random vector
potentials are shared by the probability distribution corresponding
to {\it averaged} generalized inverse participation ratios.
In the literature it is assumed that the
strict concavity of $\tau^*(q)$ for small $q$ implies that of $\tau(q)$.
Although this assumption seems reasonable, it would be highly desirable to
verify it by reconstructing
$\{\tau(q)|q\in{\rm I}\hskip -0.08 true cm{\bf N}\}$
{}from
$\{\tau^*(q)|q\in{\rm I}\hskip -0.08 true cm{\bf N}\}$.
Moreover, by doing so it could be established whether or not the existence of
negative dimensional local operators implies multifractality of the
critical wave function.

\section{Current Algebra for Graded Symmetries: $U(1/1)$ current algebra}
\label{sec:Current Algebra for Graded Symmetries: $U(1/1)$ current algebra}

In the previous sections,
we have studied some disordered critical points through their
fixed point Lagrangians. In this section, we take a very different approach.
We want to describe new disordered critical points using conformal
and internal symmetries and the associated current algebra.
We will see how the current algebra approach allows us to access new
fixed points which cannot be reached perturbatively {}from some known
fixed point (contrast this approach to that in Refs.
\cite{Ludwig 1987,Ludwig 1990}).

Current algebra (and the associated bosonization technique) is a
powerful method for studying strongly correlated electrons in one spatial
dimension \cite{Affleckhouches}.
Current algebra is also very useful for studying critical points
in two dimensions which are characterized by certain continuous symmetries.
Effective theories for non-interacting disordered systems have
a {\it graded symmetry} (or supersymmetry) \cite{Efetov 1983}.
For example, we have seen
in section \ref{sec:AveragedGreenfunctions}
that disorder average over  products of Green functions can
equally well be represented by path integrals over
fermionic or bosonic coherent states.
In this section, we study the current algebra of graded symmetry in
two dimensions. We classify {\it all} critical points for
two-dimensional systems with {\it unbroken} $U(1/1)$ graded symmetry.
We recover the critical points corresponding to $U(1)$
vector gauge-like disorder studied in the previous sections.
We also find new critical points.
The operator content for these critical points is constructed.
In particular, we show that most critical points are characterized by
infinitely many local operators with negative dimensions.

In the current algebra approach, one first constructs the commutator and
anticommutator algebra for the conserved currents. At a critical point,
conformal symmetry allows one to calculate all correlations between the
currents {}from the commutator and anticommutator algebra.
In this way, the theory is completely specified.

\subsection{Current algebra description of a free model}
\label{Current algebra description of a free model}

We begin with the simplest model with graded $U(1/1)$ symmetry. It is
given by the partition function
\begin{mathletters}
\label{freeu11}
\begin{eqnarray}
&&
{\cal Z}^{\ }_0\ =\
\int {\cal D}[\psi   ^{\dag}_{\pm},\psi   ^{\ }_{\pm}]
\int {\cal D}[\varphi^{\dag}_{\pm},\varphi^{\ }_{\pm}]
\  e^{{\rm i}\int^{+\infty}_{-\infty} dx d\tau\ {\cal L}^{\ }_0},
\\
&&
{\cal L}^{\ }_0\ =\
{{\rm i}\over\pi}
\left(
\psi   ^{\dag}_+ \partial_{\bar z} \psi   ^{\ }_+\ +\
\varphi^{\dag}_+ \partial_{\bar z} \varphi^{\ }_+\ +\
\psi   ^{\dag}_- \partial_{     z} \psi   ^{\ }_-\ +\
\varphi^{\dag}_- \partial_{     z} \varphi^{\ }_-
\right).
\end{eqnarray}
Here, $(x,\tau)$ denotes the coordinates of two-dimensional Euclidean space,
and the fields are the chiral components of Dirac spinors introduced in
Eq. (\ref{chiralbasis}).
For convenience, we have rescaled the spinors by the factor $\sqrt{2\pi}$.
The only non-vanishing two-point Green functions are
\end{mathletters}
\begin{mathletters}
\begin{eqnarray}
\langle\psi^{\dag}_+(z) \psi^{\   }_+(0)\rangle\ &=&\
\langle\psi^{\   }_+(z) \psi^{\dag}_+(0)\rangle\ =\ {1\over z},
\\
-\langle\varphi^{\dag}_+(z) \varphi^{\   }_+(0)\rangle\ &=&\
 \langle\varphi^{\   }_+(z) \varphi^{\dag}_+(0)\rangle\ =\ {1\over z},
\\
\langle\psi^{\dag}_-(\bar z) \psi^{\   }_-(0)\rangle\ &=&\
\langle\psi^{\   }_-(\bar z) \psi^{\dag}_-(0)\rangle\ =\ {1\over\bar z},
\\
-\langle\varphi^{\dag}_-(\bar z) \varphi^{\   }_-(0)\rangle\ &=&\
 \langle\varphi^{\   }_-(\bar z) \varphi^{\dag}_-(0)\rangle\ =\ {1\over\bar z},
\end{eqnarray}
where $z=x+{\rm i}\tau$ and $\bar z=x-{\rm i}\tau$ are the independent
holomorphic and antiholomorphic coordinates, respectively.
\end{mathletters}

\subsubsection{Symmetries of ${\cal L}^{\ }_0$ }

The action in Eq. (\ref{freeu11}) is left invariant by the
global chiral transformation
\begin{equation}
\psi^{\dag}_+\ =\ e^{-{\rm i}\theta^{\psi}_+}\ \psi'^{\dag}_+,
\qquad
\psi^{\   }_+\ =\ e^{+{\rm i}\theta^{\psi}_+}\ \psi'^{\   }_+,
\end{equation}
where $\theta^{\psi}_+$ is any arbitrary real number. The associated
current
\begin{equation}
J^{\psi}_z(z)\ =\ (\psi^{\dag}_+\psi^{\ }_+)(z),
\end{equation}
is conserved, i.e., its conservation is equivalent to it being a
holomorphic function
\begin{equation}
\partial^{\ }_{\bar z}\ J^{\psi}_z(z)\ =\ 0.
\end{equation}
Similarly, there are  three additional independent
global $U(1)$ chiral symmetries
with the conserved currents
\begin{equation}
J^{\psi}_{\bar z}(\bar z)\ =\ (\psi^{\dag}_-\psi^{\ }_-)(\bar z),
\quad
J^{\varphi}_z(z)\ =\ (\varphi^{\dag}_+\varphi^{\ }_+)(z),
\quad
J^{\varphi}_{\bar z}(\bar z)\ =\ (\varphi^{\dag}_-\varphi^{\ }_-)(\bar z),
\end{equation}
respectively.
The action in Eq. (\ref{freeu11}) is also left invariant by
the four independent infinitesimal transformations
\begin{mathletters}
\label{deltaetas}
\begin{eqnarray}
\delta^{\ }_{\eta}\psi^{\dag}_{\pm}\ \equiv\
\big[\bar\theta^{\ }_{\pm}\eta^{\ }_{\pm},\psi^{\dag}_{\pm}\big]\ =\
+\varphi^{\dag}_{\pm}\bar\theta^{\ }_{\pm},
\quad
\delta^{\ }_{\eta}\varphi^{\ }_{\pm}\ \equiv\
\big[\bar\theta^{\ }_{\pm}\eta^{\ }_{\pm},\varphi^{\ }_{\pm}\big]\ =\
-\bar\theta^{\ }_{\pm}\psi^{\ }_{\pm},
\end{eqnarray}
and
\begin{eqnarray}
\delta^{\ }_{\bar\eta}\varphi^{\dag}_{\pm}\ \equiv\
\big[\theta^{\ }_{\pm}\bar\eta^{\ }_{\pm},\varphi^{\dag}_{\pm}\big]\ =\
-\psi^{\dag}_{\pm}\theta^{\ }_{\pm},
\quad
\delta^{\ }_{\bar\eta}\psi^{\ }_{\pm}\ \equiv\
\big[\theta^{\ }_{\pm}\bar\eta^{\ }_{\pm},\psi^{\ }_{\pm}\big]\ =\
+\theta^{\ }_{\pm}\varphi^{\ }_{\pm},
\end{eqnarray}
respectively. The four infinitesimal Grassmann numbers
$\bar\theta^{\ }_+$, $\bar \theta^{\ }_-$, $\theta^{\ }_+$, and
$\theta^{\ }_-$,
are independent. The associated conserved fermionic currents are
\begin{equation}
\eta^{\ }_+\ =\ \varphi^{\dag}_+\psi^{\ }_+,\quad
\eta^{\ }_-\ =\ \varphi^{\dag}_-\psi^{\ }_-,\quad
\bar\eta^{\ }_+\ =\ \psi^{\dag}_+\varphi^{\ }_+,\quad
\bar\eta^{\ }_+\ =\ \psi^{\dag}_-\varphi^{\ }_-.
\label{etacurrents}
\end{equation}
Taken together, these eight conserved currents, of which four are bosonic
and four are fermionic, generate a graded $U(1/1)\times U(1/1)$
symmetry. The bosonic generators of the diagonal subgroup
$U(1/1)$ are the currents $J^{\ }_{\mu}$ and $j^{\ }_{\mu}$ with
\end{mathletters}
\begin{equation}
{1\over2}
\pmatrix
{
J^{\ }_x-{\rm i}J^{\ }_{\tau}\cr
J^{\ }_x+{\rm i}J^{\ }_{\tau}\cr
}\ \equiv\
\pmatrix
{
J^{\ }_z\cr
J^{\ }_{\bar z}\cr
}
\ \equiv\
\pmatrix
{
J^{\psi}_z+J^{\varphi}_z\cr
J^{\psi}_{\bar z}+J^{\varphi}_{\bar z}\cr
},
\qquad
{1\over2}
\pmatrix
{
j^{\ }_x-{\rm i}j^{\ }_{\tau}\cr
j^{\ }_x+{\rm i}j^{\ }_{\tau}\cr
}\ \equiv\
\pmatrix
{
j^{\ }_z\cr
j^{\ }_{\bar z}\cr
}
\ \equiv\
\pmatrix
{
J^{\psi}_z-J^{\varphi}_z\cr
J^{\psi}_{\bar z}-J^{\varphi}_{\bar z}\cr
}.
\label{Jandj}
\end{equation}
The fermionic generators of the diagonal subgroup $U(1/1)$ are
the currents $\eta^{\ }_{\mu}$ and $\bar\eta^{\ }_{\mu}$ with
\begin{equation}
{1\over2}
\pmatrix
{
\eta^{\ }_x-{\rm i}\eta^{\ }_{\tau}\cr
\eta^{\ }_x+{\rm i}\eta^{\ }_{\tau}\cr
}\ \equiv\
\pmatrix
{
\eta^{\ }_z\cr
\eta^{\ }_{\bar z}\cr
}
\ \equiv\
\pmatrix
{
\varphi^{\dag}_+\psi^{\ }_+\cr
\varphi^{\dag}_-\psi^{\ }_-\cr
},
\qquad
{1\over2}
\pmatrix
{
\bar\eta^{\ }_x-{\rm i}\bar\eta^{\ }_{\tau}\cr
\bar\eta^{\ }_x+{\rm i}\bar\eta^{\ }_{\tau}\cr
}\ \equiv\
\pmatrix
{
\bar\eta^{\ }_z\cr
\bar\eta^{\ }_{\bar z}\cr
}
\ \equiv\
\pmatrix
{
\psi^{\dag}_+\varphi^{\ }_+\cr
\psi^{\dag}_-\varphi^{\ }_-\cr
}.
\label{etaandbareta}
\end{equation}
Each component of the bosonic and fermionic currents generating
$U(1/1)$ is separately conserved due to the chiral invariance.

\subsubsection{Symmetries of ${\cal Z}^{\ }_0$ }

The global $U(1/1)\times U(1/1)$ symmetry of ${\cal L}^{\ }_0$ is
shared by the measure in the partition function. Hence, ${\cal Z}^{\ }_0$
is also left invariant by global $U(1/1)\times U(1/1)$ transformations.
The correlation function
$
\langle
A^{\ }_{1+} \cdots A^{\ }_{m+}
B^{\ }_{1-} \cdots B^{\ }_{n-}
\rangle
$,
where, say, $\{A^{\ }_{i+}\}$ are defined in terms of
$\psi^{\dag}_{+}$,$\psi^{\   }_{+}$,
$\varphi^{\dag}_{+}$,$\varphi^{\ }_{+}$,
factorizes according to
\begin{equation}
\langle
A^{\ }_{1+} \cdots A^{\ }_{m+}
B^{\ }_{1-} \cdots B^{\ }_{n-}
\rangle
\ =\
\langle
A^{\ }_{1+} \cdots A^{\ }_{m+}
\rangle
\langle
B^{\ }_{1-} \cdots B^{\ }_{n-}
\rangle,
\end{equation}
due to the decoupling of the $+$ and the $-$ sectors.
Moreover, the correlation function corresponding to the $+$ subscript must
be holomorphic, whereas the correlation function corresponding to the
$-$ subscript must be antiholomorphic. Without loss of generality, we
will only consider the holomorphic sector of the theory. The subscript
$+$ will be suppressed {}from now on.

\subsubsection{Operator product expansion}

We now study the operator product algebra with the help of the
operator product expansion (OPE) \cite{OPE 1969}.
Consider two operators, $A(z)$ and $B(z)$.
The OPE of the product $A(z_1)\ B(z_2)$ is the Laurent series expansion
\begin{equation}
A(z_1)\ B(z_2) \ =\
{1\over(z_1 - z_2)^n}O_n(z_2)
\ +\
\cdots
\ +\
{1\over z_1 - z_2} O_1(z_2)
\ +\
{\cal O}(1).
\end{equation}
Only the poles in the limit of $z_1\to z_2$ are kept in the OPE.
In the following, we will always omit the non-diverging term
${\cal O}(1)$ in the OPE. For example,
the operators $\psi^{\dag},\cdots,\varphi$ have the following OPE:
\begin{eqnarray}
\psi^{\dag}(z)\psi(0) \ =\
\psi(z) \psi^{\dag}(0) \ =\
\varphi(z)\varphi^{\dag}(0)\ =\
-\varphi^{\dag}(z)\varphi(0)\ =\ {1\over z}.
\end{eqnarray}
Another example of an OPE is that between the currents
$(J,j,\eta,\bar\eta)$ generating the $U(1/1)$ symmetry.
However before calculating them, the quantum nature of the currents must
be dealt with care. In particular, we must remove any divergences
which arise when multiplying operators at the same point.
In the context of the currents, such divergences are removed by
subtracting {}from the naive definition of the classical currents
in Eqs. (\ref{Jandj},\ref{etaandbareta}) the poles in the OPE between the
$\psi^{\dag},\cdots, \varphi$. For example,
the precise definition of $j(z)$ is {\it not} given by
Eq. (\ref{Jandj}), but by
\begin{eqnarray}
j(z)\ &=&\
\lim_{\epsilon\to 0}
\left[
\psi^{\dag} (z + \epsilon) \psi(z)\ -\
\varphi^{\dag}(z + \epsilon) \varphi (z)\ -\
{2\over \epsilon}
\right]
\nonumber\\
&\equiv&
:\psi   ^{\dag}(z)\psi(z)   :\ -\
:\varphi^{\dag}(z)\varphi(z):.
\end{eqnarray}
The normal ordering $:\cdots:$
extracts the finite contribution {}from the operator product
defining the current $j$, since one can show that
the limit $\epsilon\to 0$ in the correlation function
\begin{equation}
\left\langle
\left[
\psi^{\dag}(z + \epsilon) \psi(z) \ -\
\varphi^{\dag}(z + \epsilon)\varphi(z)
\right]
\prod_i O_i(z_i)
\right\rangle\ -\
{2\over\epsilon}
\left\langle \prod_i O_i (z_i) \right\rangle
\end{equation}
is finite and thus {\it defines} the correlation function
$\langle j(z) \prod_i O_i(z_i)\rangle$.
With this definition of the currents
and with the help of the OPE between $\{\psi^{\dag},\cdots,\varphi\}$,
we find the OPE between the currents to be given by
\begin{mathletters}
\label{freecurrentope}
\begin{eqnarray}
&&
J(z)      J(0)=0,           \quad
J(z)      j(0)={2\over z^2},\quad
J(z)   \eta(0)=0,           \quad
J(z)\bar\eta(0)=0,
\\
&&
j(z)       J(0)={2\over z^2},           \quad
j(z)       j(0)=0,                      \quad
j(z)    \eta(0)=-{2\over z}\eta(0),     \quad
j(z)\bar\eta(0)=+{2\over z}\bar\eta(0),
\\
&&
\eta(z)      J(0)=0,                           \quad
\eta(z)      j(0)=+{2\over z}\eta(0),          \quad
\eta(z)   \eta(0)=0,                           \quad
\eta(z)\bar\eta(0)=-{1\over z^2}+{1\over z}J(0),
\\
&&
\bar\eta(z)      J(0)=0,                           \quad
\bar\eta(z)      j(0)=-{2\over z}\bar\eta(0),      \quad
\bar\eta(z)   \eta(0)=+{1\over z^2}+{1\over z}J(0),\quad
\bar\eta(z)\bar\eta(0)=0.
\end{eqnarray}
An alternative definition of the OPE for the
currents can be found in appendix \ref{sec:U(1/1) Suguwara construction}.
\end{mathletters}

The operator algebra specified by the OPE of Eq. (\ref{freecurrentope})
is very powerful.  Indeed, {}from the OPE alone, i.e.,
without any reference to ${\cal Z}^{\ }_0$, we
can calculate all the correlation functions between the currents
$(J,j,\eta,\bar\eta)$. Here, we {\it assume} that expectation values
of all currents vanish
\begin{equation}
\langle J(z)       \rangle\ =\
\langle j(z)       \rangle\ =\
\langle \eta(z)    \rangle\ =\
\langle \bar\eta(z)\rangle\ =\ 0.
\end{equation}
In this sense, the OPE of Eq. (\ref{freecurrentope})
completely specifies and defines our model.
To illustrate the power of the OPE,
let us consider the correlation function
$\langle j(z_1) \bar\eta (z_2)\eta (z_3)\rangle$.
We know that this three-point function is a holomorphic function of
$z_i$'s.  We may view it as a function of $z_1$.
If so, this function must have poles at
$z_2$ and $z_3$ and must vanish asymptotically as $z_1\to \infty$.
The poles are then completely fixed by the OPE:
the three-point function has a first order pole at $z_2$ with residue
$2\langle \bar\eta (z_2) \eta (z_3)\rangle$, and a first order pole at
$z_3$ with residue $- 2\langle\bar\eta (z_2) \eta (z_3)\rangle$.  This
completely determines
\begin{equation}
\big\langle j (z_1) \bar\eta (z_2) \eta (z_3)\big\rangle \ =\
{2\over z_1 - z_2}\big\langle\bar\eta (z_2) \eta (z_3)\big\rangle\ -\
{2\over z_1 - z_3}\big\langle\bar\eta (z_2) \eta (z_3)\big\rangle.
\label{123}
\end{equation}
With the help of
$\big\langle \bar\eta (z_2) \eta (z_3)\big\rangle={1\over (z_2 - z_3)^2}$,
we find that the three-point function is given by
\begin{equation}
\big\langle j(z_1) \bar\eta (z_2) \eta (z_3)\big\rangle =
{2\over (z_1 - z_2) (z_1 - z_3) (z_2 - z_3)}.
\label{jbaretaeta}
\end{equation}

It is important to realize that not every possible OPE gives rise
to a consistent theory.  For example, to calculate the three-point
function in Eq. (\ref{jbaretaeta}), we expressed it in terms of
two-point functions whose coefficients are determined by the
OPE between the currents.
But by so doing, we chose a specific decomposition path
corresponding to viewing the three-point function as a function of $z_1$.
One needs to show that different decomposition paths
always lead to the same correlation function.  In the above example we may
instead first view
$\langle j(z_1) \bar\eta (z_2) \eta (z_3)\rangle$
as a function of $z_2$,
which has a first order pole at $z_1$ with residue
$- 2 \langle\bar\eta (z_1) \eta (z_3)\rangle$,
a second order pole at $z_3$ with residue
$\langle j(z_1)\rangle$,
and a first order pole at $z_3$ with residue
$\langle j (z_1) J (z_3)\rangle$.
Thus,
\begin{equation}
\big\langle j(z_1) \bar\eta (z_2) \eta (z_3)\big\rangle =
-{2\over z_2 - z_1}
\big\langle\bar\eta (z_1) \eta (z_3)\big\rangle +
{1\over z_2 - z_3}
\big\langle j(z_1) J(z_3)\big\rangle.
\label{213}
\end{equation}
Comparing Eq. (\ref{123}) with Eq. (\ref{213}), we see
that the three-point function
can be decomposed into different Laurent series (OPE).
Thus, the OPE between the currents
need to satisfy certain self-consistent (associativity)
conditions insuring that
all current correlation functions are uniquely defined.
In our example, this is the case.
The self-consistency of the OPE in Eq. (\ref{freecurrentope})
follows {}from it being derived {}from the partition function
${\cal Z}^{\ }_0$.

\subsection{General $U(1/1)$ current algebra for interacting models}
\label{General $U(1/1)$ current algebra for interacting models}

We are now ready to describe the philosophy behind the current algebra
approach.  We know that the model specified by ${\cal Z}^{\ }_0$
in Eq. (\ref{freeu11}) describes a critical point characterized
by algebraic decaying correlation functions.
Moreover, the representation of ${\cal Z}^{\ }_0$ in terms of the free
fields $\psi^{\dag}_+,\cdots,\varphi^{\ }_-$ allows us to easily calculate
{\it all} properties of the critical point.  On the other hand,
for a generic interacting theory such as
\begin{mathletters}
\label{interactingu11}
\begin{eqnarray}
&&
{\cal Z}^{\ }\ =\
\int D[\psi   ^{\dag}_{\pm},\psi   ^{\ }_{\pm}]
\int D[\varphi^{\dag}_{\pm},\varphi^{\ }_{\pm}]
\  e^{{\rm i}\int^{+\infty}_{-\infty} dx d\tau\
\left({\cal L}^{\ }_0+{\cal L}^{\ }_{\rm int}\right)},
\\
&&
{\cal L}^{\ }_{\rm int}\ =\
{\rm i} {g^{\ }_A\over\pi^2}
\left(\psi^{\dag}_+\psi^{\ }_++\varphi^{\dag}_+\varphi^{\ }_+\right)
\left(\psi^{\dag}_-\psi^{\ }_-+\varphi^{\dag}_-\varphi^{\ }_-\right)
\ +\
{\rm i} {g^{\ }_M\over4\pi^2}
\left(\psi^{\dag}_+\psi^{\ }_--\varphi^{\dag}_-\varphi^{\ }_+\right)^2
\ +\ \cdots,
\end{eqnarray}
the interactions drive the system to a new fixed point.
The interaction with coupling constant $g^{\ }_A>0$
can be thought of as being
induced when integrating over Abelian vector gauge-like disorder
in the effective action $S^{(1,0)}_{\rm cr}$ of Eq. (\ref{s(10)cr}).
Similarly,
the interaction with coupling constant $g^{\ }_M>0$ corresponds to
integrating over a randomly distributed mass for Dirac fermions.
However, {}from the point of view of the current algebra, $g^{\ }_{A,M}$ can
be negative as we will see later on.
Finally, ``$\cdots$'' represents other interaction terms.
We would like to know what are the critical
properties of the new fixed point.
The difficulty is that the fields
$\psi^{\dag}_+,\cdots,\varphi^{\ }_-$
do not provide a convenient description of the new fixed point,
making it extremely difficult to calculate its critical properties
in terms of $\psi^{\dag}_+,\cdots,\varphi^{\ }_-$.
However, it is sometimes possible to take advantage of the fact
that the interaction ${\cal L}^{\ }_{\rm int}$ has a $U(1/1)$ graded symmetry.
At a critical point of the interacting model, the $U(1/1)$ symmetry
is promoted to the $U(1/1)\times U(1/1)$ symmetry in the low energy sector,
due to the decoupling of the holomorphic and antiholomorphic sectors
at low energy.
In this case, many critical properties of the new
fixed point can then be derived solely {}from the knowledge of
$U(1/1)\times U(1/1)$ current algebra.
The current algebra approach to a critical point in two dimensions is
based on two assumptions: $i)$ the critical point has a conformal symmetry
(which is responsible for promoting the $U(1/1)$ to  $U(1/1)\times U(1/1)$
symmetry),
$ii)$ the conserved currents have algebraic decaying correlations, i.e.,
they belong to gapless sectors.
\end{mathletters}

To construct the critical points with $U(1/1)\times U(1/1)$ symmetry,
we denote the infinitesimal generators of the $U(1/1)$ graded symmetry in the
holomorphic sector by
$(I^{\ }_1,I^{\ }_2,I^{\ }_3,I^{\ }_4)(z)=(J,j,\eta,\bar\eta)(z)$.
Without loss of generality, we only work in the holomorphic sector.
Since conserved currents cannot develop anomalous dimensions,
the conformal weights of $I^{\ }_a$, $a=1,\cdots,4$ are $(1,0)$.
Consequently, the leading pole in the OPE among the $\{I^{\ }_a\}$
must be of second order, and the most general OPE between the
holomorphic components of the generators of the $U(1/1)$ graded
algebra takes the form
\begin{equation}
I^{\ }_a (z) I^{\ }_b (0)\ =\
{k^{\ }_{ab}\over z^2}\ +\ {f^{\ }_{abc}\over z} I^{\ }_c\ +\ {\cal O}(1),
\quad a,b=1,\cdots,4.
\label{genericu11kacmoody}
\end{equation}
The coefficients $k^{\ }_{ab}$
of the second order pole contain information about
the dynamics of the critical point. We will see below that they are
severely restricted by the condition that  $k^{\ }_{ab}$
be an invariant second rank tensor with respect
to the $U(1/1)$ symmetry.
The coefficients $f^{\ }_{abc}$ are the structure constants of the
graded Lie algebra $U(1/1)$.

To see this last point, we introduce the conserved charges
\begin{equation}
Q_a (\tau)\ =\
{1\over 2\pi{\rm i}}\int_{-\infty}^{+\infty} dx\ I_a (x,\tau),
\quad a=1,\cdots,4,
\end{equation}
which generate finite $U(1/1)$ transformations.
Recalling that correlation functions defined by
a path integral corresponds to
$\tau$-ordered correlation functions in the
operator formalism \cite{Ginsparghouches},
we find ($a,b=1,\cdots,4$)
\begin{eqnarray}
\left\langle
[Q^{\ }_a(0), Q^{\ }_b(0)]^{\ }_{\pm} \prod_i O(x_i, \tau_i)
\right\rangle\ =\
\left({1\over2\pi{\rm i}}\right)^2
\left\langle
\int dz \oint_C dz' I^{\ }_a (z') I^{\ }_b (z)\
\prod_i O(z_i)
\right\rangle,
\label{comQQ}
\end{eqnarray}
where the integration contour $C$ is a small circle centered at $z$.
We have used the fact that
$I^{\ }_a$ is a holomorphic function of $z$ in the last line of
Eq. (\ref{comQQ}).
Care has to be taken in Eq. (\ref{comQQ}) to use the
anticommutator $[\ \cdot\ ,\cdot\ ]^{\ }_+ $
whenever $Q^{\ }_a$ and $Q^{\ }_b$
are both fermionic, whereas for all other cases
the commutator $[\ \cdot\ ,\cdot\ ]^{\ }_- $
is to be used. Substitution of the OPE between currents,
Eq. (\ref{genericu11kacmoody}),
into Eq. (\ref{comQQ}) yields the desired result
\begin{equation}
[Q^{\ }_a, Q^{\ }_b]^{\ }_{\pm} = f^{\ }_{abc} Q^{\ }_c,
\quad a,b=1,\cdots,4.
\label{QaQb}
\end{equation}
We see that the first order poles in the OPE of the conserved current
are determined solely by the symmetry group, and thus do not depend on
the interaction. For the $U(1/1)$ symmetry, $f^{\ }_{abc}$
can be read off {}from the first
order poles in Eq. (\ref{freecurrentope}).
Moreover,
we can generalize the calculation leading to
Eq. (\ref{QaQb}) by introducing
\begin{equation}
Q^f_a(\tau)\ =\
{1\over 2\pi{\rm i}}\int_{-\infty}^{+\infty} dx\ I_a (x,\tau)\ f(z),
\quad a=1,\cdots,4,
\end{equation}
for an arbitrary holomorhic function $f(z)$.  We find
\begin{mathletters}
\begin{equation}
\left[Q^f_a (\tau), I^{\ }_b (x,\tau)\right]^{\ }_{\pm}\ =\
f'(z) k^{\ }_{ab}\ +\ f(z)f^{\ }_{abc} I^{\ }_c(z),
\quad a,b=1,\cdots,4,
\end{equation}
or equivalently
\begin{equation}
\left[ I^{\ }_a(x',\tau), I^{\ }_b(x,\tau)\right]^{\ }_{\pm}\ =\
-\delta'(x'-x) k^{\ }_{ab}\ +\ \delta (x' -x)f^{\ }_{abc} I^{\ }_c(z),
\quad a,b=1,\cdots,4.
\end{equation}
\end{mathletters}

To be consistent with the $U(1/1)$ symmetry,
$k^{\ }_{ab}$ must be an invariant tensor of the symmetry group.
This constraint alone requires $k^{\ }_{ab}$ to depend on only two
(real) parameters $(k,k^{\ }_j)$, which
are uniquely determined by the $U(1/1)$ symmetric interaction.
To see this, we consider first an infinitesimal transformation generated by
$\bar\theta\eta+\theta\bar\eta$. Under this transformation, the currents
$(J, j, \eta, \bar\eta)$ change by the infinitesimal amounts
\begin{equation}
(\delta^{\ }_{\eta}+\delta^{\ }_{\bar\eta}) J     \ =\
0,\quad
(\delta^{\ }_{\eta}+\delta^{\ }_{\bar\eta})j      \ =\
2\bar\eta\theta + 2\bar\theta \eta,\quad
(\delta^{\ }_{\eta}+\delta^{\ }_{\bar\eta})\eta    \ =\
\theta J,\quad
(\delta^{\ }_{\eta}+\delta^{\ }_{\bar\eta})\bar\eta\ =\
\bar\theta.
\label{deltaJ,j,eta,bareta}
\end{equation}
Here, as a result of current conservation, the transformation law obeyed
by the currents is independent of the interaction,
and thus can be derived {}from the free
theory ${\cal Z}^{\ }_0$ of Eq. (\ref{freeu11}).
Now, by requiring that $k^{\ }_{ab}$ be invariant under $U(1/1)$, it
follows that any infinitesimal change $\delta(I^{\ }_a(z)I^{\ }_b(0))$
generated by the currents
does not contain poles higher than first order in $z$.
For example, {}from Eq. (\ref{deltaJ,j,eta,bareta}) one finds
\begin{eqnarray}
\big(\delta^{\ }_{\eta}+\delta^{\ }_{\bar\eta}\big)
\left(\eta(z)j(0)\right)\ &=&\
\theta J(z)j(0)\ +\
2\eta(z)
\left(\bar\eta(0) \theta\ +\ \bar\theta \eta(0)\right)
\nonumber\\
&=&\
{\theta \left(k^{\ }_{Jj}\ +\ 2 k^{\ }_{\eta\bar\eta}\right)\over z^2}
\ -\
{2\bar\theta k^{\ }_{\eta\eta}\over z^2}
\ +\
{\cal O}( {1\over z}),
\end{eqnarray}
which implies that $k^{\ }_{Jj}+2 k^{\ }_{\eta\bar\eta}=0$
and
$k^{\ }_{\eta\eta}=0$.
Repeating this argument for all
$\delta\big(I^{\ }_a(z)I^{\ }_b(0)\big)$,
we find that the most general OPE between
the currents generating $U(1/1)$ transformations
in the holomorphic sector are given by
\begin{mathletters}
\label{u11currentope}
\begin{eqnarray}
&&
J(z)      J(0)=0,            \quad
J(z)      j(0)={2k\over z^2},\quad
J(z)   \eta(0)=0,            \quad
J(z)\bar\eta(0)=0,
\\
&&
j(z)       J(0)={2k\over z^2},           \quad
j(z)       j(0)={k^{\ }_j\over z^2},     \quad
j(z)    \eta(0)=-{2\over z}\eta(0),      \quad
j(z)\bar\eta(0)=+{2\over z}\bar\eta(0),
\\
&&
\eta(z)      J(0)=0,                           \quad
\eta(z)      j(0)=+{2\over z}\eta(0),          \quad
\eta(z)   \eta(0)=0,                           \quad
\eta(z)\bar\eta(0)=-{k\over z^2}+{1\over z}J(0),
\\
&&
\bar\eta(z)      J(0)=0,                           \quad
\bar\eta(z)      j(0)=-{2\over z}\bar\eta(0),      \quad
\bar\eta(z)   \eta(0)=+{k\over z^2}+{1\over z}J(0),\quad
\bar\eta(z)\bar\eta(0)=0.
\end{eqnarray}
We show in appendix
\ref{sec:Lagrangian realization of U(1/1) x U(1/1) current algebra}
that $k=1$, $k^{\ }_j>0$ corresponds to the current algebra
derived {}from the  effective
partition function ${\cal Z}^{(1,0)}_{\rm cr}$
of Eq. (\ref{s(10)cr})
with $U(1)$ vector gauge-like disorder. In other words,
the $U(1/1)\times U(1/1)$ current algebra given by
Eq. (\ref{u11currentope})
with $k=1,k^{\ }_j>0$
describes a two-dimensional random $U(1)$ vector potential model
at criticality.
It is also shown that the OPE between the currents
is self-consistent (associative)
for all real values of $(k,k^{\ }_j)$. In other words,
all correlation functions for the currents are uniquely defined by their
OPE in Eq. (\ref{u11currentope}).
\end{mathletters}

We close this subsection with the construction of the energy-momentum tensor
associated with the OPE in Eq. (\ref{u11currentope}). The
holomorphic component $T$ of the energy-momentum tensor (EM) describes
important properties of the fixed point. We sketch here
how the energy-momentum tensor $T$ in the holomorphic sector
is uniquely specified by the requirements that:
$i)$    $T$ is Hermitean and is solely constructed {}from the currents
$I^{\ }_a$,
$ii)$   $T$ has conformal weights $(2,0)$,
$iii)$  $T$ transforms like a singlet under $U(1/1)$.
At the classical level, conditions $i)$ and $iii)$ are met by the Ansatz
\begin{equation}
T(z)\ =\ \sum_{a,b=1}^4\ \big(I^{\ }_a T^{\ }_{ab} I^{\ }_b\big)(z),
\end{equation}
where $T^{\ }_{ab}$ is Hermitean. The condition that $T$ be a
$U(1/1)$ singlet is purely classical.
Imposing $iii)$ yields
\begin{equation}
T(z)\ =\
\kappa^{\ }_1\left(Jj\ +\ \eta\bar\eta-\bar\eta\eta\right)(z)
\ +\
\kappa^{\ }_2 \big(JJ\big)(z),
\end{equation}
where $\kappa^{\ }_{1,2}$ are real.
At the quantum level,
conditions $i)$ and $ii)$ must be treated with care.
The quantum definition of $T$ involves taking the limit
\begin{eqnarray}
T(z)\ &=&\
\lim_{\epsilon\to 0}
\Bigg\{
\kappa^{\ }_1
\left[
{1\over 2}J(z+\epsilon)j(z)
+
{1\over 2}j(z+\epsilon)J(z)
+
\eta(z+\epsilon)\bar\eta(z)
-
\bar\eta(z+\epsilon)\eta(z)
\right]
\nonumber\\
&+&\
\kappa^{\ }_2 J(z+\epsilon)J(z)
\Bigg\},
\label{quantumEM}
\end{eqnarray}
which defines the product of operators at the same point.  In general, we
need to subtract out some infinities, i.e., the poles that appear
in the OPE of the currents,
in order for the EM tensor to be finite in the limit $\epsilon\rightarrow0$.
However, one can show {}from the OPE in Eq. (\ref{u11currentope}) that
$T$, as defined by Eq. (\ref{quantumEM}), is
already finite so that additional subtractions are not needed.
Also note that the right-hand side of Eq. (\ref{quantumEM}) is a $U(1/1)$
singlet even for finite values of $\epsilon$.
The two additional coefficients $\kappa^{\ }_1$ and $\kappa^{\ }_2$
can be determined {}from the OPE of the EM tensor.  In a two-dimensional
conformal field theory \cite{Belavin 1984},
the OPE of the EM tensor must have the form
\begin{equation}
T(z) T(0)\ =\
{{c\over 2} \over z^4}\ +\ {2\over z^2} T(0)\ +\
{1\over z} \partial_z T(0).
\label{EMOPE}
\end{equation}
The residue of the fourth order pole determines the so-called Virasoro
central charge $c$.
The residue of the second order pole
determines the conformal weight of the energy momentum tensor, namely
two. The constraint $ii)$ is equivalent to imposing Eq. (\ref{EMOPE}).
One can show {}from the OPE in Eq. (\ref{u11currentope})
and the definition Eq. (\ref{quantumEM})
that Eq. (\ref{EMOPE}) holds if
\begin{equation}
\kappa^{\ }_1\ =\ {1\over2k},
\qquad
\kappa^{\ }_2\ =\ {4-k^{\ }_j\over8k^2},
\end{equation}
in which case the Virasoro central charge vanishes: $c=0$. One can also show
that the currents $(J,j,\eta,\bar\eta)$ have conformal weights $(1,0)$,
since their OPE with $T$ is given by:
\begin{eqnarray}
&&
T(z)I^{\ }_a(0)\ =\
{1\over z^2} I^{\ }_a(0)\ +\ {1\over z} \partial^{\ }_zI^{\ }_a(0),
\quad a=1,\cdots,4.
\label{OPETcurrents}
\end{eqnarray}
Conversely, we show in appendix
\ref{sec:U(1/1) Suguwara construction}
that, requiring Eq (\ref{OPETcurrents}) to hold together with the definition
Eq (\ref{quantumEM}),
uniquely defines the energy momentum tensor.

\subsection{Operators in one- and two-dimensional representations of
the $U(1/1)$ Lie algebra}

The $U(1/1)$ current algebra contains many operators apart {}from
the four currents $(J, j, \eta, \bar\eta)$.  For example, {}from the fact
that the current $J$ is a singlet of $U(1/1)$,
we can use $J$ to construct operators
which form a one-dimensional representation of the $U(1/1)$ Lie algebra.
Let us consider the following operator
\begin{mathletters}
\label{fq}
\begin{equation}
f^{\ }_q(z)\ \equiv\ e^{-{\rm i}{q\over 2 k} \theta^{\ }_J (z)},\quad
\end{equation}
where $q$ is real, and $\theta_J (z)$ is the solution of
\begin{equation}
\partial^{\ }_z\theta^{\ }_J (z)\ \equiv\  {\rm i} J(z).
\end{equation}
We use the OPE between the currents to construct the $U(1/1)$ charges of
$f^{\ }_q$ and its (holomorphic) conformal weight $h^{\ }_{f^{\ }_q}$.
One verifies that $f^{\ }_q(z)$ has the following OPE with the currents
\end{mathletters}
\begin{equation}
J(z) f^{\ }_q (0)\ =\ 0,
\quad
j(z) f^{\ }_q (0)\  =\
-{q\over z} f^{\ }_q (0),
\quad
\eta (z) f^{\ }_q (0)\ =\
\bar\eta (z) f^{\ }_q (0)\ =\ 0.
\label{OPEfq}
\end{equation}
The OPE in Eq. (\ref{OPEfq}) gives the $U(1/1)$ charges
$(J,j,\eta,\bar\eta)=(0,-q,0,0)$ of $f^{\ }_q$.
Clearly, $f^{\dag}_qf^{\ }_q$ is then a $U(1/1)$ singlet, and
$f^{\ }_q$ carries a one-dimensional representation of $U(1/1)$.
The OPE in Eq. (\ref{OPEfq}) allows us to reduce any correlations
between currents and
$f^{\ }_q$'s into one that involve only $f^{\ }_q$'s.
The correlation between $f^{\ }_q$'s is given by
\begin{equation}
\langle f^{\ }_q (z^{\ }_1)\ f^{\ }_{-q} (z^{\ }_2)\rangle\ =\
{\rm constant},
\end{equation}
since $J(z)J(0)$=0.
This implies that $f^{\ }_q$ has vanishing conformal weight:
$h^{\ }_{f^{\ }_q} = 0$.  More generally, the
$n$-point correlation function for $f^{\ }_{q}$ is given by
\begin{equation}
\left\langle\prod_{i=1}^n f^{\ }_{q^{\ }_i} (z^{\ }_i)\right\rangle\ =\
\cases
{
0        & if $\sum_{i=1}^n q^{\ }_i, \ne 0$,\cr
\ &\ \cr
{\rm constant}    & if $\sum_{i=1}^n q^{\ }_i = 0.$
\cr}
\label{npointcorrelationfq}
\end{equation}

The $U(1/1)$ current algebra also contains operators which form
two-dimensional representations of the $U(1/1)$ algebra.
To see this, we define
\begin{equation}
\psi^{\dag}_p\ \equiv\ \varphi^{\dag(p-1)}\psi^{\dag},
\quad
\varphi^{\dag}_p\ \equiv\ {1\over\sqrt{p}}\varphi^{\dag p},
\quad
\psi^{\ }_p\ \equiv\ \psi\varphi^{(p-1)},
\quad
\varphi^{\ }_p\ \equiv\ {1\over\sqrt{p}}\varphi^p
\label{psip},
\end{equation}
where $p$ is a positive integer.
By showing that
$
\psi^{\dag}_p\psi^{\ }_p + \varphi^{\dag}_p\varphi^{\ }_p
$
is a singlet under $U(1/1)$, it follows that
$(\psi^{\ }_p,\varphi^{\ }_p)$ transforms like a doublet of $U(1/1)$.
Notice that this is certainly true when $p=1$.
The $J$ and $j$ charges of
$\{\psi^{\dag}_p,\varphi^{\dag}_p,\psi^{\ }_p,\varphi^{\ }_p\}$
can be read {}from Eq. (\ref{psip}).
They are
$\{p, p, -p, -p\}$ and
$\{2 - p, -p, p -2, p\}$,
respectively. Hence, on the one hand, the $J$ and $j$ charges of
$\psi^{\dag}_p\psi^{\ }_p$
and
$\varphi^{\dag}_p\varphi^{\ }_p$
both vanish. On the other hand,
because of the infinitesimal transformation law
\begin{mathletters}
\label{doublet1}
\begin{eqnarray}
&&
\delta^{\ }_{\eta} \psi^{\dag}_p\ =\
+\sqrt{p}\ \varphi^{\dag}_p\ \bar\theta,
\quad
\delta^{\ }_{\eta} \varphi^{\dag}_p\ =\
0,\quad
\delta^{\ }_{\eta} \psi^{\   }_p\ =\
0,\quad
\delta^{\ }_{\eta} \varphi^{\   }_p\ =\
-\sqrt{p}\ \bar\theta\ \psi^{\   }_p,
\\
&&
\delta^{\ }_{\bar\eta} \psi^{\dag}_p\ =\
0,\quad
\delta^{\ }_{\bar\eta} \varphi^{\dag}_p\ =\
-\sqrt{p}\ \psi^{\dag}_p\ \theta,\quad
\delta^{\ }_{\bar\eta} \psi^{\   }_p\ =\
+\sqrt{p}\ \bar\theta\ \psi^{\ }_p,\quad
\delta^{\ }_{\bar\eta}\varphi^{\   }_p\ =\
0,
\end{eqnarray}
which is generated by the fermionic currents
$\eta$, and $\bar\eta$ of $U(1/1)$ (compare with Eq. (\ref{deltaetas})),
only the sum of
$\psi^{\dag}_p\psi^{\ }_p$
and
$\varphi^{\dag}_p\varphi^{\ }_p$
is a $U(1/1)$ singlet.
Moreover, we can read {}from Eq. (\ref{doublet1}) and the $J$ and $j$
charges of $\psi^{\dag}_p,\varphi^{\dag}_p,\psi^{\ }_p,\varphi^{\ }_p$,
the OPE between the currents and
$\{\psi^{\dag}_p,\varphi^{\dag}_p,\psi^{\ }_p,\varphi^{\ }_p\}$.
They are
\end{mathletters}
\begin{eqnarray}
&&
J(z)       \psi   ^{\dag}_p(0)={p\over z}\psi          ^{\dag}_p(0),\;
j(z)       \psi   ^{\dag}_p(0)={2-p\over z}\psi        ^{\dag}_p(0),\;
\eta(z)    \psi   ^{\dag}_p(0)={\sqrt{p}\over z}\varphi^{\dag}_p(0),\;
\bar\eta(z)\psi   ^{\dag}_p(0)=0,
\nonumber\\
&&
J(z)       \varphi^{\dag}_p(0)={p\over z}       \varphi^{\dag}_p(0),\;
j(z)       \varphi^{\dag}_p(0)=-{p\over z}       \varphi^{\dag}_p(0),\;
\eta(z)    \varphi^{\dag}_p(0)=0,\;
\bar\eta(z)\varphi^{\dag}_p(0)=-{\sqrt{p}\over z}\psi^{\dag}_p(0),
\nonumber\\
&&
J(z)       \psi^{\   }_p(0)=-{p\over z}       \psi^{\   }_p(0),\;
j(z)       \psi^{\   }_p(0)=-{2-p\over z}     \psi^{\   }_p(0),\;
\eta(z)    \psi^{\   }_p(0)=0,\;
\bar\eta(z)\psi^{\   }_p(0)={\sqrt{p}\over z}\varphi^{\   }_p(0),
\nonumber\\
&&
J(z)       \varphi^{\   }_p(0)=-{p\over z}       \varphi^{\   }_p(0),\;
j(z)       \varphi^{\   }_p(0)={  p\over z}     \varphi^{\   }_p(0),\;
\eta(z)    \varphi^{\   }_p(0)=-{\sqrt{p}\over z}\psi   ^{\   }_p(0),\;
\bar\eta(z)\varphi^{\   }_p(0)=0.
\nonumber\\
&&
\label{OPEpsip}
\end{eqnarray}
We see that the OPE of (3.37) and (3.38) are well defined even when $p$ is
not an integer.  {}From now on we will regard $p$ to be a positive real
number.

Even more general doublet can be constructed if we
combine the $\psi^{\dag}_p,\cdots, \varphi^{\ }_p$ with $f_q$
by defining
\begin{equation}
\psi^{\dag}_{pq}\ \equiv\ f^{\dag}_{q} \psi^{\dag}_{p},
\quad
\varphi^{\dag}_{pq}\ \equiv\ f^{\dag}_{q} \varphi^{\dag}_{p},
\quad
\psi^{\ }_{pq}\ \equiv\ \psi^{\ }_{p} f^{\ }_{q},
\quad
\varphi^{\ }_{pq}\ \equiv\ \varphi^{\ }_{p} f^{\ }_{q}.
\label{psipq}
\end{equation}
Here, $p$ is a positive real number while $q$ is real.
The OPE between $(\psi_{p q}, \varphi_{p q})$ and the currents are very
similar to those between $(\psi_p, \varphi_p)$ and the currents,
since the multiplication of
$f_q$ merely shifts the quantum number of $j$. In fact, the OPE between
$(\psi_{p q}, \varphi_{p q})$ and the currents are still given by
Eq. (\ref{OPEpsip})
except for $p$ in the OPE with the current $j$
being  shifted to $p-q$. In fact,
the transformation law of Eq. (\ref{doublet1})
also applies to $(\psi^{\ }_{p q},\varphi^{\ }_{p q})$.

{}From the OPE of Eqs. (\ref{u11currentope},\ref{OPEfq},\ref{OPEpsip}),
together with the definition of
the EM tensor in Eq. (\ref{quantumEM}),
we can calculate the OPE between $T$ and
$(\psi_{p q}, \varphi_{p q})$.
It is
\begin{mathletters}
\label{TOPEpsipq}
\begin{eqnarray}
T(z) \psi^{\ }_{pq}(0)&=&
{h^{\ }_{pq}\over z^2}\psi^{\ }_{pq} (0)+
{1\over z}\left(\partial^{\ }_z\psi^{\ }_{pq}\right)(0),
\\
T(z)\varphi^{\ }_{pq} (0)&=&
{h^{\ }_{pq}\over z^2}\varphi^{\ }_{pq}(0)+
{1\over z}\left(\partial^{\ }_z\varphi^{\ }_{pq}\right)(0).
\end{eqnarray}
Eq. (\ref{TOPEpsipq})
implies that $(\psi_{pq}, \varphi_{pq})$ are {\it primary fields}
\cite{Belavin 1984} with (holomorphic) conformal weight given by
\end{mathletters}
\begin{equation}
h^{\ }_{pq}\ =\
{p\over 2k}\left(1-p+q\right)
\ +\
{p^2\over 2k^2} \left(1-{k^{\ }_j\over4}\right).
\label{hpq}
\end{equation}
For the free theory of Eq. (\ref{freeu11}), $(k, k^{\ }_j) = (1, 0)$.
In this case Eq. (\ref{hpq}) predicts
$h^{\ }_{p0} = {p\over 2}$, which agrees with the dimension of
$\psi\varphi^{p-1}$ and $\varphi^p$ calculated using
the dimension $h^{\ }_{\psi,\varphi}={1\over 2}$.
Similarly,
$(\psi^{\dag}_{pq},\varphi^{\dag}_{pq})$
have the same scaling dimension $h^{\ }_{pq}$.  This is
a consequence of the charge conjugation symmetry of the theory.

We have shown that the OPE between the currents in Eq.
(\ref{u11currentope})
completely determine all their $n$-point correlation functions.
We want to know if the same holds for the doublets
$(\psi_{pq}, \varphi_{pq})$, i.e., whether
it is possible to calculate any $n$-point
correlation function between those new operators and the currents.
We begin with the two-point correlation functions between
$\psi^{\dag}_{pq},\varphi^{\dag}_{pq},\psi^{\ }_{pq},\varphi^{\ }_{pq}$.
On the one hand, {}from Eq. (\ref{OPEpsip}),
$\psi^{\ }_{pq}\psi^{\dag}_{pq}$
and
$\varphi^{\ }_{pq}\varphi^{\dag}_{pq}$
have vanishing $(J,j)$ charges, and, therefore, can acquire non-vanishing
expectation values. We see that current conservation
together with Eq. (\ref{TOPEpsipq}) requires that
\begin{equation}
\big\langle\psi^{\ }_{pq}(z)\psi^{\dag}_{pq}(0)\big\rangle\ =\
{C^{\ }_\psi\over z^{2 h^{\ }_{pq}}},
\quad
\big\langle\varphi^{\ }_{pq}(z)\varphi^{\dag}_{pq}(0)\big\rangle\ =\
{C^{\ }_\varphi\over z^{2 h^{\ }_{pq}}}.
\label{twopointfunctionforpq}
\end{equation}
On the other hand,  Eq. (\ref{OPEpsip}) tells us that
$\psi^{\ }_{pq}\varphi^{\dag}_{pq}$
and
$\varphi^{\ }_{pq}\psi^{\dag}_{pq}$
have the $(J,j)$ charges
$(0,-2)$
and
$(0,+2)$,
respectively.
Hence, they cannot acquire non-vanishing expectation values due to
current conservation:
\begin{equation}
\big\langle\psi^{\ }_{pq}(z)\varphi^{\dag}_{pq}(0)\big\rangle\ =\
\big\langle\varphi^{\ }_{pq}(z)\psi^{\dag}_{pq}(0)\big\rangle\ =\ 0.
\label{identity1}
\end{equation}
This allows us to relate the real parameters $C^{\ }_{\psi}$ and
$C^{\ }_{\varphi}$. Indeed, if we use the transformation law
\begin{mathletters}
\label{doublet2}
\begin{eqnarray}
&&
\delta^{\ }_{\eta} \psi^{\dag}_{pq}\ =\
+\sqrt{p}\ \varphi^{\dag}_{pq}\ \bar\theta,
\quad
\delta^{\ }_{\eta} \varphi^{\dag}_{pq}\ =\
0,\quad
\delta^{\ }_{\eta} \psi^{\   }_{pq}\ =\
0,\quad
\delta^{\ }_{\eta} \varphi^{\   }_{pq}\ =\
-\sqrt{p}\ \bar\theta\ \psi^{\   }_{pq},
\\
&&
\delta^{\ }_{\bar\eta} \psi^{\dag}_{pq}\ =\
0,\quad
\delta^{\ }_{\bar\eta} \varphi^{\dag}_{pq}\ =\
-\sqrt{p}\ \psi^{\dag}_{pq}\ \theta,\quad
\delta^{\ }_{\bar\eta} \psi^{\   }_{pq}\ =\
+\sqrt{p}\ \bar\theta\ \psi^{\ }_{pq},\quad
\delta^{\ }_{\bar\eta}\varphi^{\   }_{pq}\ =\
0,
\end{eqnarray}
under infinitesimal transformations generated by $\eta$, and $\bar\eta$,
we can show that
\end{mathletters}
\begin{equation}
\left(\delta^{\ }_{\eta}+\delta^{\ }_{\bar\eta}\right)
\big\langle\varphi^{\ }_{pq} (z)\psi^{\dag}_{pq} (0)\big\rangle\ =\
\sqrt{p}\ \bar\theta\
\left(-C^{\ }_\psi + C^{\ }_\varphi\right)
{1\over z^{2h^{\ }_{pq}}}.
\label{identity2}
\end{equation}
Eqs. (\ref{identity1}) and (\ref{identity2})
then imply that $C^{\ }_\psi = C^{\ }_\varphi$.
Setting
$C^{\ }_\psi = C^{\ }_\varphi = 1$
through  rescaling of the doublet $(\psi^{\ }_{pq}, \varphi^{\ }_{pq})$,
reduces Eq. (\ref{twopointfunctionforpq}) to
\begin{equation}
\langle\psi^{\ }_{pq}(z)       \psi^{\dag}_{pq} (0)\rangle\ =\
\langle\varphi^{\ }_{pq}(z) \varphi^{\dag}_{pq} (0)\rangle\ =\
{1\over z^{2h^{\ }_{pq}}}.
\label{normalizedtwopointfunctionforpq}
\end{equation}

It is useful to consider the OPE of two doublets
$(\psi_{p_1q_1}, \varphi_{p_1q_1})$
and
$(\psi_{p_2q_2}, \varphi_{p_2q_2})$.
One can easily see that the product
$(\psi_{p_1q_1}, \varphi_{p_1q_1})\times(\psi_{p_2 q_2},\varphi_{p_2q_2})$
contains a new doublet
$(\psi_{p_3q_3},\varphi_{p_3q_3})$
with
$p_3 = p_1 + p_2$
and
$q_3 =q_1 + q_2$.  Thus, the OPE of $(\psi_{p_1 q_1}, \varphi_{p_1 q_1})$
and $(\psi_{p_2 q_2}, \varphi_{p_2 q_2})$ must be of the form
\begin{mathletters}
\label{OPEqpqp}
\begin{eqnarray}
\varphi_{p_1 q_1} (z) \varphi_{p_2 q_2} (0)\ &\sim&\
{1\over z^{h_1 + h_2 - h_3}}\varphi_{p_3 q_3}+ \cdots,
\\
\psi_{p_1q_1} (z) \varphi_{p_2 q_2} (0)     \ &\sim&\
{1\over z^{h_1 + h_2 - h_3}} \psi_{p_3 q_3} + \cdots,
\\
\varphi_{p_1 q_1} (z) \psi_{p_2 q_2} (0)     \ &\sim&\
{1\over z^{h_1 + h_2 - h_3}} \psi_{p_3 q_3} + \cdots,
\end{eqnarray}
where $h_i$ are dimensions of $(\psi_{p_i q_i}, \varphi_{p_i q_i})$.  This
formula will be used below to discuss the locality of operators.
\end{mathletters}

\subsection{Local operators and quantization of $k$}

According to the general principles of two-dimensional
conformal field theory (CFT) \cite{Belavin 1984},
any generic operator can be written as the product of operators in the
holomorphic ($z$) and antiholomorphic ($\bar z$) sectors:
\begin{equation}
O(z, \bar z) = O^{\ }_{z}(z) O^{\ }_{\bar z}(\bar z).
\label{mixedform}
\end{equation}
The OPE of such mixed operators are given by
\begin{equation}
O^{\ }_a (z, \bar z) O^{\ }_b (0,0)\ =\ \sum_c C_{ab}^c\
z^{h_c - h_a - h_b}\
\bar
z^{\bar h_c - \bar h_a - \bar h_b}\
O^{\ }_c(0,0),
\label{generalOPEinCFT}
\end{equation}
where $(h_a,\bar h_a)$ are the conformal weights of
$O_a$, and $C_{ab}^c$ are complex numbers.

The concept of a local operator is very important in our theory.
In CFT, two operators $O_a(z,\bar z)$ and $O_b(z,\bar z)$
are said to be {\it mutually local}
if their OPE
(i.e., the right-hand side of Eq. (\ref{generalOPEinCFT}))
is single-valued.
This implies that the spin difference
\begin{equation}
s^{\ }_c-s^{\ }_a-s^{\ }_b \equiv\
\big(h^{\ }_c-\bar h^{\ }_c\big)-
\big(h^{\ }_a-\bar h^{\ }_a\big)-
\big(h^{\ }_b-\bar h^{\ }_b\big)
\label{localitycondition}
\end{equation}
is an integer for all $c$ with non-vanishing $C^c_{ab}$.
A set of operators is said to be made of local operators if they are all
pairwise mutually local. We will call the set of all local operators the
``{\it center}'' of the operator product algebra.
{}From this definition we see that the correlation functions
between local operators (i.e., between operators in the {\it center}
of the operator product algebra)
are always single-valued functions.

With  these definitions in hand, we come back to the theory defined
by Eq. (\ref{interactingu11}),
assuming only that the interaction be invariant with respect to $U(1/1)$.
In the presence of a generic $U(1/1)$ symmetric interaction
${\cal L}^{U(1/1)}_{\rm int}$,
the spatial dependency of the doublets
$(\psi_+, \varphi_+)$
and
$(\psi_-, \varphi_-)$
are no longer purely holomorphic and antiholomorphic, respectively.
This is not to say that the holomorphic and antiholomorphic sectors
are absent. Any critical point of Eq. (\ref{interactingu11}) with $U(1/1)$
symmetry still possesses two decoupled sectors characterized by holomorphic
and antiholomorphic correlation functions. Rather, the doublets
$(\psi^{\ }_{\pm}, \varphi{\ }_{\pm})$
are not necessarily primary fields anymore.
Instead, they are of the mixed form in Eq. (\ref{mixedform}).
(They do, however, still transform like $U(1/1)$
doublets due to current conservation.)

It should be obvious that $\psi^{\dag}_+,\cdots,\varphi^{\ }_-$
and their products in our model Eq. (\ref{interactingu11})
are local operators provided we assume that the interaction
${\cal L}^{U(1/1)}_{\rm int}$, besides
being $U(1/1)$ symmetric, is also local
(i.e., only involves product of fields at the same point).
If so, the doublets
$(\psi^{\ }_{\pm},\varphi^{\ }_{\pm})$
and
$(\psi^{\dag}_{\pm},\varphi^{\dag}_{\pm})$
are local regardless of whether the model is at a critical point
or away {}from a critical point and regardless of the nature of
the critical point as long as it possesses the $U(1/1)$ symmetry.
Thus, any critical point associated to ${\cal L}^{U(1/1)}_{\rm int}$
must contain a set of local operators that carry the same
$U(1/1)$ quantum numbers as
$
\psi^{\dag}_{\pm},
\varphi^{\dag}_{\pm},
\psi^{\ }_{\pm},
\varphi^{\ }_{\pm}
$
and their products do.
We recall that $U(1/1)$ denotes the diagonal subgroup of $U(1/1)\times U(1/1)$,
which appears at the critical point.
This puts a strong constraint on the possible critical
points that ${\cal L}^{U(1/1)}_{\rm int}$ can have.
We now show that the constraints of locality and of current conservation
restricts the possible values of the central charge $k$ to be the inverse of
some integer $m$:
\begin{equation}
k={1\over m}, \quad m\in\ {\bf {\rm Z\hskip -0.2 true cm Z}}.
\label{k=1overm}
\end{equation}
We also show that all the critical points with $U(1/1)\times U(1/1)$
symmetry
for which $\psi^{\dag}_+,\cdots,\varphi^{\ }_-$ are local
are labelled by two integers
$l,m\in{\bf {\rm Z\hskip -0.2 true cm Z}}$
and one real number
$k^{\ }_j$.

The proof uses the components
\begin{equation}
J^{\ }_{\tau}\ =\ {\rm i}\left(J^{\ }_z-J^{\ }_{\bar z}\right),
\quad
j^{\ }_{\tau}\ =\ {\rm i}\left(j^{\ }_z-j^{\ }_{\bar z}\right),
\label{Jjtau}
\end{equation}
which were defined in Eq. (\ref{Jandj}).
Here, $J^{\ }_{\tau}$ and $j^{\ }_{\tau}$ are the bosonic generators of
the diagonal subgroup $U(1/1)$ of $U(1/1)\times U(1/1)$.
For convenience, we have listed
some $U(1/1)\times U(1/1)$ charges in table \ref{table1},
{}from which one immediately finds that the $J^{\ }_\tau$ charges
of $(\psi^{\ }_+,\varphi^{\ }_+,\psi^{\ }_-,\varphi^{\ }_-)$
are ${\rm i}(-1,-1,1,1)$,
while their $j^{\ }_\tau$ charges are ${\rm i}(-1,1,1,-1)$.
Let us assume that we are at a critical point with $U(1/1)\times U(1/1)$
symmetry characterized by the real valued
Kac-Moody central charges $k$ and $k^{\ }_j$.
We know {}from the previous subsection that there exists primary fields
$f^{\ }_{qz}$ and $f^{\ }_{q\bar z}$ ($q$ real) with
vanishing conformal weights and carrying a one-dimensional representation
of the $U(1/1)$ Kac-Moody algebra in the holomorphic ($z$)
and antiholomorphic ($\bar z$)
sectors, respectively. There also exists primary fields
$\psi^{\ }_{pqz},\varphi^{\ }_{pqz}$
and
$\psi^{\ }_{pq\bar z},\varphi^{\ }_{pq\bar z}$
($p$ real and positive)
carrying conformal weights
$(h^{\ }_{pq},0)$ and $(0,h^{\ }_{pq})$
(see Eq. (\ref{hpq})), respectively.
Moreover,
$(\psi^{\ }_{pqz},\varphi^{\ }_{pqz})$
transforms like a doublet (singlet) of the $U(1/1)$ Kac-Moody algebra
in the holomorphic (antiholomorphic) sector, while
$(\psi^{\ }_{pq\bar z},\varphi^{\ }_{pq\bar z})$
transforms like a doublet (singlet) of the $U(1/1)$ Kac-Moody algebra
in the antiholomorphic (holomorphic) sector. Because
$(\psi^{\ }_+,\varphi^{\ }_+)$ and
$(\psi^{\ }_-,\varphi^{\ }_-)$
transform like doublets under diagonal $U(1/1)$,
we must then have
\begin{equation}
\left\lgroup
\matrix
{
\psi^{\ }_+(z,\bar z)\cr
\varphi^{\ }_+(z,\bar z)\cr
}
\right\rgroup
\ \sim\
\left\lgroup
\matrix
{
\psi^{\ }_{pqz}(z)f^{\ }_{q'\bar z}(\bar z)\cr
\varphi^{\ }_{pqz}(z)f^{\ }_{q'\bar z}(\bar z)\cr
}
\right\rgroup,
\quad
\left\lgroup
\matrix
{
\psi^{\ }_-(z,\bar z)\cr
\varphi^{\ }_-(z,\bar z)\cr
}
\right\rgroup
\ \sim\
\left\lgroup
\matrix
{
\psi^{\ }_{pq\bar z}(\bar z)f^{\ }_{q'z}(z)\cr
\varphi^{\ }_{pq\bar z}(\bar z)f^{\ }_{q'z}(z)\cr
}
\right\rgroup,
\end{equation}
where $\sim$ means obeying the same transformation
law under diagonal $U(1/1)$.
The real numbers $p>0,q,q'$ are thus not arbitrary, since the
$J^{\ }_{\tau},j^{\ }_{\tau},\eta^{\ }_{\tau},\bar\eta^{\ }_{\tau}$
charges of the left-hand side must equal those of the right-hand side.
{}From Eq. (\ref{Jjtau}) and table \ref{table1}, conservation
of the $J^{\ }_{\tau}$ current yields the constraint
\begin{equation}
\left\lgroup
\matrix
{
-1\cr
-1\cr
}
\right\rgroup
\ =\
\left\lgroup
\matrix
{
-p\cr
-p\cr
}
\right\rgroup,
\end{equation}
while conservation of the $j^{\ }_{\tau}$ current yields
\begin{equation}
\left\lgroup
\matrix
{
-1\cr
+1\cr
}
\right\rgroup
\ =\
\left\lgroup
\matrix
{
p-2-q+q'\cr
p-q+q'\cr
}
\right\rgroup.
\end{equation}
Current conservation fixes $p=1$ and only leaves $q=q'$ undetermined.
We thus have found that
\begin{equation}
\left\lgroup
\matrix
{
\psi^{\dag}_+,\
\varphi^{\dag}_+,\
\psi^{\ }_+,\
\varphi^{\ }_+\cr
\psi^{\dag}_-,\
\varphi^{\dag}_-,\
\psi^{\ }_-,\
\varphi^{\ }_-\cr
}
\right\rgroup
\ \sim\
\left\lgroup
\matrix
{
\psi^{\dag}_{1qz}f^{\dag}_{q\bar z},\quad
\varphi^{\dag}_{1qz}f^{\dag}_{q\bar z},\quad
\psi^{\ }_{1qz}f^{\ }_{q\bar z},\quad
\varphi^{\ }_{1qz}f^{\ }_{q\bar z}
\cr
\psi^{\dag}_{1q\bar z}f^{\dag}_{qz},\quad
\varphi^{\dag}_{1q\bar z}f^{\dag}_{qz},\quad
\psi^{\ }_{1q\bar z}f^{\ }_{qz},\quad
\varphi^{\ }_{1q\bar z}f^{\ }_{qz}
\cr
}
\right\rgroup.
\label{psisintermsprimaryfields}
\end{equation}
Again, $\sim$ means obeying the same transformation
law under diagonal $U(1/1)$.
Having derived the correlation functions of the primary fields on the
right-hand side of Eq. (\ref{psisintermsprimaryfields}), the
correlation functions among $\{\psi^{\dag}_+,\cdots,\varphi^{\ }_-\}$
can be calculated.
But since
$\{\psi^{\dag}_+,\cdots,\varphi^{\ }_-\}$
are local operators,
their correlation functions must be single-valued.
The condition of locality then further limits the real valued $q$.
For example,
\begin{equation}
\big\langle\varphi^{\ }_+(z,\bar z)\varphi^{\dag}_+(0,0)\big\rangle
\ =\
z^{-2h^{\ }_{1q}}
\end{equation}
is only single-valued if
\begin{equation}
2h^{\ }_{1q}\ =\ l,
\end{equation}
where $l$ is an integer. In other words,
using Eq. (\ref{hpq}) we find that $q$ is parameterized by an integer $l$
\begin{equation}
q\ =\
lk\ -\ {1\over k}\left(1-{k^{\ }_j\over4}\right),
\quad l\in{\bf {\rm Z\hskip -0.2 true cm Z}}.
\label{qofl}
\end{equation}
One can verify that Eq. (\ref{qofl}) insures that all correlation
functions among
$\{\psi^{\dag}_+,\cdots,\varphi^{\ }_-\}$
and powers thereof are single-valued.

Finally, we show that the constraint of locality on the OPE among
$\{\psi^{\dag}_+,\cdots,\varphi^{\ }_-\}$
forces the Kac-Moody central charge $k$ to be quantized.
Without loss of generality, we only consider the OPE between
$\varphi^{\ }_+(z,\bar z)$ and $\varphi^{\ }_+(0,0)$.
Except for the residues, it is the same as the OPE between
$\varphi^{\ }_{1qz}(z)$ and $\varphi^{\ }_{1qz}(0)$
due to Eq. (\ref{npointcorrelationfq}).
It must therefore be of the form (see Eq. (\ref{OPEqpqp})):
\begin{equation}
\varphi^{\ }_+(z,\bar z)
\varphi^{\ }_+(0,0)\
\ =\
z^{h^{\ }_{2(2q)}-2h^{\ }_{1q}}\
\varphi^{\ }_{2(2q)z}(0)f^{\ }_{(2q)\bar z}(0)
+\cdots.
\end{equation}
Locality again requires that
\begin{equation}
h^{\ }_{2(2q)}-2h^{\ }_{1q}\ =\
l +{1\over k},
\end{equation}
be an integer, i.e., $k$ must be the inverse of some integer
$m\in{\bf {\rm Z\hskip -0.2 true cm Z}}$.
We thus see that the $U(1/1)$ critical point of ${\cal L}^{U(1/1)}_{\rm int}$
must be labelled by the triplet $(l,m,k^{\ }_j)$ with
$l,m\in{\bf {\rm Z\hskip -0.2 true cm Z}}$
and $k^{\ }_j$ being a real number.

\subsection{ Operator content of the $U(1/1)\times U(1/1)$ critical points}

We assume that we are at the critical point labelled by the triplet
$(l,m,k^{\ }_j)$. The full local operator content of this critical point
can be obtained {}from the operator algebra generated by the currents
$(J,j,\eta,\bar\eta)$ and the local operators
$\{\psi^{\dag}_+,\cdots,\varphi^{\ }_-\}$.
In the following we want to calculate the conformal weights
of some local composite operators.
We begin with
\begin{equation}
\Phi^{\ }_{+n}\ =\ \varphi^{|n|}_+,\quad
\Phi^{\ }_{-n}\ =\ \varphi^{|n|}_-,\quad
n\in{\bf {\rm Z\hskip -0.2 true cm Z}}.
\label{Phin}
\end{equation}
With a calculation along the lines of the previous subsection,
one verifies that the conformal weights $(h^{\ },0)$ of
$\Phi^{\ }_{+n}$ are the same as the conformal weights
$(h^{\ }_{p'q'},0)$
of $\varphi^{\ }_{p'q'z}$,
where $p'=n,q'=nq$. Hence, with the help of
Eqs. (\ref{k=1overm}), (\ref{hpq}) and (\ref{qofl}),
\begin{equation}
h\ =\ {1\over 2}m|n| +{1\over 2}n^2( l-m ),
\quad n\in{\bf {\rm Z\hskip -0.2 true cm Z}}.
\label{hn}
\end{equation}
Similarly, $\Phi^{\ }_{-n}$ has $(0,h)$ for conformal weights.
More generally, we introduce (compare with Eq. (\ref{Psi}))
\begin{equation}
\Phi^{\ }_{n^{\ }_1n^{\ }_2}\ =\
\cases{
\varphi^{n^{\ }_1}_+ \varphi^{n^{\ }_2}_-,&$n^{\ }_1>0,n^{\ }_2>0$,\cr
\varphi_+^{\dag-n_1} \varphi^{n^{\ }_2}_-,&$n^{\ }_1<0,n^{\ }_2>0$,\cr
\varphi^{n_1}_+      \varphi_-^{\dag-n_2},&$n^{\ }_1>0,n^{\ }_2<0$,\cr
\varphi_+^{\dag-n_1} \varphi_-^{\dag-n_2},&$n^{\ }_1<0,n^{\ }_2<0$,\cr}
\label{Phin1n2}
\end{equation}
where $n_1,n_2$ are integers. The conformal weights
$(h,\bar h)$
of $\Phi^{\ }_{n^{\ }_1n^{\ }_2}$ are equal to: $(i)$ those of
$\varphi^{\ }_{p'q'z}$ in the holomorphic sector with
$p'=|n_1|,q'=(|n_1|\pm|n_2|)q$,
and $(ii)$ those of $\varphi^{\ }_{p'q'\bar z}$ in the antiholomorphic sector
with $p'=|n_2|,q'=(|n_2|\pm|n_1|)q$.
Thus, according to
Eqs. (\ref{k=1overm}), (\ref{hpq}) and (\ref{qofl})
\begin{mathletters}
\label{hn1n2}
\begin{eqnarray}
h\ &=&\
{1\over2} m |n^{\ }_1|\ +\
{1\over2} n^2_1(l-m)\ +\
{1\over2} n^{\ }_1 n^{\ }_2\left[l-m^2\left(1-{k^{\ }_j\over4}\right)\right],
\\
\bar h\ &=&\
{1\over2} m |n^{\ }_2|\ +\
{1\over2} n^2_2(l-m)\ +\
{1\over2} n^{\ }_1n^{\ }_2\left[l-m^2\left(1-{k^{\ }_j\over4}\right)\right].
\end{eqnarray}
Hence, the scaling dimension
$\Delta_{n_1n_2}=h+\bar h$
and spin
$s_{n_1n_2}=h-\bar h$
of $\Phi^{\ }_{n^{\ }_1n^{\ }_2}$
for $n^{\ }_1,n^{\ }_2\in{\bf {\rm Z\hskip -0.2 true cm Z}}$ are
\end{mathletters}
\begin{mathletters}
\label{dimenionandspinn1n2}
\begin{eqnarray}
&&
\Delta^{\ }_{n^{\ }_1n^{\ }_2}\ =\
{1\over 2}m(|n^{\ }_1|+|n^{\ }_2|)\ +\
{1\over 2}
(n^2_1+n^2_2)(l-m)\ +\
n^{\ }_1n^{\ }_2\left[l-m^2\left(1-{k^{\ }_j\over4}\right)\right],
\label{Deltan1n2}
\\
&&
s^{\ }_{n^{\ }_1n^{\ }_2}\ \ =\
{1\over 2}m(|n^{\ }_1|-|n^{\ }_2|)\ +\
{1\over 2} (n^2_1-n^2_2)(l-m),
\label{sn1n2}
\end{eqnarray}
respectively. Notice that we recover the conformal weights of
Eq. (\ref{weightsofPsi}) if $k^{\ }_j={4g^{\ }_A\over\pi}$.
\end{mathletters}

We see {}from Eqs. (\ref{hn},\ref{sn1n2}) that the spins of
$\Phi_{+n},\Phi_{-n}$, and $\Phi_{n_1n_2}$ are either integers or
half-integers depending on whether the labels
$l,m\in{\bf {\rm Z\hskip -0.2 true cm Z}}$
of the  critical point are even or odd.
The composite operators
$\Phi_{+n},\Phi_{-n}$, and $\Phi_{n_1n_2}$ are thus local
for all integer valued $n,n_1$ and $n_2$,
according to Eq. (\ref{localitycondition}).

It should be noted that for any critical point
$(l,m,k_j)$, the condition that $\psi_\pm$ and $\varphi_\pm$
have half-integer spin (although not necessarily ${1\over2}$)
must hold. This is so because the transformation corresponding to a
rotation in the $(\tau,x)$ plane by $2\pi$ changes the sign of the spinors
$\psi^{\ }_\pm$ and $\varphi^{\ }_\pm$, while it
leaves the impurity potential unchanged.
Hence, this property should be preserved at the interacting fixed point.
It then follows {}from Eq. (\ref{sn1n2}) that $l$ is an odd integer.
Other constraints on the possible critical points $(l,m,k_j)$
require additional symmetries. For example by assuming that
the interaction ${\cal L}^{U(1/1)}_{\rm int}$ is invariant
under rotations of the Euclidean plane $(x,\tau)$, it follows
that only $l=1$ is allowed. Indeed, at the free theory critical point
$(1,1,0)$, $\psi_\pm$ and $\varphi_\pm$ have spin $\pm{1\over2}$.
By assumption, this quantum number is unchanged  at the new
critical point $(l,m,k_j)$. {}From Eq. (\ref{sn1n2}),
the condition that, say $s_{10}={1\over2}$, thus requires
$l=1$.

In summary, the possible critical points
${\cal L}^{\ }_0+{\cal L}^{U(1/1)}_{\rm int}$
with $U(1/1)\times U(1/1)$
symmetry are labelled by the odd integer $l$, the integer
$m$ and a real number $k_j$: $(l, m, k^{\ }_j)$.
Many correlation functions can be calculated exactly {}from the OPE.
Although $(l,m,k_j)$ enumerates all
possible critical points with currents obeying
a $U(1/1)\times U(1/1)$ Kac-Moody
algebra, it is not clear whether or not all these critical points
can be realized by a theory with a Lagrangian solely constructed
{}from the fermionic spinors
$\psi^{\dag}_\pm,\psi^{\ }_\pm$ and the  bosonic spinors
$\varphi^{\dag}_\pm,\varphi_\pm$.
We were only able to identify the line $(1,1,k_j)$, $k_j>0$,
with a Lagrangian realization of the random $U(1)$ vector gauge-like model.
Finally, the same approach can be used to identify all critical points
with currents obeying the $G\times G$ Kac-Moody algebra, where $G$
is a graded Lie algebra.
This we have done explicitly for the compact group $U(2/2)$
and the non-compact group $U(1,1/1,1)$.
In both cases, all critical points are labelled by
the triplet $(l,m,k_j)$ and Eq. (\ref{hn1n2})
holds.

\section{Discussion}

We have studied a model of massless Dirac fermions in two spatial
dimensions and in the presence of static random vector potentials
which are distributed according to a white noise probability distribution.
We have shown that this model is an example of a non-unitary
two-dimensional CFT with a highly unusual operator content,
since there exists an infinite hierarchy of negative dimensional operators.
The physical origin of this hierarchy of negative dimensional operators
is the fact that the single-particle Green function
is log-normal distributed with respect to realizations of the randomness.
In turn, the non-selfaveraging nature of the single-particle Green function is
related to the multifractal nature of a typical wave function with
vanishing energy, although it is still an open question of how to calculate
the multifractal scaling exponents associated to the critical
(i.e., vanishing energy) typical
(i.e., for a fixed realization of the disorder)
wave function {}from the field theory.
We have argued that such disordered critical points
are unstable in a very special way, if perturbations corresponding to, say,
randomness in the mass are present in the lattice regularization of
Dirac fermions. Indeed, in addition to the strength of the mass-like
randomness, the shape of its probability distribution induces
infinitely many relevant perturbations through all the cumulants
of the probability distribution.

On the one hand,
it would appear {}from our study of disordered critical points with
$U(1/1)\times U(1/1)$ graded Kac-Moody algebra that the existence
of infinite hierarchies of negative dimensional operators is
related to the multifractality of a typical critical wave function,
and leads to a very special instability of the critical point.
On the other hand,
there are numerical evidences that the plateau transition in the integer
quantum Hall effect is also characterized by negative dimensional operators
\cite{dhLee 1995}.
If so, they clearly do not destabilize the disordered critical point
corresponding to the plateau transition.
The most obvious difference between the plateau transition and
Dirac fermions with random vector potentials is that the averaged
density of states is non-critical in the former. This is very suggestive of
requiring spontaneous symmetry breaking of the chiral symmetry of
disordered Dirac fermions (there is no Mermin-Wagner theorem
preventing this {}from happening in the context of a non-unitary CFT).
Hence, it might be too restrictive to assume as we did that there is
{\it no spontaneous symmetry breaking} of the $U(1/1)$ graded symmetry.

A disordered system related to Dirac fermions in the presence
of random vector potentials is the two-dimensional $XY$ model
with quenched random bond coupling studied by
Rubinstein et al. \cite{Rubinstein 1982}.
Indeed, one can use the equivalence
between the massive Thirring model and the Sine-Gordon model \cite{Thirring},
on the one hand, and the equivalence between the Sine-Gordon model and
the Coulomb gas \cite{Zinn-Justin 1989}, on the other hand,
to relate the random $XY$ model to {\it interacting} Dirac fermions
in the presence of random vector potentials.
For a Gaussian distribution of the randomness with small enough strength
(variance) $0<\sigma<\sigma_{\rm cr}$, Rubinstein et al.
predicted a phase with algebraic decaying correlations at finite
temperatures
for the disordered system.
Korshunov \cite{Korshunov 1993} has argued that this
intermediate phase
with quasi-long range order is not stable.
The source of the instability in Korshunov's analysis is very
reminiscent of the special instability induced by large powers of the
mass operator for our Dirac spinors.
It would thus be extremely useful to understand in the context
of the replica trick how to describe the special
instability of non-interacting Dirac fermions in the presence
of vector gauge-like disorder.

\section*{\acknowledgements}

We would like to thank B. Altshuler, H. Carruzzo, E. Fradkin, M. Janssen,
M. Kardar, Y. B. Kim, P. A. Lee, and Z. Q. Wang
for sharing their insights.
This work was supported by NSF grants  DMR-9411574 (XGW) and
DMR-9400334 (CCC).
CM acknowledges a fellowship {}from the Swiss Nationalfonds and XGW
acknowledges the support {}from A.P. Sloan Foundation.

\appendix

\section{U(1/1) Suguwara construction}
\label{sec:U(1/1) Suguwara construction}

\subsubsection{Semi-direct algebra}

We assume the formal construction \cite{Sugawara 1968}
\begin{equation}
T(z)\  =\ \kappa^{\ }_1
\left(
Jj\ +\
\eta\bar\eta\ -\
\bar\eta\eta
\right)(z)
\ +\
\kappa^{\ }_2\left(JJ\right)(z)
\label{formalenergymomentumtensor}
\end{equation}
for the holomorphic component of the
energy momentum tensor out of the holomorphic components
of the currents $J,j,\eta,\bar\eta$
obeying a $U(1/1)$ Kac-Moody algebra. Above definition is meaningless
until what is meant by the products of currents at the same point is
specified. To this end we define the Laurent series
\begin{equation}
I^a(z)\ =\
\sum^{\ }_{n\in\ {\bf {\rm Z\hskip -0.2 true cm Z}} }
z^{-n-1}I^a_n,
\quad
I^a\in\{J,j,\eta,\bar\eta\}.
\end{equation}
The algebra obeyed by the normal modes $I^a_n$ of the currents $I^a(z)$ is
\begin{equation}
[I^a_m\ , \ I^b_n\ ]^{\ }_{\pm}\equiv\
k^{ab}m\delta^{\ }_{m,-n}
\ +\
f^{abc} I^c_{m+n},
\end{equation}
where the bracket $[A,B]^{\ }_+$ is the anticommutator
$\{A,B\}$ for $A,B$ two fermionic operators, and the bracket
$[A,B]^{\ }_-$ is the commutator $[A,B]$ for all other cases.
According to Eq. (\ref{u11currentope}), the
only non-vanishing central extensions $k^{ab}$ and
structure constants $f^{abc}$
are
\begin{eqnarray}
&&
k^{12}= k^{21}
      = 2k      ,\quad
k^{22}= k^{\ }_j,\quad
k^{34}=-k^{43}
      =-k       ,\quad
\nonumber\\
&&
f^{233}=-f^{323}
       =-2,\quad
f^{244}=-f^{424}
       =+2,\quad
f^{341}=f^{431}
       =+1.
\label{kadandfabc}
\end{eqnarray}
{}From the algebraic point of view, $k$ and $k^{\ }_j$ can be taken to be
arbitrary complex numbers.
The 12 commutators and 4 anticommutators for the generators of the $U(1/1)$
Kac-Moody algebra are explicitly given by
\begin{mathletters}
\begin{eqnarray}
&&
[J^{\ }_m,J^{\ }_n]=0,\quad
[J^{\ }_m,j^{\ }_n]=2km\delta^{\ }_{m,-n},\quad
[J^{\ }_m,\eta^{\ }_n]=0,\quad
[J^{\ }_m,\bar\eta^{\ }_n]=0,
\label{Jcommutators}\\
&&
[j^{\ }_m,J^{\ }_n]=2km\delta^{\ }_{m,-n},\;
[j^{\ }_m,j^{\ }_n]=k^{\ }_jm\delta^{\ }_{m,-n},\;
[j^{\ }_m,\eta^{\ }_n]=-2\eta^{\ }_{m+n},\;
[j^{\ }_m,\bar\eta^{\ }_n]=+2\bar\eta^{\ }_{m+n},
\label{jcommutators}\\
&&
[\eta^{\ }_m,J^{\ }_n]=0,\quad
[\eta^{\ }_m,j^{\ }_n]=+2\eta^{\ }_{m+n},\quad
\{\eta^{\ }_m,\eta^{\ }_n\}=0,\quad
\{\eta^{\ }_m,\bar\eta^{\ }_n\}=-km\delta^{\ }_{m,-n}+J^{\ }_{m+n},
\label{etacommutators}\\
&&
[\bar\eta^{\ }_m,J^{\ }_n]=0,\;
[\bar\eta^{\ }_m,j^{\ }_n]=-2\bar\eta^{\ }_{m+n},\;
\{\bar\eta^{\ }_m,\eta^{\ }_n\}=+km\delta^{\ }_{m,-n}+J^{\ }_{m+n},\;
\{\bar\eta^{\ }_m,\bar\eta^{\ }_n\}=0.
\label{baretacommutators}
\end{eqnarray}
\end{mathletters}

Having defined the Kac-Moody algebra $U(1/1)$,
we introduce the convolutions $L^{\ }_m$ of the currents through
\begin{eqnarray}
L^{\ }_m\ &=&\
\kappa^{\ }_1
\sum_{m'\in\ {\bf {\rm Z\hskip -0.2 true cm Z}} }
:
\left(
J^{\ }_{m+m'}j^{\ }_{-m'}+
\eta^{\ }_{m+m'}\bar\eta^{\ }_{-m'}-
\bar\eta^{\ }_{m+m'}\eta^{\ }_{-m'}
\right)
:
+
\kappa^{\ }_2
\sum_{m'\in\ {\bf {\rm Z\hskip -0.2 true cm Z}} }
:
J^{\ }_{m+m'}J^{\ }_{-m'}
:
\nonumber\\
&=&\
\kappa^{\ }_1
\sum_{m+m'<0}
\left(
J^{\ }_{m+m'}j^{\ }_{-m'}+
\eta^{\ }_{m+m'}\bar\eta^{\ }_{-m'}-
\bar\eta^{\ }_{m+m'}\eta^{\ }_{-m'}
\right)
+
\kappa^{\ }_2
\sum_{m+m'<0}
J^{\ }_{m+m'}J^{\ }_{-m'}
\nonumber\\
&+&\
\kappa^{\ }_1
\sum_{m+m'\geq0}
\left(
j^{\ }_{-m'}J^{\ }_{m+m'}-
\bar\eta^{\ }_{-m'}\eta^{\ }_{m+m'}+
\eta^{\ }_{-m'}\bar\eta^{\ }_{m+m'}
\right)
+
\kappa^{\ }_2
\sum_{m+m'\geq0}
J^{\ }_{-m'}J^{\ }_{m+m'}
{}.
\end{eqnarray}
The parameters $\kappa^{\ }_{1,2}$ are complex and to be determined below.
Here, the normal ordering is defined by the second line on the right-hand-side.
We now show that the property
\begin{equation}
[L^{\ }_m,I^a_n]\ =\ -n\ I^a_{m+n},
\quad\forall m,n\in\ {\bf {\rm Z\hskip -0.2 true cm Z}},
\label{semidirectalgebra}
\end{equation}
holds if and only if
\begin{equation}
\kappa^{\ }_1\ =\ {1\over2k},
\quad
\kappa^{\ }_2\ =\ {4-k^{\ }_j\over4k}\ \kappa^{\ }_1.
\label{kappas}
\end{equation}
Notice that it follows {}from Eq. (\ref{semidirectalgebra})
that all $L^{\ }_m$ are left invariant by the generators
$I^a_0$, $a=1,\cdots,4$.

\noindent
{\bf PROOF:}
Let $m,n$ be arbitrary integers. Decompose $L^{\ }_m$ into three
contributions:
\begin{mathletters}
\begin{eqnarray}
L^{1 }_m\ &=&\
\kappa^{\ }_1
\sum_{m'\in\ {\bf {\rm Z\hskip -0.2 true cm Z}} }
:
J^{\ }_{m+m'}j^{\ }_{-m'}
:,
\\
L^{2 }_m\ &=&\
\kappa^{\ }_1
\sum_{m'\in\ {\bf {\rm Z\hskip -0.2 true cm Z}} }
:
\left(
\eta^{\ }_{m+m'}\bar\eta^{\ }_{-m'}-
\bar\eta^{\ }_{m+m'}\eta^{\ }_{-m'}
\right)
:,
\\
L^{3 }_m\ &=&\
\kappa^{\ }_2
\sum_{m'\in\ {\bf {\rm Z\hskip -0.2 true cm Z}} }
:
J^{\ }_{m+m'}J^{\ }_{-m'}
:.
\end{eqnarray}
The case $a=1$ is straightforward. It yields a constraint on
$\kappa^{\ }_1$. Indeed, due to Eq. (\ref{Jcommutators}),
$[L^{\ }_m,J^{\ }_n]=[L^{1 }_m,J^{\ }_n]$. Now,
\end{mathletters}
\begin{eqnarray}
[L^{1 }_m,J^{\ }_n]&=&\
\kappa^{\ }_1
\sum_{m+m'<0}
\left(
J^{\ }_{m+m'}[j^{\ }_{-m'},J^{\ }_n]
+
[J^{\ }_{m+m'},J^{\ }_n]j^{\ }_{-m'}
\right)
\nonumber\\
&+&\
\kappa^{\ }_1
\sum_{m+m'\geq0}
\left(
j^{\ }_{-m'}[J^{\ }_{m+m'},J^{\ }_n]
+
[j^{\ }_{-m'},J^{\ }_n]J^{\ }_{m+m'}
\right)
\nonumber\\
&=&\
\kappa^{\ }_1
\sum_{m+m'<0}
J^{\ }_{m+m'}2k(-m')\delta^{\ }_{-m',-n}
\ +\
\kappa^{\ }_1
\sum_{m+m'\geq0}
2k(-m')\delta^{\ }_{-m',-n}J^{\ }_{m+m'}
\nonumber\\
&=&\
-n\ 2k\kappa^{\ }_1\ J^{\ }_{m+n}.
\end{eqnarray}

The case $a=2$ is more involved. It yields a constraint on $\kappa^{\ }_1$
and $\kappa^{\ }_2$.
The first contribution is
\begin{eqnarray}
[L^{1 }_m,j^{\ }_n]\ &=&\
\kappa^{\ }_1
\sum_{m+m'<0}
\left(
J^{\ }_{m+m'}[j^{\ }_{-m'},j^{\ }_n]
+
[J^{\ }_{m+m'},j^{\ }_n]j^{\ }_{-m'}
\right)
\nonumber\\
&+&\
\kappa^{\ }_1
\sum_{m+m'\geq0}
\left(
j^{\ }_{-m'}[J^{\ }_{m+m'},j^{\ }_n]
+
[j^{\ }_{-m'},j^{\ }_n]J^{\ }_{m+m'}
\right)
\nonumber\\
&=&\
\kappa^{\ }_1
\sum_{m+m'<0}
\left(
J^{\ }_{m+m'}k^{\ }_j(-m')\delta^{\ }_{-m',-n}
\ +\
2k(m+m')\delta^{\ }_{m+m',-n}j^{\ }_{-m'}
\right)
\nonumber\\
&+&\
\kappa^{\ }_1
\sum_{m+m'\geq0}
\left(
j^{\ }_{-m'}2k(m+m')\delta^{\ }_{m+m',-n}
\ +\
k^{\ }_j(-m')\delta^{\ }_{-m',-n}J^{\ }_{m+m'}
\right)
\nonumber\\
&=&\
-n2k\kappa^{\ }_1j^{\ }_{m+n}
\ -\
nk^{\ }_j\kappa^{\ }_1 J^{\ }_{m+n}.
\end{eqnarray}
The second contribution is
\begin{eqnarray}
&&
[L^{2 }_m,j^{\ }_n] =
\nonumber\\
&&+
\kappa^{\ }_1
\sum_{m+m'<0}
\left(
\eta^{\ }_{m+m'}[\bar\eta^{\ }_{-m'},j^{\ }_n]
+
[\eta^{\ }_{m+m'},j^{\ }_n]\bar\eta^{\ }_{-m'}
-
\bar\eta^{\ }_{m+m'}[\eta^{\ }_{-m'},j^{\ }_n]
-
[\bar\eta^{\ }_{m+m'},j^{\ }_n]\eta^{\ }_{-m'}
\right)
\nonumber\\
&&-
\kappa^{\ }_1
\sum_{m+m'\geq0}
\left(
\bar\eta^{\ }_{-m'}[\eta^{\ }_{m+m'},j^{\ }_n]
+
[\bar\eta^{\ }_{-m'},j^{\ }_n]\eta^{\ }_{m+m'}
-
\eta^{\ }_{-m'}[\bar\eta^{\ }_{m+m'},j^{\ }_n]
-
[\eta^{\ }_{-m'},j^{\ }_n]\bar\eta^{\ }_{m+m'}
\right)
=
\nonumber\\
&&
-2\kappa^{\ }_1
\left[
\sum_{m+m'<0}
\left(
\eta^{\ }_{m+m'}\bar\eta^{\ }_{n-m'}
+
\bar\eta^{\ }_{m+m'}\eta^{\ }_{n-m'}
\right)
-\!\!\!
\sum_{m+m'<0}
\left(
\eta^{\ }_{m+n+m'}\bar\eta^{\ }_{-m'}
+
\bar\eta^{\ }_{m+n+m'}\eta^{\ }_{-m'}
\right)
\right]
\nonumber\\
&&+
2\kappa^{\ }_1
\left[
\sum_{m+m'\geq0}
\left(
\bar\eta^{\ }_{n-m'}\eta^{\ }_{m+m'}
+
\eta^{\ }_{n-m'}\bar\eta^{\ }_{m+m'}
\right)
-\!\!\!
\sum_{m+m'\geq0}
\left(
\bar\eta^{\ }_{-m'}\eta^{\ }_{m+n+m'}
+
\eta^{\ }_{-m'}\bar\eta^{\ }_{m+n+m'}
\right)
\right].
\end{eqnarray}
It is very important to account for the fermionic sign in the normal ordering
of $L^{2 }_m$. This allows us to reduce the infinite sum to a finite sum.
To see this, introduce the integer
\begin{equation}
I\ =\ \min\{-m,-m+n\},
\end{equation}
and change variables in the second sums in the square brackets:
\begin{eqnarray}
&&
[L^2_m,j^{\ }_n]=
\nonumber\\
&&
-2\kappa^{\ }_1
\left[
\sum_{m'<-m}
\left(
\eta^{\ }_{m+m'}\bar\eta^{\ }_{n-m'}
+
\bar\eta^{\ }_{m+m'}\eta^{\ }_{n-m'}
\right)
-\!\!
\sum_{m''<-m+n}
\left(
\eta^{\ }_{m+m''}\bar\eta^{\ }_{n-m''}
+
\bar\eta^{\ }_{m+m''}\eta^{\ }_{n-m''}
\right)
\right]
\nonumber\\
&&+
2\kappa^{\ }_1
\left[
\sum_{m'\geq-m}
\left(
\bar\eta^{\ }_{n-m'}\eta^{\ }_{m+m'}
+
\eta^{\ }_{n-m'}\bar\eta^{\ }_{m+m'}
\right)
-\!\!
\sum_{m''\geq-m+n}
\left(
\bar\eta^{\ }_{n-m''}\eta^{\ }_{m+m''}
+
\eta^{\ }_{n-m'}\bar\eta^{\ }_{m+m''}
\right)
\right]=
\nonumber\\
&&
2\kappa^{\ }_1{n\over|n|}
\left[
\sum_{m'=I}^{I+|n|-1}
\left(
\eta^{\ }_{m+m'}\bar\eta^{\ }_{n-m'}
+
\bar\eta^{\ }_{m+m'}\eta^{\ }_{n-m'}
\right)
+\!\!
\sum_{m'=I}^{I+|n|-1}
\left(
\bar\eta^{\ }_{n-m'}\eta^{\ }_{m+m'}
+
\eta^{\ }_{n-m'}\bar\eta^{\ }_{m+m'}
\right)
\right].
\end{eqnarray}
We are thus left with the finite sum
\begin{eqnarray}
[L^2_m,j^{\ }_n]\ &=&\
2\kappa^{\ }_1{n\over|n|}
\sum_{m'=I}^{I+|n|-1}
\left(
\{\eta^{\ }_{m+m'},\bar\eta^{\ }_{n-m'}\}
+
\{\bar\eta^{\ }_{m+m'},\eta^{\ }_{n-m'}\}
\right)
\nonumber\\
&=&\
2\kappa^{\ }_1{n\over|n|}
\sum_{m'=I}^{I+|n|-1}
\Big(
-k(m+m')\delta^{\ }_{m+m',-n+m'}+J^{\ }_{m+n}
\nonumber\\
&+&\
k(m+m')\delta^{\ }_{m+m',-n+m'}+J^{\ }_{m+n}
\Big)
\nonumber\\
&=&
+n4\kappa^{\ }_1J^{\ }_{m+n}.
\end{eqnarray}
Finally, the third contribution is
\begin{eqnarray}
[L^{3 }_m,j^{\ }_n]\ &=&\
\kappa^{\ }_2
\sum_{m+m'<0}
\left(
J^{\ }_{m+m'}[J^{\ }_{-m'},j^{\ }_n]
+
[J^{\ }_{m+m'},j^{\ }_n]J^{\ }_{-m'}
\right)
\nonumber\\
&+&\
\kappa^{\ }_2
\sum_{m+m'\geq0}
\left(
J^{\ }_{-m'}[J^{\ }_{m+m'},j^{\ }_n]
+
[J^{\ }_{-m'},j^{\ }_n]J^{\ }_{m+m'}
\right)
\nonumber\\
&=&\
2k\kappa^{\ }_2
\sum_{m+m'<0}
\left(
J^{\ }_{m+m'}(-m')\delta^{\ }_{-m',-n}
\ +\
(m+m')\delta^{\ }_{m+m',-n}J^{\ }_{-m'}
\right)
\nonumber\\
&+&\
\kappa^{\ }_2
\sum_{m+m'\geq0}
\left(
J^{\ }_{-m'}(m+m')\delta^{\ }_{m+m',-n}
\ +\
(-m')\delta^{\ }_{-m',-n}J^{\ }_{m+m'}
\right)
\nonumber\\
&=&\
-n4k\kappa^{\ }_2 J^{\ }_{m+n}.
\end{eqnarray}
Collecting all three contributions yield
\begin{equation}
[L^{\ }_m,j^{\ }_n]\ =\
-n\ 2k\kappa^{\ }_1\ j^{\ }_{m+n}
\ -\
n\left(k^{\ }_j\kappa^{\ }_1-4\kappa^{\ }_1+4k\kappa^{\ }_2\right)\
J^{\ }_{m+n}.
\end{equation}

Without loss of generality, we only consider the case $a=3$, the case $a=4$
being similar. We show that $\kappa^{\ }_1$ and $\kappa^{\ }_2$ are
not overdetermined. We only need the commutators of $\eta^{\ }_n$ with
$L^{1 }_m$ and $L^{2 }_m$ since $J^{\ }_m$ commutes with $\eta^{\ }_n$.
The first commutator is
\begin{eqnarray}
[L^{1 }_m,\eta^{\ }_n]&=&\
\kappa^{\ }_1
\sum_{m+m'<0}
\left(
J^{\ }_{m+m'}[j^{\ }_{-m'},\eta^{\ }_n]
+
[J^{\ }_{m+m'},\eta^{\ }_n]j^{\ }_{-m'}
\right)
\nonumber\\
&+&\
\kappa^{\ }_1
\sum_{m+m'\geq0}
\left(
j^{\ }_{-m'}[J^{\ }_{m+m'},\eta^{\ }_n]
+
[j^{\ }_{-m'},\eta^{\ }_n]J^{\ }_{m+m'}
\right)
\nonumber\\
&=&\
-2\kappa^{\ }_1
\left[
\sum_{m+m'<0}
J^{\ }_{m+m'}\eta^{\ }_{n-m'}
\ +\
\sum_{m+m'\geq0}
\eta^{\ }_{n-m'}J^{\ }_{m+m'}
\right]
\nonumber\\
&=&\
-2\kappa^{\ }_1\sum_{m'\in\ {\bf {\rm Z\hskip -0.2 true cm Z}} }
J^{\ }_{m+m'}\eta^{\ }_{n-m'}.
\end{eqnarray}
The second commutator is
\begin{eqnarray}
&&
[L^{2 }_m,\eta^{\ }_n] =
\nonumber\\
&&+
\kappa^{\ }_1
\sum_{m+m'<0}
\left(
\eta^{\ }_{m+m'}\{\bar\eta^{\ }_{-m'},\eta^{\ }_n\}
-
\{\eta^{\ }_{m+m'},\eta^{\ }_n\}\bar\eta^{\ }_{-m'}
-
\bar\eta^{\ }_{m+m'}\{\eta^{\ }_{-m'},\eta^{\ }_n\}
+
\{\bar\eta^{\ }_{m+m'},\eta^{\ }_n\}\eta^{\ }_{-m'}
\right)
\nonumber\\
&&-
\kappa^{\ }_1
\sum_{m+m'\geq0}
\left(
\bar\eta^{\ }_{-m'}\{\eta^{\ }_{m+m'},\eta^{\ }_n\}
-
\{\bar\eta^{\ }_{-m'},\eta^{\ }_n\}\eta^{\ }_{m+m'}
-
\eta^{\ }_{-m'}\{\bar\eta^{\ }_{m+m'},\eta^{\ }_n\}
+
\{\eta^{\ }_{-m'},\eta^{\ }_n\}\bar\eta^{\ }_{m+m'}
\right)
=
\nonumber\\
&&+
\kappa^{\ }_1\!\!
\sum_{m+m'<0}
\left(
\eta^{\ }_{m+m'}k(-m')\delta^{\ }_{-m',-n}
+
\eta^{\ }_{m+m'}J^{\ }_{n-m'}
+
k(m+m')\delta^{\ }_{m+m',-n}\eta^{\ }_{-m'}
+
J^{\ }_{m+n+m'}\eta^{\ }_{-m'}
\right)
\nonumber\\
&&+
\kappa^{\ }_1\!\!
\sum_{m+m'\geq0}
\left(
k(-m')\delta^{\ }_{-m',-n}\eta^{\ }_{m+m'}
+J^{\ }_{n-m'}\eta^{\ }_{m+m'}
+\eta^{\ }_{-m'}k(m+m')\delta^{\ }_{m+m',-n}
+\eta^{\ }_{-m'}J{\ }_{m+n}
\right)
=
\nonumber\\
&&
-n\ 2k\kappa^{\ }_1\ \eta^{\ }_{m+n}
\ +\
2\kappa^{\ }_1\sum_{m'\in\ {\bf {\rm Z\hskip -0.2 true cm Z}} }
J^{\ }_{m+m'}\eta^{\ }_{n-m'}.
\end{eqnarray}
Collecting all terms, we find
\begin{equation}
[L^{\ }_m,\eta^{\ }_n]\ =\
-n\ 2k\kappa^{\ }_1\ \eta^{\ }_{m+n}.
\end{equation}
This concludes the proof.

\subsubsection{Virasoro algebra}


We now show that the set
$\{L^{\ }_m\}^{\ }_{m\in\ {\bf {\rm Z\hskip -0.2 true cm Z}}}$
generates a Virasoro algebra with vanishing central charge,
provided $\kappa^{\ }_1$ and $\kappa^{\ }_2$ are chosen as in
Eq. (\ref{kappas}):
\begin{equation}
[L^{\ }_m,L^{\ }_n]\ =\
(m-n)L^{\ }_{m+n}.
\end{equation}

\noindent{\bf PROOF}: Let $m,n$ be arbitrary integers.
The first commutator to be considered is
\begin{eqnarray}
[L^{\ }_m,L^{1 }_n]\ &=&\
\kappa^{\ }_1\sum_{n+n'<0}
[L^{\ }_m,J^{\ }_{n+n'}j^{\ }_{-n'}]
+
\kappa^{\ }_1\sum_{n+n'\geq0}
[L^{\ }_m,j^{\ }_{-n'}J^{\ }_{n+n'}]
\nonumber\\
&=&\
\kappa^{\ }_1
\sum_{n+n'<0}
n'J^{\ }_{n+n'}j^{\ }_{m-n'}
-
\kappa^{\ }_1
\sum_{n+n'<0}
(n+n')J^{\ }_{m+n+n'}j^{\ }_{-n'}
\nonumber\\
&+&\
\kappa^{\ }_1
\sum_{n+n'\geq0}
n'j^{\ }_{m-n'}J^{\ }_{n+n'}
-
\kappa^{\ }_1
\sum_{n+n'\geq0}
(n+n')j^{\ }_{-n'}J^{\ }_{m+n+n'}.
\end{eqnarray}
With the help of the change of variables $n'=m+n''$ in the first
summation on the last two lines and with the introduction of
\begin{equation}
I=\min\{-n,-n-m\},
\end{equation}
we can rewrite the right-hand-side as
\begin{eqnarray}
[L^{\ }_m,L^{1 }_n]\ &=&\
\kappa^{\ }_1\!\!\!
\sum_{n+m+n''<0}\!\!\!
(m+n'')J^{\ }_{n+m+n''}j^{\ }_{-n''}
-\kappa^{\ }_1\!\!\!
\sum_{m+n+n'<0}\!\!\!
(n+n')J^{\ }_{m+n+n'}j^{\ }_{-n'}
\nonumber\\
&+&\
\kappa^{\ }_1\!\!\!
\sum_{n+m+n''\geq0}\!\!\!
(m+n'')j^{\ }_{-n''}J^{\ }_{n+m+n''}
-
\kappa^{\ }_1\!\!\!
\sum_{m+n+n'\geq0}\!\!\!
(n+n')j^{\ }_{-n'}J^{\ }_{m+n+n'}
\nonumber\\
&-&\
\kappa^{\ }_1{m\over|m|}
\sum_{n'=I}^{I+|m|-1}
(n+n')J^{\ }_{m+n+n'}j^{\ }_{-n'}
+
\kappa^{\ }_1{m\over|m|}
\sum_{n'=I}^{I+|m|-1}
(n+n')j^{\ }_{-n'}J^{\ }_{m+n+n'}
\nonumber\\
&=&\
(m-n)\ L^{1 }_{m+n}
\ -\
\kappa^{\ }_1\
{m\over|m|}\sum_{n'=I}^{I+|m|-1}
(n+n')[J^{\ }_{m+n+n'},j^{\ }_{-n'}]
\nonumber\\
&=&\
(m-n)\ L^{1 }_{m+n}
\ -\
2k\kappa^{\ }_1\
{m\over|m|}\sum_{n'=I}^{I+|m|-1}
(n+n')(m+n+n')\delta^{\ }_{m+n+n',n'}
\nonumber\\
&=&\
(m-n)\ L^{1 }_{m+n}
\ -\
2k\kappa^{\ }_1\
{m\over|m|}\delta^{\ }_{m,-n}\sum_{n'=I}^{I+|m|-1}
(n+n')n'.
\end{eqnarray}
Because the commutators of $J^{\ }_p$ with $J^{\ }_q$ vanishes for any
integers $p$ and $q$, one immediately obtains {}from $[L^{\ }_m,L^{1 }_n]$
that
\begin{equation}
[L^{\ }_m,L^{3 }_n]\ =\
(m-n)\ L^{3 }_{m+n}.
\end{equation}
The proof that the Virasoro central charge vanishes is then established by
showing that
\begin{equation}
[L^{\ }_m,L^{2 }_n]\ =\
(m-n)\ L^{2 }_{m+n}
\ +\
2k\kappa^{\ }_1\
{m\over|m|}\delta^{\ }_{m,-n}\sum_{n'=I}^{I+|m|-1}
(n+n')n'.
\label{comll2}
\end{equation}

To show this, start {}from
\begin{eqnarray}
&&
[L^{\ }_m,L^{2 }_n]=
\kappa^{\ }_1\!\!\!\sum^{\ }_{n+n'<0}
[
L^{\ }_m,
\left(
    \eta^{\ }_{n+n'}\bar\eta^{\ }_{-n'}-
\bar\eta^{\ }_{n+n'}    \eta^{\ }_{-n'}
\right)
]
-
\kappa^{\ }_1\!\!\!\sum^{\ }_{n+n'\geq0}
[
L^{\ }_m,
\left(
\bar\eta^{\ }_{-n'}    \eta^{\ }_{n+n'}-
    \eta^{\ }_{-n'}\bar\eta^{\ }_{n+n'}
\right)
]=
\nonumber\\
&&
\kappa^{\ }_1\!\!\!\sum^{\ }_{n+n'<0}
n'
\left(
    \eta^{\ }_{n+n'}\bar\eta^{\ }_{m-n'}-
\bar\eta^{\ }_{n+n'}    \eta^{\ }_{m-n'}
\right)
-\kappa^{\ }_1\!\!\!\sum^{\ }_{n+n'<0}
(n+n')
\left(
    \eta^{\ }_{m+n+n'}\bar\eta^{\ }_{-n'}-
\bar\eta^{\ }_{m+n+n'}    \eta^{\ }_{-n'}
\right)
\nonumber\\
&&-
\kappa^{\ }_1\!\!\!\sum^{\ }_{n+n'\geq0}\!\!\!
n'
\left(
\bar\eta^{\ }_{m-n'}    \eta^{\ }_{n+n'}-
    \eta^{\ }_{m-n'}\bar\eta^{\ }_{n+n'}
\right)
+
\kappa^{\ }_1\!\!\!\sum^{\ }_{n+n'\geq0}\!\!\!
(n+n')
\left(
\bar\eta^{\ }_{-n'}    \eta^{\ }_{m+n+n'}-
    \eta^{\ }_{-n'}\bar\eta^{\ }_{m+n+n'}
\right).
\end{eqnarray}
With the help of the substitution $n'=m+n''$ in the first summation on
the last two lines, we can again extract $(m-n)L^{2 }_{m+n}$. The reminder is
a finite sum over anticommutators due to the fermionic normal ordering:
\begin{equation}
[L^{\ }_m,L^{2 }_n]=
(m-n)L^{2 }_{m+n}-
\kappa^{\ }_1{m\over|m|}\!
\sum_{n'=I}^{I+|m|-1}\!\!
(n+n')
\left(
\{    \eta^{\ }_{m+n+n'},\bar\eta^{\ }_{-n'}\}
-
\{\bar\eta^{\ }_{m+n+n'},    \eta^{\ }_{-n'}\}
\right).\nonumber
\end{equation}
The term $J^{\ }_{m+n}$ cancels {}from the difference of the two
anticommutators
and we are left with twice the contribution {}from the central charge $k$ to
the
OPE between $\eta$ and $\bar\eta$ as given in Eq. (\ref{comll2}).
This concludes the proof.

\section{Lagrangian realization of $U(1/1)\times U(1/1)$ current algebra}
\label{sec:Lagrangian realization of U(1/1) x U(1/1) current algebra}

In this appendix, we show that the $U(1/1)\times U(1/1)$ Kac-Moody
algebra of Eq. (\ref{kadandfabc}) or Eq. (\ref{u11currentope})
with real central charges $(k,k_j)$ is associative.
To prove associativity, it is sufficient to construct a
Lagrangian realization of a free field theory,
whose currents obey the algebra of Eq. (\ref{u11currentope})
with $(k,k_j)=(1,k_j)$, $k_j\geq0$. Associativity for arbitrary values
of $(k,k_j)$, where $k\neq0$, then follows on purely algebraic
grounds by independent rescalings of the currents $(J,j,\eta,\bar\eta)$.
We have already seen a Lagrangian realization of the current algebra
with central charges $(k,k_j)=(1,0)$ in Eq. (\ref{freeu11}).
We now construct a Lagrangian realization of
$U(1/1)\times U(1/1)$ Kac-Moody algebra with central charges
$(k,k_j)=(1,k_j)$, $k_j>0$.
(For a similar trick, see Ref. \cite{Zamolodchikov 1985})
Consider the partition function
\begin{mathletters}
\label{u11k=1kj}
\begin{eqnarray}
&&
{\cal Z}^{(1,k^{\ }_j)}_0\ =\
\int {\cal D}[\phi]
\int {\cal D}[\psi   ^{\dag}_{\pm},\psi   ^{\ }_{\pm}]
\int {\cal D}[\varphi^{\dag}_{\pm},\varphi^{\ }_{\pm}]
\  e^{-\int^{+\infty}_{-\infty} dx d\tau\
({\cal L}^{\ }_0+{\cal L}^{\ }_{k^{\ }_j})},
\\
&&
{\cal L}^{\ }_0\ =\
{1\over\pi}
\left(
\psi   ^{\dag}_+ \partial_{\bar z} \psi   ^{\ }_+\ +\
\varphi^{\dag}_+ \partial_{\bar z} \varphi^{\ }_+\ +\
\psi   ^{\dag}_- \partial_{     z} \psi   ^{\ }_-\ +\
\varphi^{\dag}_- \partial_{     z} \varphi^{\ }_-
\right),
\\
&&
{\cal L}^{\ }_{k^{\ }_j}\ =\
{1\over2}\sqrt{1\over8\pi}
\left(\partial^{\ }_{     z}\phi\right)
\left(\partial^{\ }_{\bar z}\phi\right).
\end{eqnarray}
Define the holomorphic and antiholomorphic components of the free real
scalar field $\phi(z,\bar z)$ by
(compare with Eq. (\ref{scalingdimprimaryfieldsphi1}))
\end{mathletters}
\begin{equation}
\phi(z,\bar z)=
\phi(z)+\bar\phi(\bar z),\quad
\langle\phi(z)\phi(0)\rangle=-2\ln z,\quad
\langle\bar\phi(\bar z)\bar\phi(0)\rangle=-2\ln \bar z,\quad
\langle\phi(z)\bar\phi(0)\rangle=0.
\end{equation}
The real scalar field $\phi$ is taken to commute with all spinors.
Without loss of generality, we only consider the holomorphic sector.
Define the holomorphic component of the currents
\begin{mathletters}
\begin{eqnarray}
&&
J(z)\ =\
\left(\psi^{\dag}_+\psi^{\ }_+\right)(z)
\ +\
\left(\varphi^{\dag}_+\varphi^{\ }_+\right)(z),
\\
&&
j(z)\ =\
\left(\psi^{\dag}_+\psi^{\ }_+\right)(z)
\ -\
\left(\varphi^{\dag}_+\varphi^{\ }_+\right)(z)
\ +\
{\rm i}\sqrt{{k^{\ }_j\over2}}\ \partial^{\ }_z\phi(z),
\\
&&
\eta(z)\ =\ \left(\varphi^{\dag}_+\psi^{\ }_+\right)(z),
\\
&&
\bar\eta(z)\ =\ \left(\psi^{\dag}_+\varphi^{\ }_+\right)(z).
\end{eqnarray}
The structure constants $\{f^{\ }_{abc}\}$ for the OPE of these currents
are those of $U(1/1)$, since $\phi$ commutes with the spinors.
By assuming that currents do not acquire vacuum expectation values,
the only change in the algebra takes place in the OPE
$j(z)j(0)={k^{\ }_j\over z^2}$.
Note the similarity between Eq. (\ref{u11k=1kj})
and the effective critical theory defined by
$S^{(1,0)}_{\rm cr}$ in Eq. (\ref{s(10)cr}).

\end{mathletters}

\begin{table}
\caption
{
$U(1/1)\times U(1/1)$ charges for the generators
$J^{\ }_{z},j^{\ }_{z},J^{\ }_{\bar z},j^{\ }_{\bar z}$
}
\begin{tabular}{ccccccccccccccc}
 &
$\psi^{\dag}_+$ &
$\varphi^{\dag}_+$ &
$\psi^{\ }_+$ &
$\varphi^{\ }_+$ &
$f^{\dag}_{qz}$ &
$f^{\ }_{qz}$ &
$\psi^{\dag}_{pz}$ &
$\varphi^{\dag}_{pz}$ &
$\psi^{\ }_{pz}$ &
$\varphi^{\ }_{pz}$ &
$\psi^{\dag}_{pqz}$ &
$\varphi^{\dag}_{pqz}$ &
$\psi^{\ }_{pqz}$ &
$\varphi^{\ }_{pqz}$
\\
\hline
\hfill&\hfill&\hfill&\hfill&\hfill&\hfill&\hfill&\hfill&
\hfill&\hfill&\hfill&\hfill&\hfill&\hfill\\
$J^{\ }_z$ &
$1$& $1$& $-1$& $-1$&
$0$&$0$&
$p$& $p$& $-p$& $-p$&
$p$& $p$& $-p$& $-p$
\\ \hline
\hfill&\hfill&\hfill&\hfill&\hfill&\hfill&\hfill&\hfill&
\hfill&\hfill&\hfill&\hfill&\hfill&\hfill\\
$j^{\ }_z$ &
$1$& $-1$& $-1$& $1$&
$q$& $-q$&
$2-p$& $-p$& $p-2$& $p$&
$2-p+q$& $q-p$& $p-2-q$& $p-q$
\\ \hline
\hfill&\hfill&\hfill&\hfill&\hfill&\hfill&\hfill&\hfill&
\hfill&\hfill&\hfill&\hfill&\hfill&\hfill\\
$J^{\ }_{\bar z}$ &
0&0&0&0&0&0&0&0&0&0&0&0&0&0
\\ \hline
\hfill&\hfill&\hfill&\hfill&\hfill&\hfill&\hfill&\hfill&
\hfill&\hfill&\hfill&\hfill&\hfill&\hfill\\
$j^{\ }_{\bar z}$ &
0&0&0&0&0&0&0&0&0&0&0&0&0&0
\\
\end{tabular}
\label{table1}
\end{table}

\end{document}